\documentclass[aps,pre,twocolumn,floatfix,superscriptaddress]{revtex4-2}
\usepackage{hyperref}
\usepackage{xcolor}

\hypersetup{%
  breaklinks=true,
  colorlinks=true,
}

\usepackage{calligra}

\usepackage{amsmath}  
\usepackage{amsfonts} 
\usepackage{graphicx} 
\usepackage{gensymb}
\usepackage{physics}
\usepackage[version=4]{mhchem}

\usepackage{soul}
\usepackage{cases}

\newcommand{\RR}{\mathbb{R}}
\newcommand{\ZZ}{\mathbb{Z}}
\newcommand{\I}{\text{I}}
\newcommand{\II}{\text{II}}
\DeclareMathOperator{\inv}{inv}
\DeclareMathOperator{\cl}{cl}
 
\DeclareMathOperator{\HH}{H}

\DeclareMathAlphabet{\mathcalligra}{T1}{calligra}{m}{n} 
\DeclareFontShape{T1}{calligra}{m}{n}{<->s*[2.2]callig15}{} 
\newcommand{\scr}{k}

\usepackage[resetlabels,labeled]{multibib}
\newcites{Sup}{Supplementary References}

\begin{document}

\title{Spiking at the edge}

\author{Colin Scheibner}
\affiliation{Department of Physics and The James Franck Institute, The University of Chicago, Chicago, IL 60637}

\author{Hillel Ori}
\affiliation{Department of Chemistry and Chemical Biology, Harvard University, Cambridge, MA 02138}

\author{Adam E. Cohen}
\affiliation{Department of Chemistry and Chemical Biology, Harvard University, Cambridge, MA 02138}
\affiliation{Department of Physics, Harvard University, Cambridge, MA 02138}

\author{Vincenzo Vitelli}
\affiliation{Department of Physics and The James Franck Institute, The University of Chicago, Chicago, IL 60637}
\affiliation{Kadanoff Center for Theoretical Physics, The University of Chicago, Chicago, IL 60637}

\begin{abstract}
Excitable media, ranging from bioelectric tissues and chemical oscillators to forest fires and competing populations, are nonlinear, spatially extended systems capable of spiking. 
Most investigations of excitable media consider situations where the amplifying and suppressing forces necessary for spiking coexist at every point in space. 
In this case, spiking requires a fine-tuned ratio between local amplification and suppression strengths.
But, in Nature and engineered systems, these forces can be segregated in space, forming structures like interfaces and boundaries. 
Here, we show how boundaries can generate and protect spiking if the reacting components can spread out:
even arbitrarily weak diffusion can cause spiking at the edge between two non-excitable media.
This edge spiking is a robust phenomenon that can occur even if the ratio between amplification and suppression does not allow spiking when the two sides are homogeneously mixed. 
We analytically derive a spiking phase diagram that depends on two parameters: (i) the ratio between the system size and the characteristic diffusive length-scale, and (ii) the ratio between the amplification and suppression strengths.   
Our analysis explains recent experimental observations of action potentials at the interface between two non-excitable bioelectric tissues.
Beyond electrophysiology, we highlight how 
edge spiking emerges in predator-prey dynamics and in oscillating chemical reactions. 
Our findings provide a theoretical blueprint for a class of interfacial excitations in reaction-diffusion systems, 
with potential implications for spatially controlled chemical reactions, nonlinear waveguides and neuromorphic computation, as well as spiking instabilities, such as cardiac arrhythmias, that naturally occur in heterogeneous biological media.  
\end{abstract}

\maketitle

A spike is a large nonlinear excursion in a dynamical system followed by a time of latency known as the refractory period. 
Protecting the ability to spike is crucial for a wide range of biological functions, from cardiac pacemaking~\cite{Winfree1994Electrical, Bers2002Cardiac,tusscher2004model,Cheng1993Calcium,Stern1992Theory,Witkowski1998Spatiotemporal} to neural information processing~\cite{rieke1997spikes}, while in other contexts, such as forest fires~\cite{Drossel1992Self} and disease outbreaks~\cite{Anderson1979Population,Murray1986spatial,Anderson1981Population,Rohani1999Opposite},
spiking must be avoided. 
In a spatially extended medium, the ability to spike gives rise to distinctive spatiotemporal patterns~\cite{Kondo2010Reaction,Bourret1969Fungal,Loose2008Spatial,Tompkins2014Testing,Rotermund1990imaging,Anisotropy1995Science,Vinson1997Control,Fuseya2021Nanoscale,Tan2020Topological,Winfree1994Persistent} appearing in processes ranging from morphogenesis~\cite{Turing1952chemical, Nakamasu2009Interactions,kondo1995reaction,Newman1979Dynamics,Mitchell2022Visceral,Wigbers2021hierarchy,Talia2022Waves,Vergassola2018Mitotic,Lechleiter1991Spiral,McFadden2018excitable,Change2013mitotic} to spiral waves observed in electrograms of the heart~\cite{Davidenko1992Stationary,Gray1998Spatial}.  
While analytical studies have revealed important features of excitable media whose properties are spatially homogeneous~\cite{Halatek2018Rethinking,Cross1993Pattern,Kim2001Controlling,Brauns2020Phase,Alonso2003Taming},
less is understood about abrupt heterogeneities such as sample edges or interfaces~\cite{McNamara2020bioelectrical,Eckstein2020Experimental,VidalHenriquez2017Convective,Ni1995location,Bub2002Spiral,Mainen1996influence, Wigbers2020pattern,Brauns2021Diffusive,Bub200Spiral,Agladze1994Rotating,Staddon2022Pulsatile}. As is often the case with wave mechanics, edges and interfaces can have properties that differ qualitatively from those of the bulk medium~\cite{Murugan2017Topologically,Kane2014Topologically,Hasan2010Colloquium,Shankar2022Topological,chen2014Nonlinear,Mao2018Maxwell,Huber2016Topological}.

\begin{figure*}[ht!]
    \centering
    \includegraphics[width=0.8\textwidth]{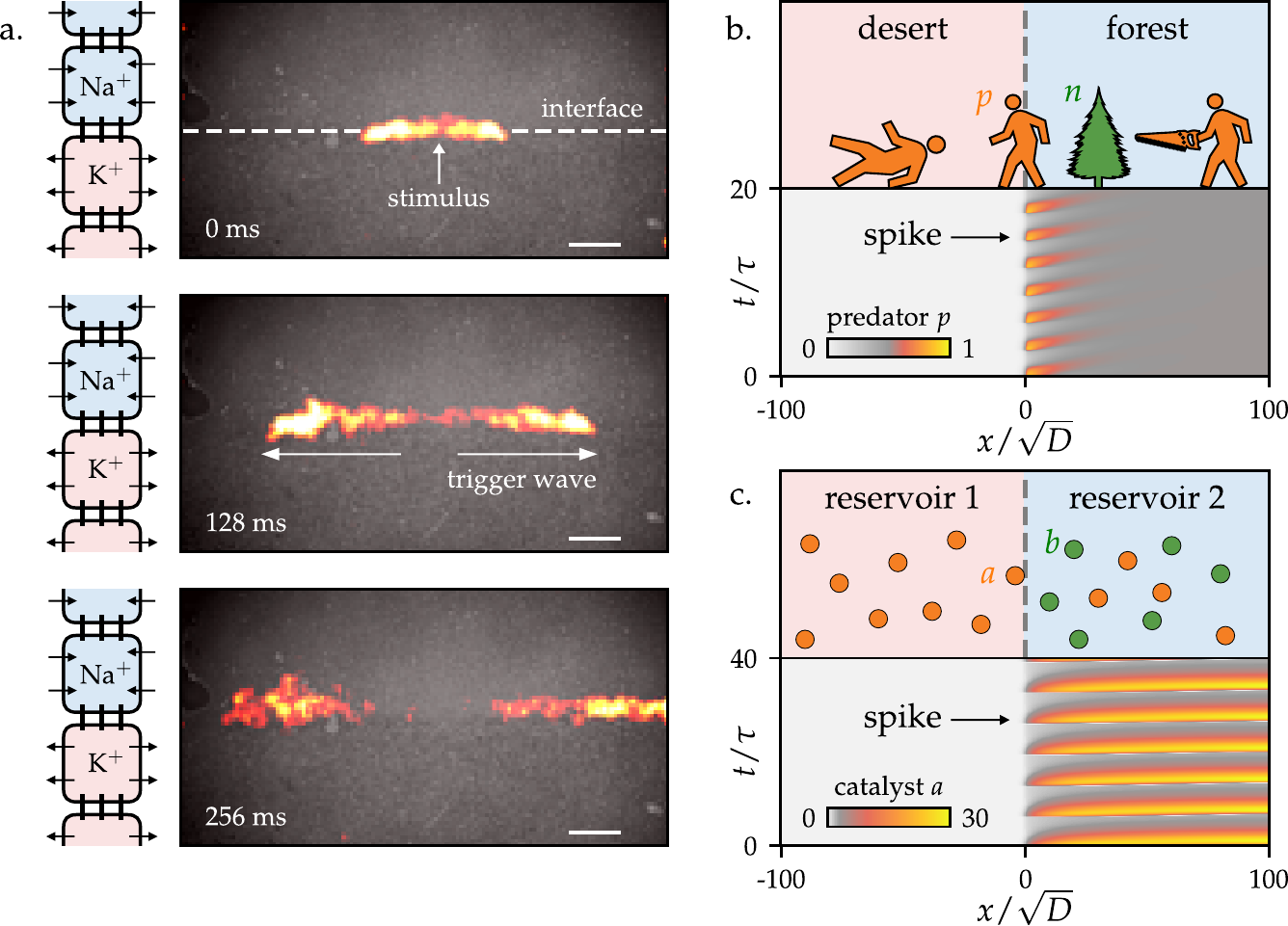}
    \caption{{\bf Edge spiking in electrophysiology, population dynamics, and chemistry.}~{\bf a.}~Experiments from Ref.~\cite{Ori2023Observation} in which an action potential propagates along a tissue interface, as revealed by a voltage sensitive red dye. See Supplementary Video 1. Scale bar 1 mm. The left column shows a schematic vertical cross-section of the interface: the top tissue features sodium ion channels (inward arrows), while the bottom tissue features potassium ion channels (outward arrows). The vertical lines represent gap junction coupling between the cells.
    {\bf b.}~A fast diffusing predator (lumberjacks) and relatively immobile prey (trees) are described by an interfacial Lotka-volterra model [Eqs.~(\ref{eq:lumber1}-\ref{eq:lumber2})].  A kymograph generated by the model reveals spikes in the lumberjack population generated at the interface. {\bf c.} An interface between two chemical reservoirs, neither of which are capable of oscillating, is described by Eqs.~(\ref{eq:ybulk}-\ref{eq:zbulk}).~A kymograph of the fast, mobile catalyst $a$ reveals repeated spikes generated at the interface.   
    }
    \label{fig:overview}
\end{figure*}

\begin{figure*}
    \centering
    \includegraphics[width=0.98\textwidth]{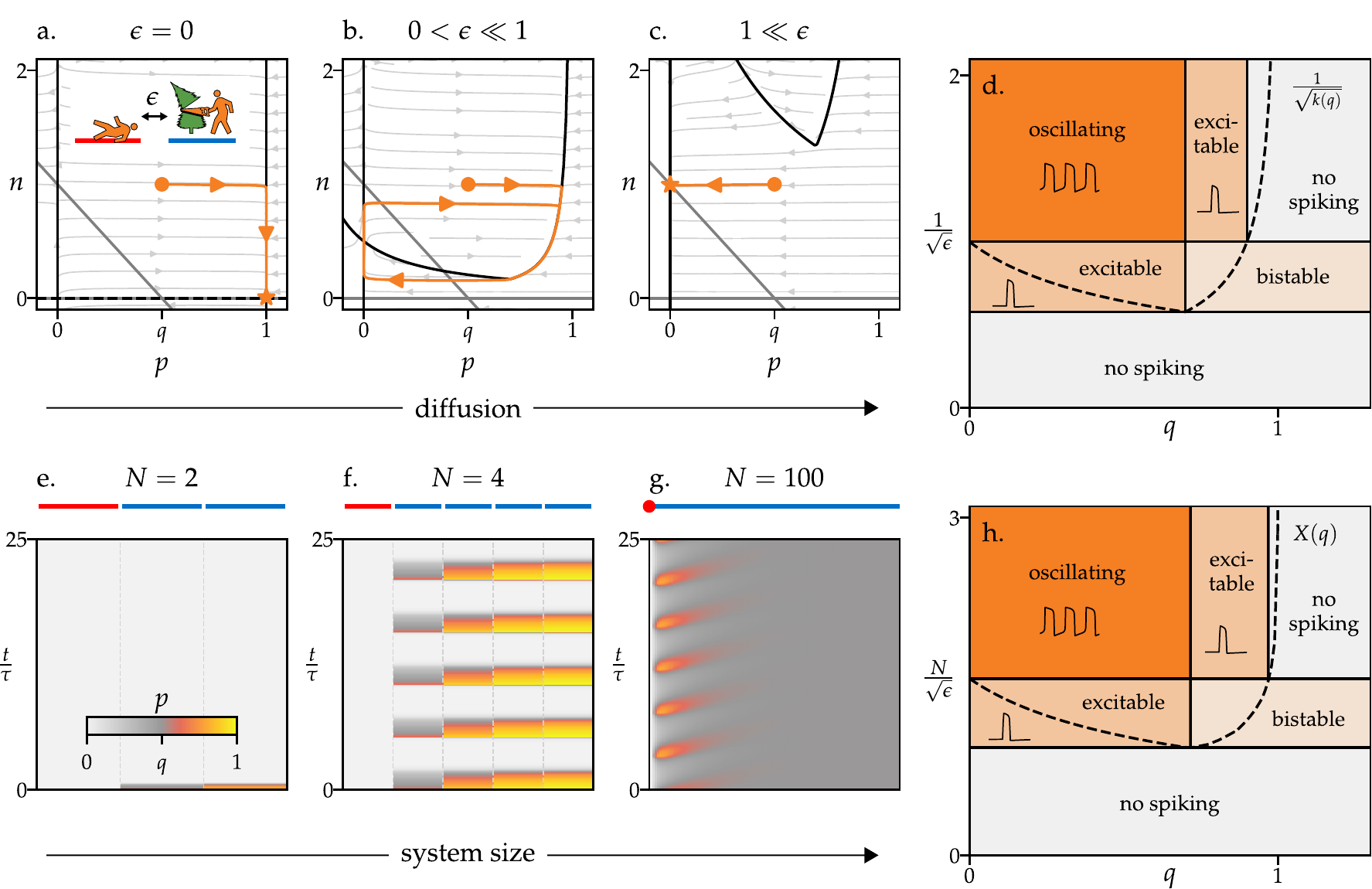}
    \caption{ {\bf Spikes induced by weak diffusion and large system size.}~{\bf a.}~(inset)~A predator-prey system with a desert (red) and a forest (blue) described by Eqs.~(\ref{eq:pred1}-\ref{eq:pred2}). In the forest, the predators (lumberjacks, $p$) consume the prey (trees, $n$) with predation rate $1/q$. The predators cross from the forest to the desert (and subsequently perish) with hopping rate $\epsilon$. 
    {\bf a-c.}~Three phase portraits, for $\epsilon=0$, $0<\epsilon \ll 1 $, and $1 \ll \epsilon$, illustrate the role of diffusion across a boundary:  
    For $\epsilon=0$ (a), the lumberjacks reach their carrying capacity and the trees go extinct. For $\epsilon \gg 1$ (c), the lumberjacks go extinct and the trees reach their carrying capacity. Spikes can only occur in the intermediate range  $0<\epsilon \ll 1 $ (b), in which the effective death rate due to hopping is present but not overpowering. Orange curves are example trajectories. The $\dot p=0$ and $\dot n=0$ nullclines are denoted by black and grey lines, respectively.
    {\bf d.}~A phase diagram for Eqs.~(\ref{eq:pred1}-\ref{eq:pred2}) summarizes the possible behaviors: if the lumberjack hopping rate $\epsilon$ is too large,  the lumberjack population cannot spike. The phase boundaries are determined by the consumption nonlinearity $k(q)$ in Eqs.~(\ref{eq:pred1}-\ref{eq:pred2}). 
 {\bf e-g.}~A chain of $N$ forests (blue lines) are coupled to a desert (red line) by a large hopping rate $\epsilon = 4$. Kymographs for systems with $N=2$, $N=4$, and $N=100$ exemplify a distinctive transition: oscillation onset is driven by increasing system size, even as $\epsilon$ is held constant. (See Methods~\S\ref{sec:pop} for simulation details.)
     {\bf h.}~A phase diagram for Eqs.~(\ref{eq:lumber1}-\ref{eq:lumber2}), applicable for $N \gg 1$, reveals a crucial distinction between the spatially and non-spatially extended systems: the vertical axis in (h) features $N/\sqrt{\epsilon}$, implying that spiking occurs for a much larger range of $\epsilon$ in the spatially extended limit. 
     The curve $X(q)$ determines the locations of the phase boundaries and is given in Eq.~(\ref{eq:xdef}). 
    }
    \label{fig:coupling}
\end{figure*}

For instance, 
Fig.~\ref{fig:overview}a shows a recent experiment in which 
human embryonic kidney (HEK293) cells were genetically modified to express either sodium (Na\textsubscript{V}1.5) or potassium (K\textsubscript{ir}2.1) channels~\cite{Ori2023Observation}. 
Usually, 
a cell containing both potassium and sodium channels 
spikes via the following mechanism, which is representative of excitable systems: 
The potassium channels favor a low membrane potential while the sodium channels favor a high membrane potential. 
Given a suitably large voltage stimulation, the membrane potential (a fast variable) spikes upward towards the value set by the sodium channels.
 The sodium channels then gradually shut due to open-state inactivation (a slow variable), causing the membrane potential to fall towards the value set by the potassium channels.  The sodium channels then take some time to recover their strength (the refractory period). 
Because the competition between the two channels is essential, neither sodium nor potassium channels alone are sufficient for an individual cell to spike. Furthermore, even when both channels coexist in a single cell, spikes only occur when they have the appropriate ratio of open-state conductances (i.e.\ channel strengths).

\begin{figure*}
    \centering
    \includegraphics[width=0.9\textwidth]{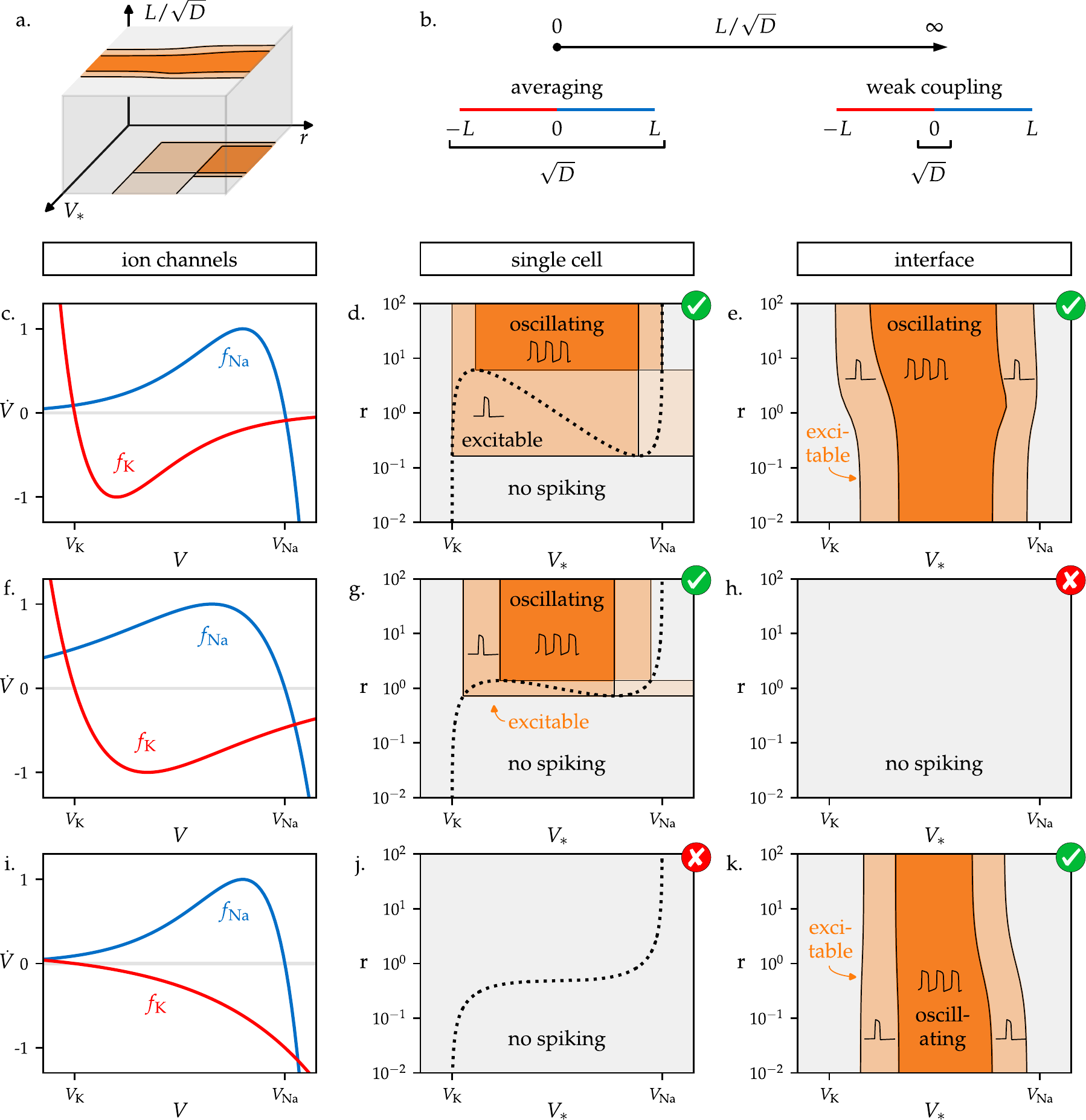}
    \caption{ {\bf A spiking interface is more than the sum of its parts.}~{\bf a.}~A spiking phase diagram is shown for the bioelectric interface in Eqs.~(\ref{eq:big1}-\ref{eq:big2}). Here, $L/\sqrt{D}$ is the ratio of the system size to the diffusion strength and $r$ and $V_*$ appear in $h_\infty(V)= r \, \Theta (V_*-V)$. The parameter $r$ is the ratio of the amplifier (sodium channel) strength to the suppressor (potassium channel) strength.~{\bf b.} At small $L/\sqrt{D}$, diffusion forces the membrane potential to be approximately constant across the entire tissue, creating an effective single cell with both ion channels. For large $L/\sqrt{D}$, the coupling is weak, so the dynamics are spatially heterogeneous.  
    {\bf c-k.}~A table comparing the $L/\sqrt{D} \to 0$ (single cell) and the $L/\sqrt{D} \to \infty$ (interfacial) limits. The left column (c, f, i) shows three examples of voltage-current curves for potassium (red) and sodium (blue) ion channels. Their reversal potentials are denoted $V_{\ce{K}}$ and $V_{\ce{Na}}$, respectively. The center column (d, g, j) contains phase diagrams for $L/\sqrt{D} \to 0$, corresponding to a single cell with both ion channels. The phase boundaries are determined by the dashed line $h_\text{eq} = - f_{\ce{K}}/f_{\ce{Na}}$. In order of brightness, the regions are oscillating, excitable, bistable, and no spiking. If $h_\text{eq}$ is monotonic, then the single cell does not exhibit spikes. The rightmost column (e, h, k) contains phase diagrams in the limit $L/\sqrt{D} \to \infty$, describing interfacial dynamics.  
     Row 1 (c-e) shows an example of ion channels in which both the single cell and the interfacial phase diagrams exhibit spiking, but boundaries differ. Row 2 (f-h) shows an example of ion channels for which the single-cell phase diagram contains spiking while the interfacial phase diagram does not. 
     Row 3 (i-k) shows an example of ion channels for which the interface exhibits spiking, but no ratio $r$ of the amplifier and suppressor give rise to spiking in a single cell. The green checks and red crosses indicate phase diagrams that do or do not contain spiking, respectively.  
    }
    \label{fig:phase}
\end{figure*}

Something visually striking happens when two distinct and non-excitable tissues (composed of the two cell types) are placed in contact and weakly coupled by gap junctions, which allow voltage diffusion. When stimulated at the interface, a voltage spike (i.e.\ an action potential) emerges and robustly propagates along the interface, see Fig.~\ref{fig:overview}a and Supplementary Video 1. 
Crucially, these interfacial spikes persist for a much wider range of open-state conductances than for a single cell~\cite{Ori2023Observation}. 
This observation suggests that spikes generated at an interface may have a distinct, and possibly more robust, dynamical origin than those in a homogenized system.  
Here, we reveal the underlying dynamical mechanism behind this phenomenon and demonstrate that it is not limited to electrophysiology. For instance, we provide examples from 
population dynamics (Fig.~\ref{fig:overview}b) in which a fast, mobile predator (lumberjacks) consume a slow, sedentary prey (trees) while diffusing across an environmental (forest-desert) boundary; and from chemical reaction networks in which a fast catalyst diffuses between two chemically distinct reservoirs (Fig.~\ref{fig:overview}c). 
In all these examples, the interfacial spiking does not result from merely superimposing the two halves. 
In fact, coupling the two halves too strongly can destroy spiking.

The basic notion of an edge spike involves two distinct processes: transport across two domains and transport within the domains themselves. To illustrate the former, consider a two-compartment model of predator-prey dynamics shown in Fig.~\ref{fig:coupling}a (inset). The model features a population $p$ of lumberjacks (the predators) that consumes a population $n$ of trees (the prey). 
The rightmost compartment, the forest (blue bar), acts as a lumberjack amplifier in which tree consumption elevates the lumberjack population. By contrast, the desert (red bar), is an infinitely strong suppressor in which any lumberjack that enters dies instantly. 
Lumberjacks from the forest wander into the adjacent desert  
with a hopping rate $\epsilon$.
The populations evolve according to the following Lotka-Volterra equation:
\begin{align}
    \dot p =& - \epsilon  \, p+ n \, p \, \scr(p)  \label{eq:pred1} \\
    \dot n =& \frac{n} \tau \qty( 1-n -  \frac{p}{q} ) \label{eq:pred2}
\end{align}
where $1/q$ is the predation rate, $\tau \gg 1$  is a long time scale implying that the tree population changes slowly, and $\scr(p)$ is a nonlinearity that encodes a lumberjack carrying capacity. A normalization has been chosen so that all variables in Eq.~(\ref{eq:pred1}-\ref{eq:pred2}) are dimensionless and the carrying capacities of the lumberjacks and trees are set to 1, see Methods~\S\ref{sec:pop}. 
Here, the lumberjack population plays the same role as the cell-membrane potential in the electrophysiology experiment (a fast, diffusing variable), the tree population corresponds to the gating variable of the sodium channels (an immobile, slow variable), while the desert and the forest correspond to cells with potassium and sodium channels, respectively. 

In this model, the ability to spike depends sensitively on the hopping rate $\epsilon$.  
If $\epsilon=0$ (a), the two halves are decoupled and the lumberjack population cannot spike: the lumberjacks will quickly return to their carrying capacity regardless of the perturbation.  
However, when $\epsilon$ is small but nonzero (b), the dynamics change dramatically: The lumberjacks can now spike because the motion into the desert depletes the lumberjack population when trees are sparse and tree consumption overpowers diffusion when trees are abundant. Crucially, though, when $\epsilon$ becomes too large (c), the desert and forest become well mixed, and the lumberjack population cannot spike because the suppressor (desert) is infinitely strong. In the two-compartment model described by Eq.~(\ref{eq:lumber1}-\ref{eq:lumber2}), the hopping rate $\epsilon$ 
 can be reinterpreted as an effective suppression strength: even though the desert itself is infinitely strong, the finite entrance rate attenuates its effect.  
 The phase diagram in Fig.~\ref{fig:coupling}d illustrates a basic mechanism: an amplifier and a suppressor need to be suitably well balanced for spikes to occur---attenuating a strong suppressor through weak diffusion across an interface helps achieve this balance.

Yet, this simplified model lacks a basic feature: the forest itself can be spatially extended.   
In Fig. \ref{fig:coupling}e-g the desert is now connected to a chain of $N$ compartments comprising the forest, each of which is coupled to its neighbors by a hopping rate $\epsilon$. The size of the forest dramatically influences the dynamics. When $N=2$ and $\epsilon =4$, the lumberjacks rapidly go extinct in all compartments (e). 
Yet, when $N=4$, the lumberjack population not only begins to survive, but undergoes large oscillations (f). 
A window into the relationship between $N$ and $\epsilon$ can be obtained in the large $N$ limit (g,h). In this limit, the dynamics can be described by a continuum reaction-diffusion equation 
\begin{align}
   \dot p = & D \nabla^2 p + n \, p \, \scr(p) \label{eq:lumber1} \\
   \dot n = & \frac{n}\tau \qty( 1-n - \frac{p}{q} ) \label{eq:lumber2}
\end{align}
where $L=Nd$ is the system size and $d$ is the lattice spacing~\footnote{Notice that Eqs.~(\ref{eq:lumber1}-\ref{eq:lumber2}) do not contain advective transport, which has also been shown to give rise oscillations near Dirichlet boundaries, for example in models of and experiments on \emph{Dictyostelium discoideum}~\cite{Eckstein2020Experimental,VidalHenriquez2017Convective}. 
}. The parameter $D = \epsilon \, d^2$  denotes the diffusion coefficient times the characteristic time scale used to nondimensionalize $\epsilon$ (see Methods~\S\ref{sec:pop}). 
The lumberjack population obeys the following boundary conditions: $\partial_x p=0$ at $x=L$ and, because of the infinitely strong desert, $p=0$ at $x=0$. 
The basic effect of spatial extent can be obtained by dimensional analysis: Only $\sqrt{D}$ and $L$ have units of length,  
so any change in qualitative behavior must depend on the dimensionless ratio $L/\sqrt{D} = N/\sqrt{\epsilon}$. 
Therefore, in the continuum, increasing $N$ is equivalent to decreasing $\epsilon$.
This collapse is physically consequential because the diffusion $D$ is an intrinsic property of the material while $L$ is an extrinsic property, so the two can often be tuned independently.  
Notably, by increasing $L$ a system can support spiking over a wider range of $D$~\footnote{For simplicity, in this example we are using the same hopping rate $\epsilon$ within the forest as between the forest and desert.  This distinction becomes irrelevant in the continuum limit (large $\epsilon$ and large $N$), because this subextensive heterogeneity is absorbed into the Dirichlet boundary condition at an edge or into the continuity requirements across an interface.}.

\begin{figure*}[t!]
    \centering
    \includegraphics[width=0.98\textwidth]{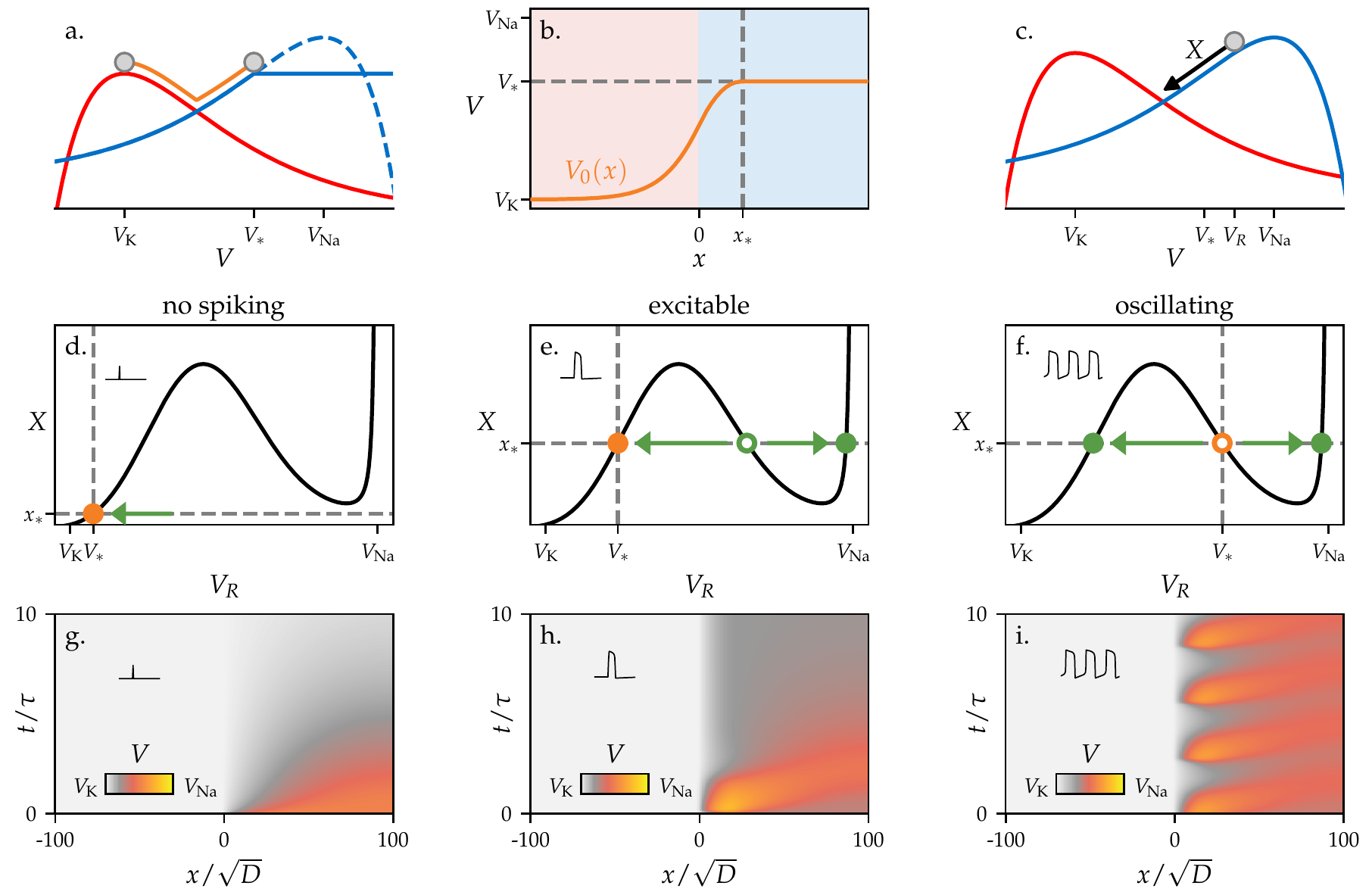}
    \caption{ 
    {\bf A geometric construction for interfacial spiking.}~{\bf a.}~Starting from Eqs.~(\ref{eq:big1}-\ref{eq:big2}), the antiderivatives of $f_{\ce K}(V)$ (solid red line) and  $h_\infty (V) f_{\ce{Na}} (V)$ (solid blue line) are visualized as hills. The dashed blue line is the antiderivative of $r f_{\ce{Na}} (V)$. To construct the stationary solution with no-flux boundary conditions, consider letting a ball roll from the top of one hill to the other (orange curve). 
    {\bf b.}~The stationary voltage solution $V_0(x)$ in space corresponds to the trajectory of a ball (in time) rolling across the potentials in (a). 
    {\bf c.}~The ability to spike is determined by the number and stability of critical points of $\Phi$ in Eq.~(\ref{eq:broad3}). To determine the critical points, we release a ball from a voltage $V_R$ and measure the ``time" $X$ it takes to reach the intersection.  {\bf d.}~We plot $X$ as a function of $V_R$, and three cases emerge. When $x_*$ intersects $X$ once, the interface is unable to spike. {\bf e.}~When $x_*$ intersects $X$ three times and $V_*$ corresponds to an increasing branch of $X$, the interface is excitable. {\bf f.}~When $x_*$ intersects 3 times and $V_*$ corresponds to the decreasing branch, the voltage at the interface oscillates.  
    {\bf g.-h.}~Kymographs illustrating no spiking, excitability, and oscillating at the interface. See the Supplementary Information for simulation details. 
    }
    \label{fig:isochrones}
\end{figure*}

The dynamics is even richer when the suppressor (e.g. the desert) is no longer infinitely strong. In this case, the Dirichlet boundary becomes an interface, and spikes can arise both in the limit $L /\sqrt{D} \to 0$ and $L/\sqrt{D} \to \infty$.
To illustrate this behavior, we consider a one-dimensional (1D) model for the electrophysiology experiment of Ref.~\cite{Ori2023Observation} which takes the form of an interfacial Fitzhugh-Nagumo equation~\cite{fitzhugh1961impluses}:
\begin{align}
    \dot V =& D \nabla^2 V +
    \begin{cases} 
    f_{\ce K}(V)  & x \in [-L, 0]  \\
    h \, f_{ \ce{Na}} (V) & x \in ( 0, L]
    \end{cases} \label{eq:big1}  \\
    \dot h =& \frac{h_\infty(V) -h}{\tau } \label{eq:big2}
\end{align}
Here, $x$ is the coordinate transverse to the interface (see Fig.~\ref{fig:overview}a), $V$ is the voltage, and $f_{\ce K} (V)$ and $f_{\ce{Na}} (V) $ capture the effect of the potassium and sodium channels, respectively. 
The sodium channels are modulated by a gating variable $h$ that slowly approaches the function $h_\infty(V)$ on a long time scale $\tau$. 
The term $D\nabla^2 V$ arises from direct cell-to-cell current flow via gap junctions. 
Like the predator prey system, a normalization is chosen such that the quantities $\sqrt{D}$ and $x$ have units of length, while all others are dimensionless (see Methods~\S\ref{sec:FN}). The system is modeled by no-flux boundary conditions at both ends, $ \eval{\partial_x V}_{\pm L}=0$, while the voltage $V$ and its first derivative $\partial_x V$ are required to be continuous across the interface.

The gating switch $h_\infty(V)$ is reasonably well approximated by a step function $h_\infty(V) = r\, \Theta(V_*-V)$, where $\Theta$ is a Heaviside step function and $V_*$ is a crossover voltage that turns off the sodium channels~\cite{tusscher2004model}. The parameter $r$ is the ratio of the open-state conductances of the sodium to the potassium ion channels. Therefore, $r$ can be interpreted as the relative strength of the amplifier (sodium) and suppressor (potassium).
When $r \ll 1$, the potassium ion channels are so strong that the interface effective becomes a Dirichlet boundary of the type 
 considered in the predator-prey system.  
 When $r \gg 1$, both sides of the interface are dynamic.

In Fig.~\ref{fig:phase}a-b, we sketch a three dimensional phase diagram spanned by the parameters $L/\sqrt{D}$, $V_*$, and $r$. When $L/\sqrt{D} \to 0$, diffusion forces the voltage to be approximately constant across the entire system, so we can think of the system as an effective single cell with both ion channels. 
By contrast, when $L/\sqrt{D} \to \infty$, the coupling is weak and the spatial heterogeneity plays a crucial role. 
To illustrate the independence of these two limits, in Fig.~\ref{fig:phase}c-k we consider three different realizations of $f_{\ce{Na}}$ and $f_{\ce{K}}$~\cite{Xu2020Computational,Payandeh2011crystal}.
For each realization, we show two cross-sections of the phase diagram: one for $L/\sqrt{D} \to 0$ and one for  $L /\sqrt{D} \to \infty$.   
Fig.~\ref{fig:phase}d-f shows an example of ion channels for which the effective single cell ($L/\sqrt{D} \to 0$) exhibits spikes but the weakly coupled interface ($L/\sqrt{D} \to \infty$) does not. 
Moreover, Fig.~\ref{fig:phase}g-i shows an example in which the interface exhibits spikes for all values of $r$, yet no ratio of the amplifier and suppressor give rise to spiking in a single cell.

\begin{figure*}[t]
    \centering
    \includegraphics[width=0.9\textwidth]{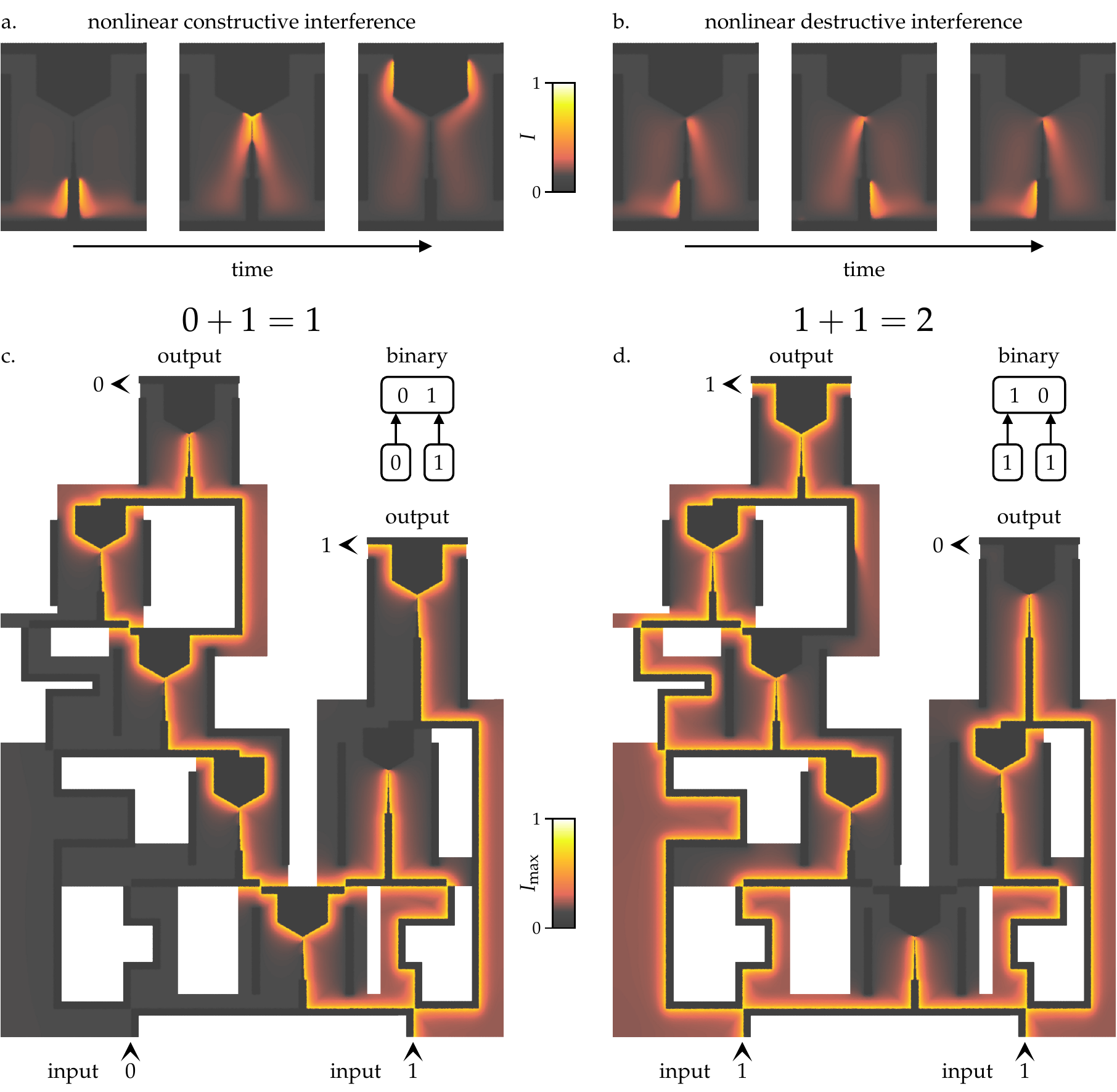}
    \caption{ {\bf Nonlinear waveguides from interfacial spiking.}~
    {\bf a-b.}~Two distinct non-spiking materials (light and dark grey) are patterned for form a four way junction of excitable interfaces. When two nonlinear wave trains arrive at a junction in-phase (a), they propagate through uninterrupted. When the two wave trains arrive out-of-phase, they annihilate at the junction.    
    The color corresponds to the intensity $I$ of the fast, diffusively coupled variable. {\bf c-d.}~A network of excitable interfaces acts as a binary half adder, which takes the sum of two 0 or 1 inputs. Here, the presence of a wave indicates the value 1 while the absence of a wave indicates the value 0. The color $I_\text{max}$ is the maximum value of $I$ over time when the network has reached steady state.  See Methods~\S\ref{sec:trigger}-\ref{sec:adder} and Supplementary Video 2 for more information, and the Supplementary Information for simulation details.
    }
    \label{fig:waveguide}
\end{figure*}

In both the interfacial and boundary systems, the presence of spikes is associated with topologically robust features of the underlying dynamical system governed by their respective reaction-diffusion equations. 
Both Eqs.~(\ref{eq:lumber1}-\ref{eq:lumber2}) and Eqs.~(\ref{eq:big1}-\ref{eq:big2}) take the form 
\begin{align}
    \dot A =& D \nabla^2 A + f(A, B, x) \label{eq:broad1} \\
    \dot B =& \frac{1}{\tau} g(A, B)  \label{eq:broad2}
\end{align}
where $A(x,t)$ is a fast field and $B(x,t)$ is a slow field.  
We will call $[A_0(x), B_0(x)]$ a stationary solution of Eqs.~(\ref{eq:broad1}-\ref{eq:broad2}) if they satisfy 
 $\dot A=\dot B=0$. Each stationary solution comes paired with a functional $\Phi$: 
\begin{align}
    \Phi[A] = \int D \qty( \partial_x A )^2 -U(A,x) \,  \dd x 
    \label{eq:broad3}
\end{align}
where $U(A,x) = \int_0^A f(A',B_0(x), x) \, \dd A'$. The meaning of $\Phi$ is as follows. If the system is prepared at the stationary solution and the variable $A$ is perturbed, then on short time scales $\dot A \approx - \delta \Phi/ \delta A$.

The number of stationary solutions and the critical points of their associated functionals 
encode the ability of a system to spike. For instance, suppose Eqs.~(\ref{eq:broad1}-\ref{eq:broad2}) permit only one stationary solution, $[A_0(x), B_0(x)]$, and the associated functional $\Phi$ only has one minimum [namely $A_0(x)$]. Then the system will not exhibit spikes because any perturbation to $A(x)$ quickly relaxes to $A_0(x)$. However, if $\Phi$ permits a second minimum $A_1(x)$ in addition to $A_0(x)$, then the system is excitable: suitable perturbations to $A$  will push the system into the basin of attraction of $A_1(x)$, and only on longer times ($t \propto \tau$), will the system return to $A_0(x)$. Oscillations (i.e.\ repeated spikes) occur when $A_0(x)$ itself is a saddle, rather than a minimum, of $\Phi$. 
The number of critical points and their unstable dimensions are topologically robust quantities: these integers are unchanged under sufficiently small, generic perturbations to Eqs.~(\ref{eq:broad1}-\ref{eq:broad2}).

For certain models, such as the electrophysiology equations~(\ref{eq:big1}-\ref{eq:big2}) in the experimentally relevant limit of $h_\infty(V) = r \Theta (V_*-V)$, the stationary solutions and associated critical points are captured by a relatively simple geometric construction. 
The stationary solution for the membrane potential $V_0(x)$ is constructed as follows: first draw potentials for $f_{\ce{K}}(V)$ (red) and $h_\infty(V) f_{\ce{Na}}(V)$ (blue) and align their maxima as shown in Fig.~\ref{fig:isochrones}a. 
Treating these as hills, let a ball roll from the top of one hill to the other. 
The trajectory of the ball in time corresponds to the voltage profile $V_0(x)$ in space (Fig.~\ref{fig:isochrones}b). 
As we show in Methods~\S\ref{sec:FN}, the existence of spiking at the interface is determined by an auxiliary function $X(V)$ defined in Fig.~\ref{fig:isochrones}c: 
Place the ball at an arbitrary voltage $V_R$ and let it roll down the blue hill.
The function $X(V_R)$ is the amount of time it takes for the ball to reach the intersection. 
Each solution to the equation $X(V_*)=X(V)$ constitutes a critical point of $\Phi(V)$. 
Whenever $X(V_*) = X(V)$ has multiple solutions, the system exhibits spikes.
As shown in Fig.~\ref{fig:isochrones}d-f, 
the precise form of the spikes (excitable vs oscillatory) depends on whether the solution $V_0(x)$ is stable (excitable) or unstable (oscillatory). Using homological techniques from Conley index theory~\cite{conley1983algebraic,Mischaikow2002Conley},  we show in the Methods that the decreasing branch of $X(V)$ must always be unstable, while the increasing branches are stable. 
The function $X(V)$ can be thought of as the high dimensional counterpart of the dashed lines, $h_\text{eq}$, in Figs.~\ref{fig:phase}d,g,j that determine the phase diagrams for a single cell.
An analogous function $X(q)$ demarcates the phase boundaries for the predator-prey diagram shown in Fig.~\ref{fig:coupling}h, see Methods~\S\ref{sec:boundary}.

So far, we have considered bulk media that alone cannot spike, but exhibit excitability or oscillations when a boundary or interface is introduced. Now we show that boundaries or interfaces can cause conversions between different modes of spiking.   
As illustrated in Fig.~\ref{fig:overview}c, we consider two chemical reservoirs separated by a semi-permeable membrane. The reaction in the right chamber ($x>0$) contains two catalysts with concentrations $a(x,t)$ and $b(x,t)$ that evolve according to the Oregonator model of the celebrated Belousov-Zhabotinsky reaction~\cite{Tyson1976Belousov}. We assume that the catalyst $a$ is free to diffuse across the interface, while the catalyst $b$ is relatively immobile. In the left reservoir, the catalyst $a$ is rapidly converted into a product that exits the reaction.   
Starting from a minimal chemical reaction network and applying the law of mass action (see Methods~\S\ref{sec:chem}), we derive the following dynamical equations:
\begin{align} 
\dot a =& D \nabla^2 a + \begin{cases} 
 - a & x < 0 \label{eq:ybulk} \\
2 m_1 b - a \, [b_\infty(a)  + m_2] & x > 0 
\end{cases}\\
\dot b =&  \frac{b_\infty(a) -b}{\tau } \label{eq:zbulk}
\end{align} 
where $m_1$ and $m_2$ are parameters set by internal rate constants, and
$b_\infty(a)$ is a monotonically decaying function given in Eq.~(\ref{eq:zinf}).
For sufficiently large $m_1$ and small $m_2$, neither of the reservoirs alone can oscillate.      
The kymograph in Fig.~\ref{fig:overview}c shows that allowing catalyst $a$ to diffuse between the two reservoirs creates spontaneous oscillations at the interface. 
However, unlike the previous examples (predator-prey and electrophysiology), the chamber on the right alone is excitable (though not oscillatory) even without the interface (see Method Fig.~\ref{fig:chemphase}).
The presence of excitability for $x>0$ changes a qualitative feature of the oscillations: the interfacial spikes are no longer spatially localized. Instead of dying off at large $x$ (as in Fig.~\ref{fig:overview}b), the spikes generated at the interface propagate at constant amplitude to the far away boundary (see Fig.~\ref{fig:overview}c). 
Oscillations at chemical interfaces have been reported previously, but they often rely on a distinct mechanism in which chemicals mix at the interface to reach locally suitable conditions for oscillations~\cite{Budroni2016spatially,Duzs2019Turing}.  
Interfacial spiking, for example using gels or other tailored chemistry~\cite{semenov2016autocatalytic,Yoshida2010Self,Rabai1989systematically,Testa2021Sustained}, may serve as a promising alternative technique for spatial control of chemical reactions because the two reservoirs can remain distinct indefinitely.

In two dimensions, an interface is a 1D line. If the interface is excitable, then the 1D line can host nonlinear waves called trigger waves, as illustrated by the bioelectric experiments in Fig.~\ref{fig:overview}a. The conditions for propagation as well as the unique wave speed of these trigger waves are discussed in Methods~\S\ref{sec:trigger}. 
Geometric primitives, such as curves, corners, and junctions,  can then be used to control the nonlinear wave propagation. For instance, Fig.~\ref{fig:waveguide}a-b shows a four-way junction formed by patterning two different materials (light and dark grey). In panel (a), two trigger waves approach the junction from below. Since the trigger waves are in phase they interfere constructively and pass through the junction. However, when the pulses are sent periodically with a phase lag (b), no pulse passes through due to overlap in their refractory periods. Since the trigger waves are nonlinear, constructive interference results in outgoing waves that have the same amplitude as the incoming waves (rather than twice the amplitude). 
This modification to the superposition principle can form the basis of more complex devices, such as those capable of computation~\cite{adamatzky2005reaction,Holley2011Logical,Toth1995Logic,Steinbock1995Navigating}. As an illustration, Figure~\ref{fig:waveguide}c-d and Supplementary Video 2 show a two-dimensional (2D) surface patterned by two materials obeying equations of the form of Eqs.~(\ref{eq:big1}-\ref{eq:big2}). The network of excitable interfaces forms an effective circuit that computes the sum of two binary numbers (see Methods~\S\ref{sec:adder} for additional minimal logic gates, such as AND, OR, and NOT gates). 
Since only diffusion is required at the boundary, interfacial excitability is potentially useful as a form of wave control that does not require electronics, additional materials, or the fabrication precision necessary to explicitly construct a narrow channel or wire.

Exciting possibilities await in systems with multiple fast degrees of freedom. 
In this case, the fast dynamics need not be gradient-like, and therefore may give rise to more complex interfacial effects like bursting, in which oscillations transiently turn on and off. Likewise, we envision extensions to three-dimensional systems in which the interface is a 2D surface.
In addition to engineered waveguides, interfacial spikes may be a useful tool for constructing models of biological functions, such as intracellular chemical signaling~\cite{Talia2022Waves,Change2013mitotic,Wigbers2021hierarchy,McFadden2018excitable}, as well as pathologies  
such as atrial fibrillation~\cite{McNamara2018Geometry}, in which erroneous pacemaking emerges at the boundary of the aortal and ventral heart tissues.

\clearpage

\setcounter{figure}{0}
\renewcommand{\thefigure}{E\arabic{figure}}
\renewcommand\figurename{Fig.}

\renewcommand{\theequation}{M\arabic{equation}}
\setcounter{equation}{0}
\renewcommand{\thesection}{M\arabic{section}}

\section*{Methods}

\subsection{Spike generation in fast-slow systems without spatial extent}
\label{sec:fsreview}

\begin{figure}
    \centering
    \includegraphics[width=\columnwidth]{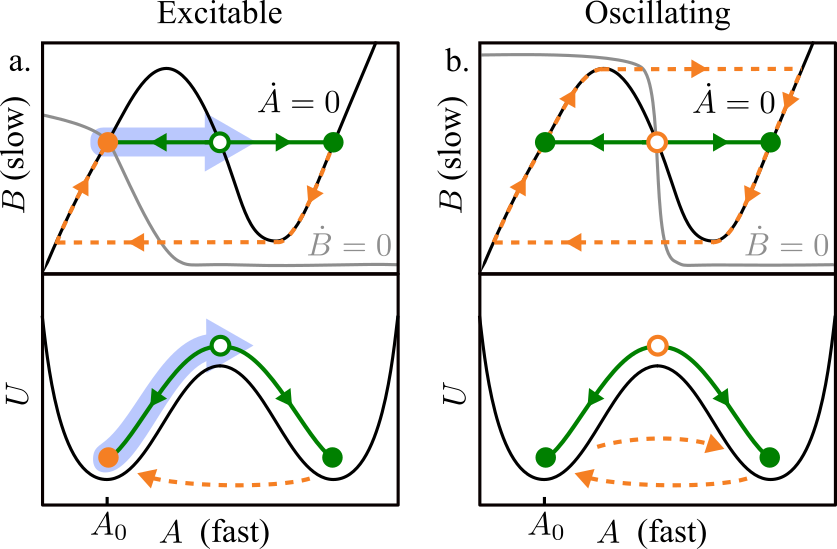}
    \caption{{\bf Spike generation in fast-slow systems without spatial extent.}~{\bf a.}~(top)~A phase portrait of an excitable fast-slow system. The black curve corresponds to $\dot A=0$, and the grey curve corresponds to $\dot B =0$. The light blue arrow represents an external perturbation. The solid green curves are the fast trajectory, and the dashed orange curve is the refractory period. The solid orange circle denotes the global fixed point. The open (closed) green circle represents an unstable (stable) fixed point of the fast dynamics. (bottom) On short time scales, the potential $U(A)$ governs the dynamics. The system is excitable since $U(A)$ has multiple minima. {\bf b.}~(top)~A phase portrait of a fast-slow system exhibiting oscillations. The dashed orange curve is a limit cycle. The open orange circle denotes an unstable global fixed point. The closed green circles are stable fixed points of the fast dynamics if initialized at the open orange circle. (bottom) On short time scales, the potential $U(A)$ governs the dynamics. The system is exhibits oscillations since since $U(A)$ has multiple minima and the orange circle (global fixed point) is not one of them.
    }
    \label{fig:theorySketch}
\end{figure}

Here we review examples of spike generation in fast-slow systems without spatial extent. We consider equations of the form
\begin{align}
    \dot A =& f(A,B) \label{eq:small1} \\
    \dot B =& \frac1{\tau}g(A,B) \label{eq:small2}
\end{align}
and we assume that $\tau \gg 1$, which implies that $A$ is a fast variable and $B$ is a slow variable. 
 Figure~\ref{fig:theorySketch}a (top) shows two curves known as nullclines, which are defined by $\dot A=0$ (black) and $\dot B=0$ (grey). The intersection of the nullclines (solid orange circle), denoted $(A_0, B_0)$, is a fixed point of Eqs.~(\ref{eq:small1}-\ref{eq:small2}). If an external stimulus (light blue arrow) pushes $A$ across a threshold value (open green circle), $A$ will evolve along the solid green line towards a high value (solid green circle) while $B$ remains approximately constant. Over a longer period of time, known as the refractory period, $A$ and $B$ will move along the dashed orange line back towards their rest position (solid orange circle). This is an example of an excitable system, in which the fast variable $A$ needs to be stimulated above a critical threshold in order to undergo a spike. Figure~\ref{fig:theorySketch}a (bottom) shows an equivalent description of the spike: when initially perturbed, $A$ will evolve according to $\dot A = -\partial_A U$, where $\partial_A U(A)=-f(A,B_0)$. From this perspective, the system is excitable because $U$ has a minimum (solid green circle) other than the one at $A_0$ (solid orange circle). 
 
 Figure~\ref{fig:theorySketch}b shows a similar example where the fixed point $(A_0, B_0)$ is unstable. Since the global fixed point is unstable, this system contains a limit cycle denoted by the dashed orange line. Such a system exhibits repeated spikes even in absence of external stimulation, which we refer to as oscillation or pacemaking. In the following sections, we use an analogous fast-slow decomposition in a high-dimensional setting to identify spiking in reaction-diffusion equations, where the potential $U$ is replaced by a functional $\Phi$ of the spatially extended fields.

\subsection{Phase diagram for spiking at a Dirichlet boundary}
\label{sec:boundary}

\begin{figure*}
    \centering
    \includegraphics[width=0.95 \textwidth]{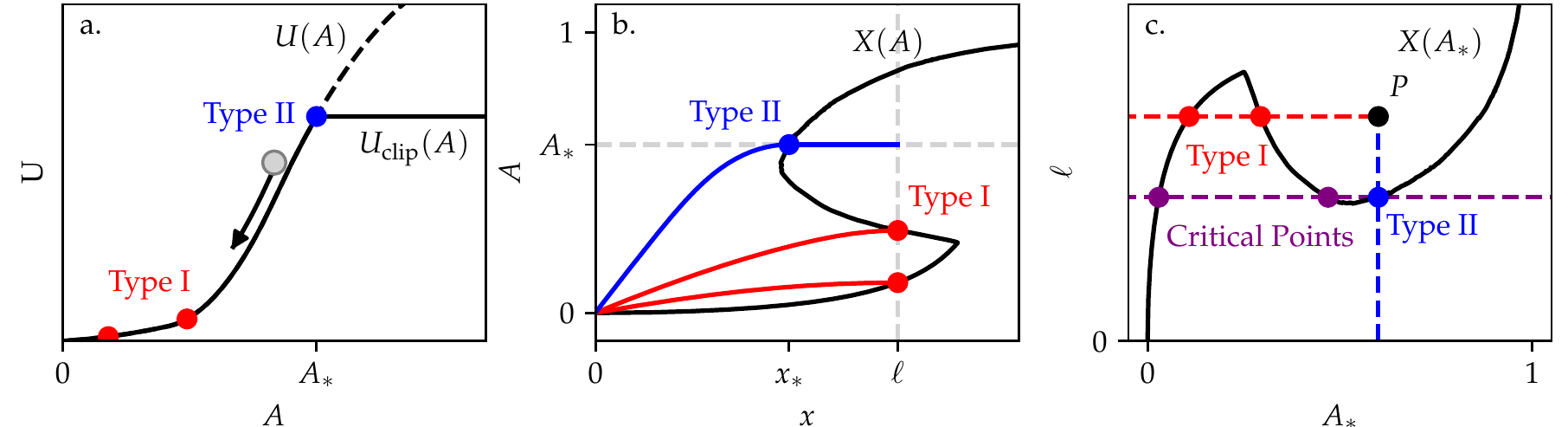}
    \caption{{\bf Stationary solutions and critical points.}~{\bf a.}~The potentials $U(A)$ and $U_\text{clip}(A)$ are depicted by dashed and solid lines, respectively.    
    The red points correspond to type I stationary solutions that lie along the non-clipped part of the potential. The blue point corresponds to a type II stationary solution, which lies at the point $A_*$. 
    The grey circle symbolizes a ball moving this 1D potential. If released from rest at either of the red points, the ball will take a ``time" $\ell$ to reach $A=0$. If the ball is released from the blue point, it will take a ``time" $X(A_*)$ to reach the origin. 
    {\bf b.}~The stationary solutions are plotted in real space. Type I solutions intersect the curve $X(A)$ along the $x=\ell$ boundary and type II solutions intersect the curve $X(A)$ along the $A =A_*$ boundary. {\bf c.} The $(A_*, \ell)$ parameter space is shown, with $X(A_*)$ plotted. A specific choice of parameters corresponds to a point $P$. Type I stationary solutions lie along the dashed red line, while type II lie along the dashed blue line. The critical points associated with the blue stationary solution are indicated with purple circles.   
    }
    \label{fig:typedef}
\end{figure*}

\subsubsection{General Setting} 

Here we derive the phase diagram featured in Fig.~\ref{fig:coupling}h. The equations we consider take the form:
\begin{align}
    \dot A =& \nabla^2 A + B \tilde f (A) \label{eq:homo1}\\
    \dot B =& \frac{1}{\tau}  g (A, B)\label{eq:homo2}
\end{align}
Notice that Eqs.~(\ref{eq:homo1}-\ref{eq:homo2}) are a specialization of Eqs.~(\ref{eq:broad1}-\ref{eq:broad2}) with $f(A,B,x) = B \tilde f (A)$. We will assume that $\tilde f (A) > 0$  for $A \in [0,1)$ and that $\tilde f$ crosses zero at $A=1$. 
Moreover, we will assume that there is a function $0 \le B_\infty(A) \le 1$ such that $g(A,B) < 0$ whenever $ B_\infty(A) < B <1 $ and $g(A,B) > 0$ whenever $0 < B< B_\infty(A) $.  
We will require the boundary conditions $A(0) =0$ and $\eval{\partial_x A}_{\ell} =0 $. 
Here $\ell = L/\sqrt{D}$ is the nondimensionalized system size.  
We will assume that the maximum value of $\tilde f$ and $g$ are of order 1 and that $\tau \gg 1$, implying that $B$ is a slow variable.
As we will illustrate with examples in subsequent sections, this form is general enough to capture a wide range of dynamical systems through suitable variable changes.

The calculations below comprise the following steps: We first find the fixed points of Eqs.~(\ref{eq:homo1}-\ref{eq:homo2}), which we refer to as stationary solutions. Setting $ \dot B =0$ in Eq.~(\ref{eq:homo2}) yields $B= B_\infty(A)$, and then setting $\dot A=0$ in Eq.~(\ref{eq:homo1}) yields the following ordinary differential equation for $A$:
\begin{align}
    \nabla^2 A = - B_\infty(A) \dv{U}{A} \label{eq:steady1} 
\end{align}
where $U$ is an antiderivative of $\tilde f$. Suppose the system is initialized to a stationary solution, given by $A_0(x)$ and $B_0(x) = B_\infty(A_0(x))$, and suppose the fast field $A$ is subject to a perturbation $A(x,t=0) = A_0(x) + \delta A(x)$, where $\delta A$ is not necessarily small. On short time scales, $B(x,t)$ will be frozen to $B_0(x)$ and $A$ will evolve according to 
\begin{align}
    \dot A = \nabla^2 A  +B_0(x) \tilde f(A) = - \fdv{\Phi}{A} \label{eq:critgen}
\end{align}
where
\begin{align}
    \Phi = \int_0^\ell [ (\nabla A)^2 - B_0 (x) U(A) ]\dd A.
\end{align}
Solutions to Eq.~(\ref{eq:critgen}) with $\dot A=0$ are critical points of $\Phi$. Notice that $A_0(x)$ is always one of the critical points. 
We will make inferences about the qualitative behavior of Eqs.~(\ref{eq:homo1}-\ref{eq:homo2}) using the structure of the stationary solutions, critical points, and orbits connecting them. Examples of such inferences are as follows:
\begin{itemize} 
\item Suppose that Eqs.~(\ref{eq:homo1}-\ref{eq:homo2}) only permit one stationary solution, and this stationary solution is linearly stable. If the associated $\Phi$ has no additional critical points beyond $A_0(x)$, then the system cannot exhibit spikes because $A(x,t)$ will quickly return to $A_0(x)$ after any perturbation. 
\item Suppose that Eqs.~(\ref{eq:homo1}-\ref{eq:homo2}) only permit one stationary solution, and this stationary solution is linearly stable. If $\Phi$ has stable critical points other than $A_0(x)$, then the system is excitable. Namely, if the initial trigger $\delta A(x)$ pushes the system into the basin of attraction of a second stable critical point, then $A(x,t)$ will be attracted to the second critical point on a fast time scale ($t \ll \tau$) and remain there for a long time ($ t \propto \tau$) until the slow variable $B(x,t)$ begins to evolve. This constitutes a spike. 
\item Suppose that Eqs.~(\ref{eq:homo1}-\ref{eq:homo2}) only permit one stationary solution, and this stationary solution is linearly unstable. Assuming chaotic behavior does not occur, the system will (generically) contain a limit cycle. The presence of this limit cycle corresponds to oscillatory activity. 
\item Equations (\ref{eq:homo1}-\ref{eq:homo2}) may have multiple stable stationary solutions. In this case, we refer to the dynamics as multistable. 
\end{itemize} 

In the next section, we specialize the form of $B_\infty (A)$ to allow for an analytical calculation of the stationary solutions and critical points, and thereby an analytical construction of a spiking phase diagram.

\begin{figure*}[t!]
    \centering
    \includegraphics[width=0.95 \textwidth]{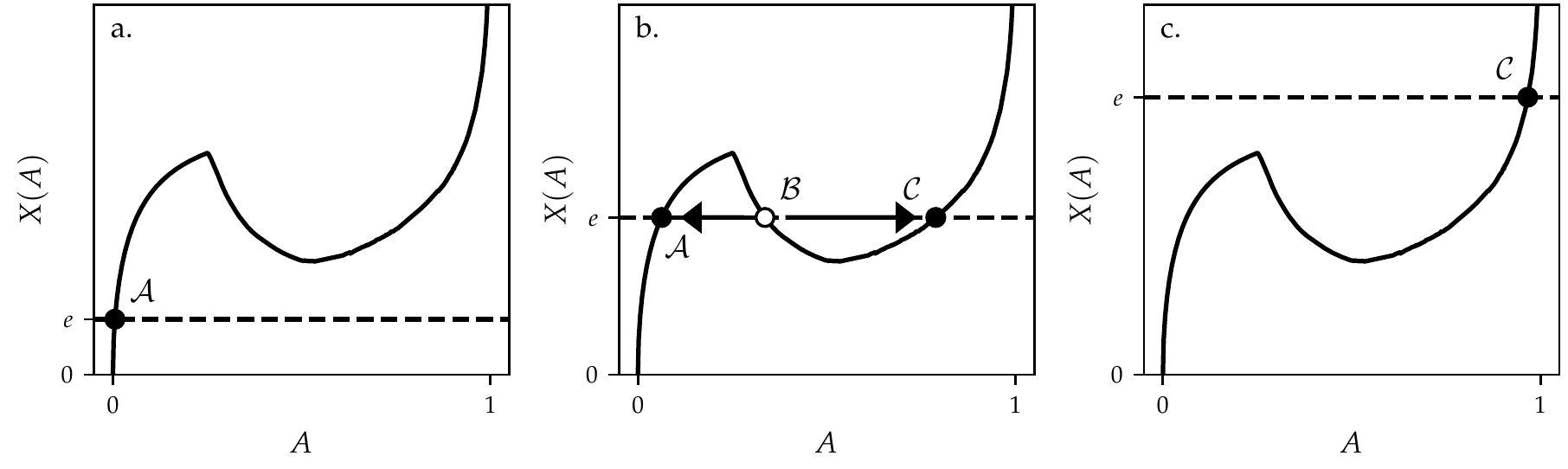}
    \caption{ {\bf Determining the stability of critical points.}~The critical points of $\Phi(A; e)$ are shown for three values of $e$. {\bf a.}~At low $e$, there exists only one critical point, $\mathcal{A}$. {\bf b.} At moderate $e$, there exist three critical points $\mathcal{A}$, $\mathcal{B}$, and $\mathcal{C}$. The solids arrows indicate heteroclinic orbits. {\bf c.} At large $e$, only one critical point ($\mathcal{C}$) remains. Critical points $\mathcal{A}$ and $\mathcal{C}$ must be minima (i.e.\ unstable dimension of 0), while $\mathcal{B}$ has an unstable dimension of 1.}
    \label{fig:morse}
\end{figure*}

\subsubsection{Construction of the phase diagram} 

In this section, we specialize the form of $B_\infty(A)$ to $B_\infty (A) = \Theta(A_*- A)$, where $\Theta$ is the Heaviside step function. Then Eq.~(\ref{eq:steady1}) becomes 
\begin{align}
    \nabla^2 A = - \Theta(A_*-A) \dv{U}{A} = \ \dv{U_\text{clip}}{A} \label{eq:stationary} 
\end{align}
where $\dv{U_\text{clip}}{A} = \Theta(A_*-A) \tilde f(A)$ defines a potential that has been clipped by the step function, see the solid black line in Fig.~\ref{fig:typedef}a. To construct solutions, notice that Eq.~(\ref{eq:stationary}) is equivalent to the equation of motion for a ball moving in a 1D potential, where $x$ corresponds to ``time" and $A$ corresponds to ``position". The boundary condition $\eval{\partial_x A}_{\ell} =0 $ is the requirement that the ball is at rest at ``time" $\ell$. Likewise, the boundary condition $A(0) =0$ is the requirement that the ball reaches ``position" $0$ at ``time" $0$. As shown in Fig.~\ref{fig:typedef}a, solutions to Eq.~(\ref{eq:stationary}) can be constructed as follows: release the ball from rest at point $A$, allow it to move through the potential, and measure the amount of ``time" it takes to reach point $A=0$. If that ``time" is equal to $\ell$, then one will have constructed a valid  solution to Eq.~(\ref{eq:stationary}). 

Fig.~\ref{fig:typedef}a demonstrates that there are two possible types of solutions. For type I (red), the ball is released along the non-clipped part of the potential ($A<A_*$). For type II, the ball is released at $A=A_*$: If the amount of ``time" $T$ it takes the ball to reach $A=0$ is less than $\ell$, then a valid solution can be constructed by letting the ball sit at rest on the flat part of the potential for a ``time" $\ell-T$ before releasing it. In Fig.~\ref{fig:typedef}b, the two types of solutions are shown in real space.  

\begin{figure*}[t!]
    \centering
    \includegraphics[width=0.95 \textwidth]{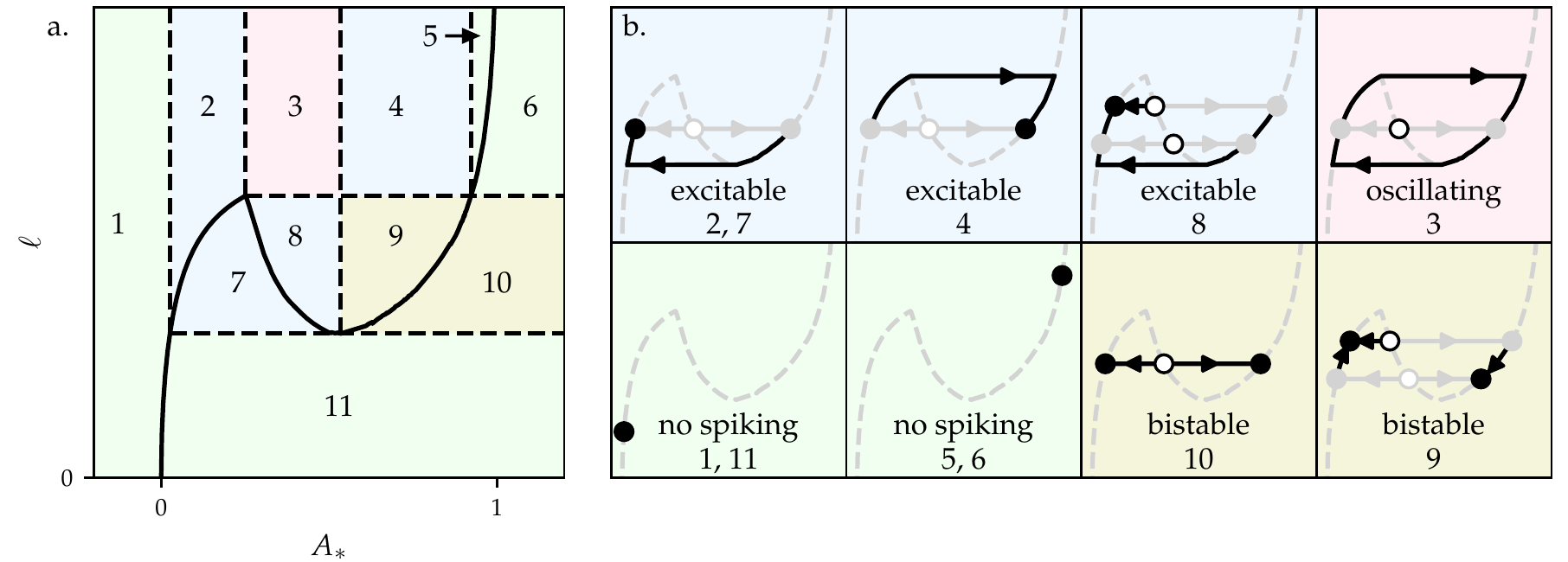}
    \caption{{\bf Phase diagram for spiking at a Dirichlet boundary}. {\bf a.}~A phase diagram in the $A_*$-$\ell$ plane is divided into 11 distinct regions. The solid black curve is $X(A_*)$. The color code indicates the qualitative behavior. Green: no spiking; olive: bistable; blue: excitable; pink: oscillating. 
    {\bf b.}~The qualitative behavior may be inferred from diagrams that summarize the topological features of the flow. In each diagram, the black circles represent stationary solutions, and the grey circles represent the critical points of the associated potentials. Solid circles are stable, and the open circles are unstable. The light grey lines represent heteroclinic orbits in the fast dynamics, and the black lines convey the evolution of the system on longer time scales.   }
    \label{fig:phasederivation}
\end{figure*}

To help count the number of solutions to Eq.~(\ref{eq:stationary}), we introduce a function $X(A)$ that corresponds to the amount of ``time"  the ball takes to reach ``position" $0$ if released from ``position" $A$. This is given by:
\begin{align}
    X(A) = \frac{1}{\sqrt 2} \int_0^{A} \frac{1}{\sqrt{ U(A) - U(a')  }} \dd a' \label{eq:xdef}
\end{align}
 Fig.~\ref{fig:typedef}b-c show an example of $X(A)$ featuring one local maximum and one local minimum. Depending on the choice of $\tilde f(A)$, the function $X(A)$ can have many local maxima and minima. Nevertheless, the assumptions that $\tilde f(1) =0$ and $\tilde f(A) > 0$ for $A< 1$ imply that $X(A) >0$, $X(0)=0$ and $X(1) =\infty$. In terms of $X(A)$, type I and type II solutions correspond to the following:  
\begin{itemize} 
\item[] \emph{Type I:}  If $a < A_*$ and $X(a) = \ell$, then there is a solution with $A(\ell) = a$. In this case, the stationary solution $A(x)$ is given by the inverse of:
\begin{align}
    x(A) = \frac{1}{\sqrt 2} \int_0^A \frac{1}{ \sqrt{ U(a) - U(a') }} \dd a'  
\end{align}
\item[] \emph{Type II:} Let $x_*=X(A_*)$. If $x_* < \ell$, then there is a solution with $A(\ell) =A_*$. In this case, the stationary solution $A(x)$ can be defined in a piecewise manner: $A(x) = A_*$ for $x \in [x_*, \ell]$; For $x < x_*$,  $A(x)$ is  the inverse of: 
\begin{align}
    x(A) = \frac{1}{\sqrt 2} \int_0^A \frac{1}{ \sqrt{ U(A_*) - U(a') }} \dd a'
\end{align}
\end{itemize} 
Now we can think of $A_*$ and $\ell$ as being parameters of our dynamical system defined by Eqs.~(\ref{eq:homo1}-\ref{eq:homo2}). Working in the $A_*$-$\ell$ plane, all the stationary solutions can be found by the graphical construction illustrated in Fig.~\ref{fig:typedef}c: 
\begin{enumerate}
\item Represent a choice of parameters $(A_*, \ell)$ as a point $P$ in the plane. 
\item Draw the curve $X(A_*)$. 
\item Draw a horizontal line extending to the left from $P$. The intersections between the horizontal line and $X(A_*)$ correspond to stationary solutions of type I. 
\item Draw a vertical line extending downward from $P$. Intersections between the vertical line and $X(A_*)$ represent stationary solutions of type II. 
\end{enumerate} 
This construction yields all the solutions to Eq.~(\ref{eq:stationary}). 
Next, we derive the stability of the stationary solutions and their consequences for spiking. 
To do so, we will specialize to the situation in which $X(A)$ is ``N"-shaped, i.e.\ it has exactly one local maximum and one local minimum. We will use the following result: consider the functional
\begin{align}
    \Phi(A; e) = \int_0^e [  \qty(\nabla A)^2 - U(A) ] \, \dd x
\end{align}
which is minimized with respect to $A$ subject to the boundary conditions $A(0) =0$ and $\eval{\partial_x A}_{e} =0$. 
Using a similar derivation to that above, one sees that the critical points of $\Phi$ correspond to the intersections between $X(a)$ and the horizontal line at $e$. As illustrated in Fig.~\ref{fig:morse}a, for sufficiently small $e$, there is only one critical point (denoted $\mathcal{A}$) and therefore this critical point must be a minimum of $\Phi$. As $e$ increases (Fig.~\ref{fig:morse}b), a bifurcation produces two new critical points, $\mathcal{B}$ and $\mathcal{C}$. As $e$ increases further, $\mathcal{B}$ and $\mathcal{A}$ annihilate (Fig.~\ref{fig:morse}c). Since $\mathcal{C}$ is now the lone remaining critical point, it must also be a minimum of $\Phi$. Conley index theory states that two critical points that emerge or annihilate  must have unstable dimensions that differ by $1$~\cite{conley1983algebraic}. Since the minima $\mathcal{A}$ and $\mathcal{C}$ have an unstable dimension of $0$, the unstable dimension of $\mathcal{B}$ is $1$. 
Moreover, the dynamical system $\dot A = - \fdv{\Phi}{A}$ must have heteroclinic orbits from $\mathcal{B}$ to $\mathcal{A}$ and $\mathcal{B}$ to $\mathcal{C}$. 
(See the Supplementary Information for a brief introduction to Conley index theory and a derivation of these facts.)

We now apply these facts to deduce the stability of the stationary solutions. For stationary solutions $A_0(x)$ of type I, $B_\infty(A_0(x)) =1$. Therefore, the fast dynamics for a type I stationary solution are governed by the equation:
\begin{align}
    \dot A = \nabla^2 A + \dv{U}{A} = - \fdv{\Phi_\I}{A}. 
\end{align}
For stationary solutions $A_0(x)$ of type II, $B_\infty(A_0(x)) = \Theta(x - x_*) $, where $x_* = X(A_*)$. Hence, the fast dynamics are governed by the equation:
\begin{align}
    \dot A = \nabla^2 A + \Theta( x-x_*)\dv{U}{A} =  - \fdv{\Phi_{\II}}{A}. 
\end{align}
Notice that the critical points of $\Phi_\I (A) $ and $\Phi_\II(A)$ can be put in correspondence with $\Phi(A; \ell)$ and $\Phi(A;x_*) $, respectively. 
Therefore, one can use the following graphical construction, illustrated in Fig.~\ref{fig:typedef}c, to find the critical points associated with each stationary solution:
\begin{enumerate}
    \item Identify the point corresponding to the stationary solution of interest. (In Fig.~\ref{fig:typedef}c, the blue stationary solution is of interest.)  
    \item Draw a horizontal line in both directions out from the point. (Dashed purple line in Fig.~\ref{fig:typedef}c.)
    \item The intersections between the horizontal line and $X(A_*)$  correspond to critical points. (The blue and purple points in Fig.~\ref{fig:typedef}c.) 
    \item The stability of each critical point is determined by which branch of $X(A_*)$ it lies on: those on an increasing branch are stable while those on a decreasing branch have an unstable dimension of 1. 
\end{enumerate}    
Notice that all stationary solutions are also critical points of their associated potential. Occasionally, critical points of one stationary solution are also stationary solutions unto themselves. 

\begin{figure*}
    \centering
    \includegraphics[width=\textwidth]{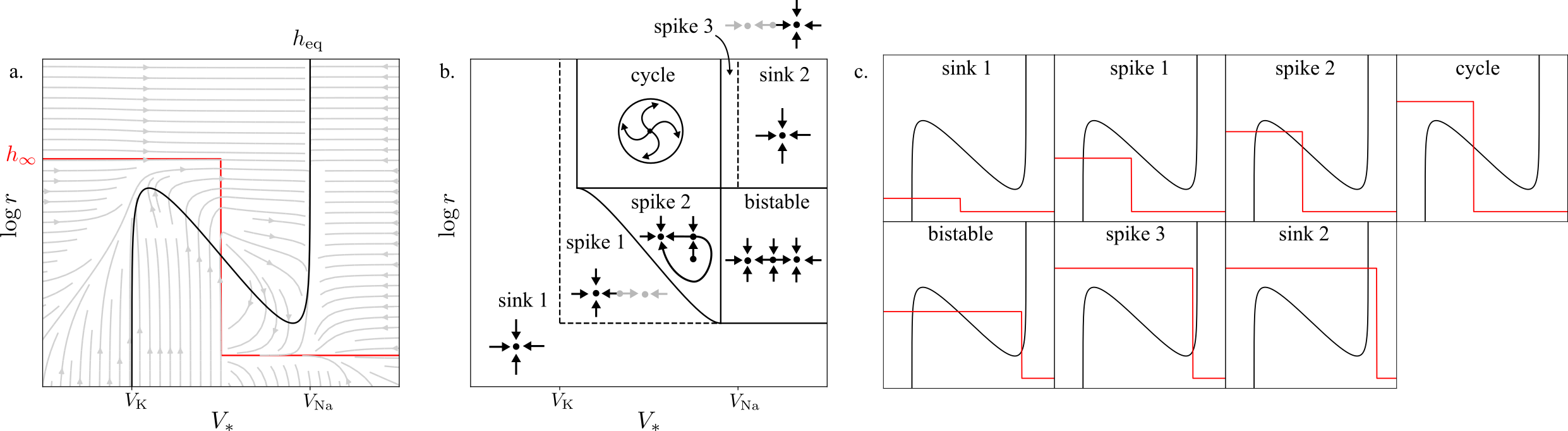}
    \caption{{\bf A spiking phase diagram for a single bioelectric cell.} {\bf a.} A representative phase portrait for Eqs.~(\ref{eq:Mfn1}-\ref{eq:Mfn2}). The fixed point (here unstable) appears at the intersection $h_\text{eq}$ (black) and $h_\infty$ (red).  
    {\bf b.} An annotated phase diagram showing the fixed points, limit cycles, and heteroclinic orbits in each phase. The light grey points and arrows represent features that are only present in the $\tau \to \infty$ dynamics.  {\bf c.} The $h_\text{eq}$ and $h_\infty$ curves for each phase in the phase diagram.   }
    \label{fig:supdiagram}
\end{figure*}

These considerations allow us to construct the phase diagram shown in Fig.~\ref{fig:phasederivation}a. The parameter space has been divided into 11 regions based on the number of type I and type II stationary solutions, and the nature of their associated critical points.  
For each region, one can construct a corresponding diagram shown in Fig.~\ref{fig:phasederivation}b.
In each diagram, the black circles denote stationary solutions while grey circles denote critical points that are not stationary solutions. Solid circles indicate stable stationary solutions/critical points, while open circles indicate unstable stationary solutions/critical points. The solid grey lines indicate heteroclinic orbits in the fast dynamics, while the solid black curves depict the evolution of the system over longer time scales.

Using these diagrams, we can then classify distinct qualitative behaviors. 
Regions 1, 5, 6, and 11 are classified as no-spiking because they feature only one stationary solution, and the functional associated with that stationary solution has only one critical point.  
Regions 2, 4, 7, and 8 are classified as excitable because they have exactly one stable stationary solution, and the potential associated with this stationary solution has multiple stable critical points. 
Region 3 is classified as oscillating because it features only one stationary solution, and this stationary solution is unstable, and hence the system exhibits a limit cycle. 
Regions 9 and 10 are classified as bistable because they feature two stable stationary solutions. 

Notice that this phase diagram differs slightly from the one shown in Fig.~\ref{fig:coupling}h, because $\tilde f$ in the lumberjack-tree setting has an additional zero at $A=0$. It will be explained in the following section how the phase diagram in Fig.~\ref{fig:phasederivation}a is mapped onto the one for population dynamics in Fig.~\ref{fig:coupling}h.

\subsection{Population dynamics}
\label{sec:pop}

In dimensionful units, the Lotka-Volterra model we consider takes the form~\cite{murray2013mathematical}:
\begin{align}
    \dot P =& -a\,  P + b\, N \, (P+g) \, R_p(P) \label{eq:dful1} \\
    \dot N =& c \, N \, R_n (N) - d \, P \, N \label{eq:dful2}
\end{align}
where $P$ is the population of the predator, $N$ is the population of the prey, and $a, b, c, d, g>0$. Here, $a$ is the dimensionful hopping rate at which the lumberjacks (predator) travel into the desert and perish; $b$ sets the benefit to the lumberjacks of consuming a tree; $c$ is the growth rate of the trees; and $d$ sets the intensity of the predation. The parameter $g$ is a regularization parameter that prevents the lumberjacks from going extinct. We will eventually be interested in the limit $g \to 0$. The function $R_n(N)$ is a nonlinearity that sets the carrying capacity $K_n$ for the prey in absence of predators. We require that $R_n(0) = 1$ and that $R_n$ only crosses zero at $K_n$. Likewise, $R_p(P)$ obeys $R_p(0) = 1$ and only crosses zero at $K_p$, the carrying capacity of the predators. 

We nondimensionalize Eqs.~(\ref{eq:dful1}-\ref{eq:dful2}) by introducing a time scale $t_0=1/b K_n $ and defining
\begin{align}
    p =& \frac{P}{K_p} & n =& \frac{N}{K_n}  \\
      \gamma =& \frac{g}{ K_p} & \epsilon =& a t_0    \\    
    \scr(p) =& R_p( p K_p )  & r_n(n) =& R_n( n K_n ) \\
      \tau =& \frac{1}{ t_0 c} & \tilde t =&  t/t_0 \\
      q =& \frac{1}{d K_p t_0}
\end{align}
yielding the equations:
\begin{align}
    \dot p =&  - \epsilon \, p +  n \, (p +\gamma) \,  \scr(p)     \label{eq:Mpred}\\
     \dot n =&  \frac{n}{\tau} \, \qty[ r_n(n) -  \frac{ p}{q} ]  \label{eq:Mprey}
\end{align}
in which all quantities are dimensionless. 
For the phase portraits in Fig.~\ref{fig:coupling}a-c, we use the following piecewise linear functions for $r_n(n)$ and $\scr(p)$: 
\begin{align} 
r_n(n) =& 1-n \\ 
\scr(p) = &\Theta(p_*-p) \qty( \frac{ p (n_*-1) }{p_*}  +1 ) + \nonumber \\
&\Theta(p-p_*) \frac{n_* (p-1)}{p_*-1} \label{eq:kp}
\end{align}
where $p_*=0.7$ and $n_*=3$. We use $\epsilon=0$, $0.5$, and $4.0$ for panels a, b, and c respectively. For all panels we use $\tau=50$ and $q=0.5$. 

Next, we detail the passage to the continuum shown in Fig.~\ref{fig:coupling}e-g. Consider a compartment model with local populations $p_i$ and $n_i$, where $-N\le i \le N$. Here, a positive index $i$ corresponds to the forest, and a non-positive $i$ corresponds to the desert. The dynamics are governed by
\begin{align}
    \dot p_i =& \epsilon (p_{i+1}-p_{i-1}- 2 p_i) \\
    &+ 
    \begin{cases}
    - \alpha p_i &  - N \le i \le  0 \\
    n_i (p_i + \gamma) \scr(p_i) & 1 \le i \le N  
    \end{cases} \label{eq:popdesc}
    \\
    \dot n_i =&   \frac{n_i}{\tau }\qty[ r_n(n_i) - \frac{p_i}{q} ] 
\end{align}
with boundary conditions $p_{-N-1} \equiv p_{-N}$ and $p_{N+1} \equiv p_{N}$. As in Eq.~(\ref{eq:lumber1}), the parameter $\epsilon$ in Eq.~(\ref{eq:popdesc})  sets the hopping rate between sites. However, in Eq.~(\ref{eq:popdesc}) we allow the desert to have an intrinsic strength $\alpha$ that is no longer necessarily infinite. One recovers Eqs.~(\ref{eq:lumber1}-\ref{eq:lumber2}) by taking $N=1$  and $\alpha \to \infty$, in which case $p_{-1}=p_0=0$ may be eliminated. Likewise, Fig.~\ref{fig:coupling}e, f, and g correspond to taking $\alpha \to \infty$ with $N=2$, $N=4$, and $N=100$ respectively. See the Supplementary Information for simulation details. 

The continuum limit applies when $N, \epsilon \gg 1$ and the discrete index $i$ is replaced by a continuous variable $x = i d$, where $d$ is the lattice spacing. 
If $\alpha \to \infty$, then $p(x<0)=0$, so the continuum equations take the form  
\begin{align}
    \dot p =& \nabla^2 p + n (p+ \gamma) \scr(p) \label{eq:p1}  \\
    \dot n =& \frac{n}{\tau}  \qty[ r_n(n) - \frac{p}{q} ] \label{eq:p2}
\end{align}
paired with the Dirichlet boundary $p(0)=0$ and the no-flux boundary $\eval{\partial_x p}_\ell =0 $. Here, we have nondimensionalized the $x$-coordinate according to $x\to x/\sqrt{D}$ and $\ell = L/\sqrt{D}$. However, if $\alpha$ is finite, then one obtains an interfacial equation of the form:
\begin{align}
    \dot p =& \nabla^2 p +\begin{cases}
        - \alpha p & x \in [-\ell, 0] \\
        n (p+ \gamma) \scr(p) & x \in [0, \ell]
    \end{cases} \label{eq:sim1bp} \\
    \dot n =& \frac{n}{\tau }\qty[ r_n(n) - \frac{p}{q} ] \label{eq:sim1bn_meth}
\end{align}
with no-flux boundaries at $x = \pm \ell$. 
We use Eq.~(\ref{eq:sim1bn_meth}) with $\alpha =1$ to produce the kymograph in Fig.~\ref{fig:overview}b. See the Supplementary Information for simulation details.

To compute the phase diagram in Fig.~\ref{fig:coupling}h, notice that Eqs.~(\ref{eq:p1}-\ref{eq:p2}) take the same form as Eqs.~(\ref{eq:homo1}-\ref{eq:homo2}). Here, $p$ is the fast variable, $n$ is the slow variable, and the function $\tilde f$ is given by $\tilde f(p) = (p+\gamma) \scr(p) $. The function $n_\infty$ (which plays the role of $B_\infty$) is determined by the functional form of $r_n$. For instance, consider the piecewise linear function 
\begin{align}
    r_n(n) = \begin{cases}
        1-m n & n \in [0, \frac{1 -q}{m} ] \\
        q & n \in [\frac{1 -q}{m}, 1-\frac q m] \\
        m (1-n) &  n \in [ 1-\frac q m , 1]
    \end{cases} \label{eq:rn}
\end{align}
If $m=1$, then $r_n(n) = 1-n$ and $n_\infty(p) = \max\{ 0, 1-p/q\}$. If $m \to \infty$, then $n_\infty(p) = \Theta(q-p) $, where $q$ plays the role of $A_*$ in Methods~\S\ref{sec:boundary}. The phase diagrams in Fig.~\ref{fig:coupling}d and Fig.~\ref{fig:coupling}h are analytically computed for $m\to \infty$. In the limit that $\gamma \to 0$, the location of the local maximum of $X(q)$ approaches $q =0$ and so the general phase diagram in Fig.~\ref{fig:phasederivation}a converges the one Fig.~\ref{fig:coupling}h.

\begin{figure*}
    \centering
    \includegraphics[width=0.9\textwidth]{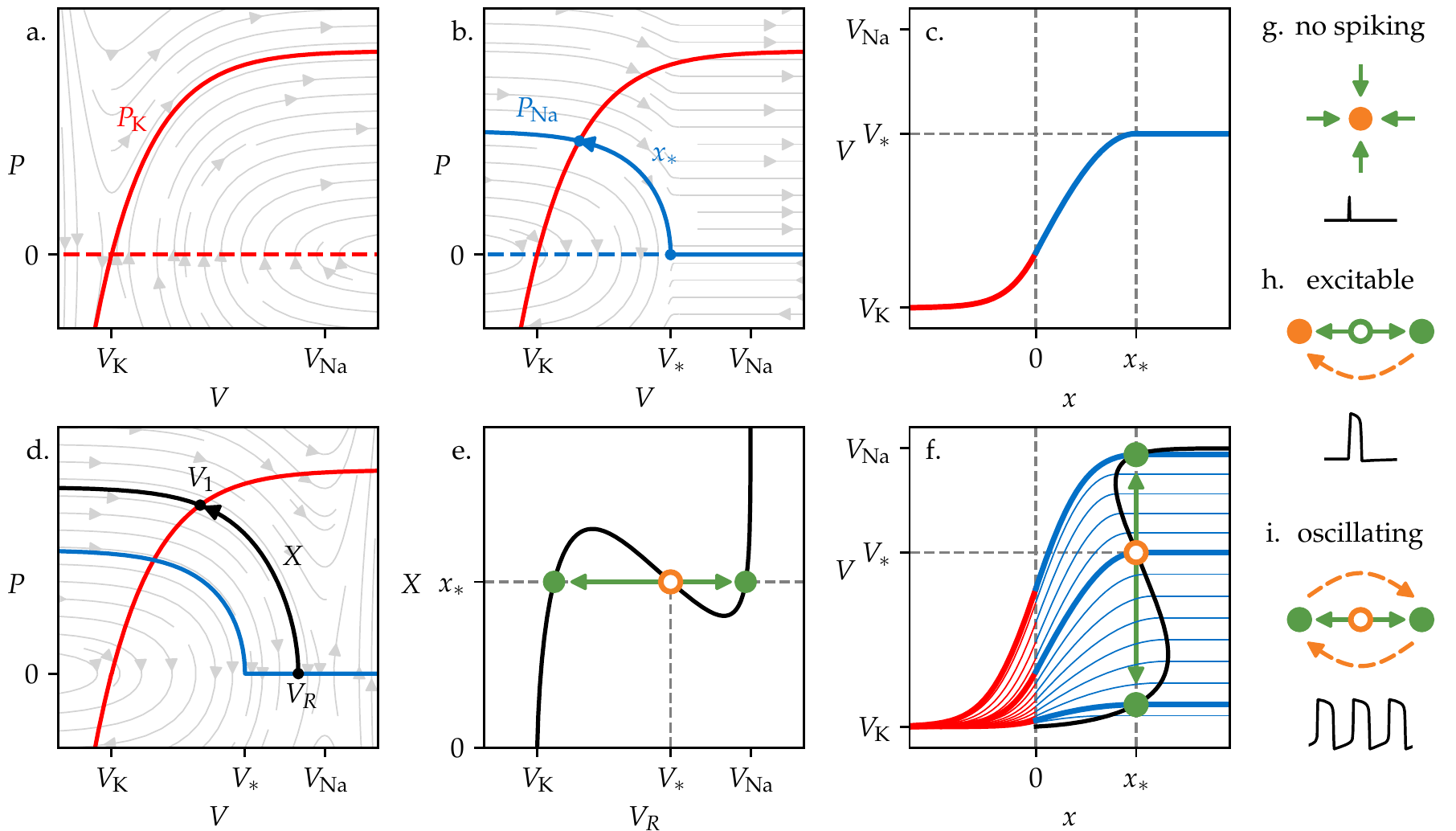}
    \caption{{\bf Bifurcation analysis of interfacial spiking.} 
    {\bf a.}~The Hamiltonian flow in Eqs.~(\ref{eq:hamil1}-\ref{eq:hamil2}) for $x<0$ is depicted. The separatrix $P_{\ce K}$ (solid blue line) results from an advection of $P=0$ (dashed blue line) through a distance $\ell \to \infty$. {\bf b.}~Similarly, the Hamiltonian flow in Eqs.~(\ref{eq:hamil1}-\ref{eq:hamil2}) for $x>0$ is depicted. The separatrix $P_{\ce{Na}}$ (solid red line) results from an advection of $P=0$ (dashed red line) through a distance $\ell \to -\infty$. Solving for the separatrices amounts to solving the differential equations on either side of the interface in Eqs.~(\ref{eq:big1}-\ref{eq:big2}). Finding the intersection of $P_{\ce{Na}}$ and $P_{\ce K}$ amounts to enforcing the continuity requirements for $V$ and $\partial_x V$ at $x=0$. {\bf c.}~In real space, composing the separatrices yields the stationary solution $V_0(x)$. The curve $V_0(x)$ matches the $x <0$ and $x>0$ solutions subject to the requirement that the right solution obtains a slope of zero at $V=V_*$.
    {\bf d.}~The curve $P_{\ce{Na}}(V, V_R)$ is shown in black. $V_R$ is the $V$ coordinate of the intersection with $P=0$, and $V_1$ marks the intersection with $P_{\ce K}$. The function $X(V_R)$ represents the $x$ distance transversed in real space between the voltages $V_R$ and $V_1(V_R)$. {\bf e.} The function $X$ is plotted as a function of $V_R$. The number of solutions to the equation $X(V_R) = x_*$ varies from 1 to 3 depending on $x_*$.  
    {\bf f.} One can visualize the function $X(V_R)$ by plotting a range of trial solutions with different slopes at their interface. The function $X(V_R)$ corresponds to the $x$ value at which each curve first attains its maximum. The thick lines correspond to the critical points of $\Phi$. 
    {\bf g.}~If $x_*$ intersects $X$ once, then the system does not exhibit spikes. {\bf h.}~If  $x_*$ intersects $X$ three times with $V_*$ corresponding to an increasing branch, then the system exhibits excitability. {\bf i.} If $V_*$ lies on the decreasing branch of $X$, then the system exhibits oscillations. In (e-i), the circles represent critical points of the functional $\Phi$ in Eqs.~(\ref{eq:tauinf}). The orange circle represents $V_0(x)$, the stationary solution of Eqs.~(\ref{eq:big1}-\ref{eq:big2}). Solid arrows represent fast heteroclinic orbits, and the dashed arrows represent slow dynamics in Eqs.~(\ref{eq:big1}-\ref{eq:big2}).  Open circles are unstable critical points, while solid circles are stable critical points of $\Phi$. 
    }
    \label{fig:topo}
\end{figure*}

\subsection{Bioelectric interfaces}
\label{sec:FN}

 \subsubsection{Conductance-based biolectric model} 
The bioelectric dynamics we consider are described by conductance-based (i.e.\ Hodgkin-Huxley type) models. Our equations focus on the role of the sodium ion channels, potassium ion channels, and gap-junction coupling between cells. If the ion channels are homogeneously distributed throughout the tissue, the dynamics are governed by
\begin{align}
     C \dot V =&  G \nabla^2 V +  g_{\ce K} \, f_{\ce K}(V) + h  g_{\ce{Na}}\, f_{\ce {Na}} (V) \label{eq:fn1dim} \\
    \dot h =& \frac{h_\infty(V) -h}{\tau } \label{eq:fn2dim}     
\end{align}
Here, $V(x,t)$ is the local membrane potential of the tissue, $C$ is the capacitance of the cell membrane, $g_{\ce{K}} \, f_{\ce{K}} (V)$ and $h\, g_{\ce{Na}}\, f_{\ce {Na}} (V)$ are the currents through the potassium and sodium channels, respectively. The constants $g_{\ce{K}}$ and $g_{\ce{Na}}$ are known as open-state conductances, and they set the relative strengths of the potassium and sodium channels. The functions $f_{\ce K}$ and $f_{\ce Na}$ are nonlinearites that control the shape of the voltage-current relationship. 
The variable $h$ is a gating variable that assumes values between $0$ and $1$, and $\tau$ is a time constant for its evolution towards a steady state value $h_\infty(V)$. Finally, voltage diffusion, modulated by the parameter $G$, arises due to gap-junction coupling. 

To nondimensionalize the equations, let $V_\text{ref}$ be a characteristic reference voltage, set $r = \frac{g_{\ce{Na}}}{g_{\ce{K}}}$ to be the relative strength of the sodium and potassium channels, and let $t_0 = \frac{V_\text{ref} C }{g_{\ce{K}}} $ denote a characteristic time of the voltage dynamics. Next, let $D=GV_\text{ref}/g_{\ce{K}}$ be the diffusion coefficient $G/C$ times the characteristic time scale $t_0$. Note that $\sqrt{D}$ has dimensions of length. We nondimensionalize the equations according to 
\begin{align}
\tilde h =& h \, r \\
\tilde t =& t/t_0 \\
\tilde \tau =& \tau/t_0 \\
\tilde V =& V/V_\text{ref} \\
\tilde x =& x/\sqrt{D}\\
\tilde f_{\ce{ K}} (\tilde V) =& f_{\ce{K}} (\tilde V V_\text{ref} ) \\
\tilde f_{\ce{ Na}} (\tilde V) =& f_{\ce{Na}} (\tilde V V_\text{ref} )
\end{align}
In the main text, we use dimensionless variables except for $x$, and hence $D$ is also retained in Eqs.~(\ref{eq:big1}-\ref{eq:big2}). In what follows, we will work in dimensionless quantities and omit the tildes.  

As shown in Fig.~\ref{fig:phase}c,f,i, we will require that $f_{\ce K}(V)$ and $f_{\ce {Na}} (V)$ each have exactly one zero crossing, at $V_{\ce{K}}$ and $V_{{\ce{Na}}}$ respectively, and that they are decreasing at this zero. 
Also, motivated by experimentally calibrated conductance models~\cite{tusscher2004model}, we take the asymptotic value of the gating variable to be a step function: $h_\infty(V) = r \, \Theta (V_*-V) $, where $V_*$ is the crossover of the step.

\begin{figure*}
    \centering
    \includegraphics[width=\textwidth]{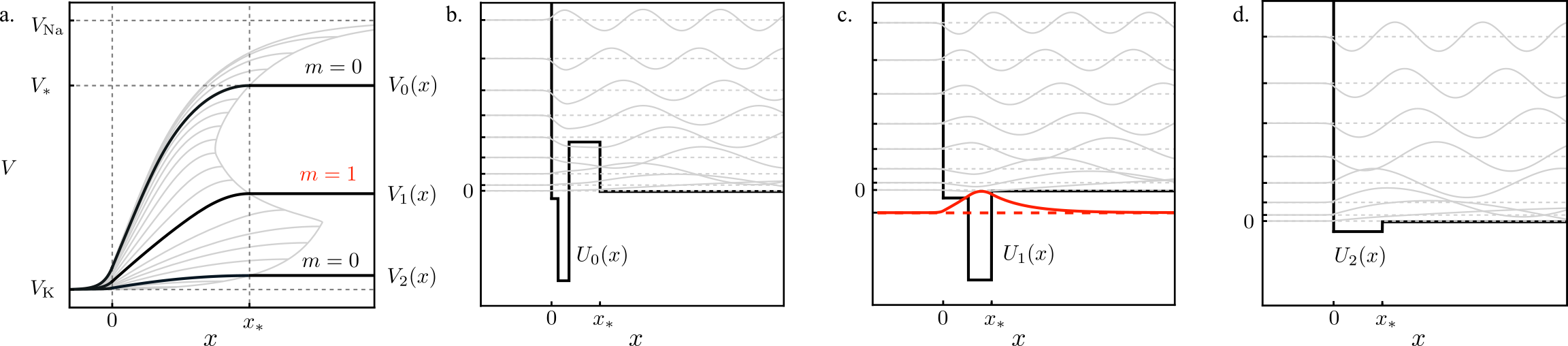}
    \caption{{\bf Linearized dynamics and interface confinement.} {\bf a.}~An example of three solutions to Eq.~(\ref{eq:tauinf}) (black lines), which constitute critical points of the functional $\Phi$. The light gray curves represent trial solutions that do not meet the boundary conditions. {\bf b-d.}~The linearized dynamics about each critical point $p$ is governed by a Schr\"odinger equation with an effective potential $U_p$. Each effective potential features a well near the interface. The unstable dimension, $m$, of each critical point corresponds to the number of negative energy states of the Hamiltonian. Because the negative energy states must be bound states, the unstable modes that drive the spiking feature a spatial profile that is localized near the interface. The sole negative energy state in (c) is highlighted in red. 
    \label{fig:linearization}}
\end{figure*}

 \subsubsection{Dynamics of a single cell} 

Before considering a spatially extended system, we consider a  the dynamics of a single cell with both sodium and potassium channels, described by the ordinary differential equation
\begin{align}
    \dot V =&  f_{\ce K}(V) + h \, f_{\ce {Na}} (V) \label{eq:Mfn1} \\
    \dot h =& \frac{h_\infty(V) -h}{\tau } \label{eq:Mfn2} 
\end{align}
The dynamics of a single cell can be understood in terms of the $V$  and $h$ nullclines. For example, in Fig.~\ref{fig:supdiagram}, the  
the solid red line indicates $h_\infty$, which is the $h$-nullcline. The solid black line indicates the $V$-nullcline, given by 
\begin{align}
    h_\text{eq} (V) = - \frac{f_{\ce K}(V)}{f_{\ce {Na}} (V)}
\end{align}
As shown in Fig.~\ref{fig:supdiagram} and Fig.~\ref{fig:phase}d,g, the $V_*$-$r$ phase diagram for single cell exhibits spiking if $h_\text{eq}$ is an ``N" shaped curve. However, if $h_\text{eq}$ is monotonically increasing (e.g., Fig.~\ref{fig:phase}j), then the $V_*$-$r$ phase diagram for a single cell does not exhibit spiking.

\subsubsection{Phase diagram for interfacial spiking}
Here we derive the phase diagram for the bioelectric interface shown in Fig.~\ref{fig:phase}. 
We will consider a 1D domain, $x \in [-\ell,\ell]$, with an interface at $x=0$. We will be particularly interested in the following two limits: $\ell \to 0$, in which we will recover the single-cell phase diagrams in Fig.~\ref{fig:phase}d,g,j; and $\ell \to \infty$, in which we obtain the interfacial phase diagrams in Fig.~\ref{fig:phase}e,h,k.  
The governing equations are:
\begin{align}
    \dot V =& \nabla^2 V +
    \begin{cases} 
    f_{\ce K}(V)  & x \in  [-\ell ,0]  \\
    h \, f_{ \ce{Na}} (V) & x \in (0,\ell]
    \end{cases} \label{eq:Mbig1}  \\
    \dot h =& \frac{h_\infty(V) -h}{\tau } \label{eq:Mbig2}
\end{align}
We will require that $V$ and $\partial_x V$ be continuous and that $\eval{\partial_x V}_{\pm \ell } =0 $. For brevity, we will write:
\begin{align}
    f(V,h, x) = f_{\ce K}(V) \Theta(-x) + h f_{ \ce{Na}} (V) \Theta(x)
\end{align}

Following the general approach from Methods~\S\ref{sec:boundary}, we first solve for the fixed points of Eqs.~(\ref{eq:Mbig1}-\ref{eq:Mbig2}), which we refer to as stationary solutions.
Setting the left-hand side of Eqs.~(\ref{eq:Mbig1}-\ref{eq:Mbig2}) to zero amounts to solving the equation:
\begin{align}
    0 = \nabla^2 V + f(V, h_\infty(V), x) \label{eq:global} 
\end{align}
which can be cast as a Hamiltonian system
\begin{align}
    \partial_x V =& P \label{eq:hamil1} \\
    \partial_x P =& - f(V,h_\infty(V),x) \label{eq:hamil2} 
\end{align}
with boundary conditions $\eval{P}_{\pm \ell} = 0 $. 
As shown in Fig.~\ref{fig:topo}, Eqs.~(\ref{eq:hamil1}-\ref{eq:hamil2}) may be solved graphically. First, construct the curve $P_{\ce{K}}(V; \ell)$ by taking the line $P=0$ and advecting it forward a distance $\ell$ according to the $x <0$ flow, as shown in Fig.~\ref{fig:topo}a. Second, construct the curve $P_{\ce{Na}}(V;-\ell)$ by advecting the line $P=0$ backwards a distance $\ell$ according to the $x >0$ flow, as shown in Fig.~\ref{fig:topo}b. The intersections $P_{\ce{K}}(V;\ell) = P_{\ce{Na}}(V;-\ell)$ correspond to stationary solutions. We will focus on two limits:

First, we consider $\ell \to 0$. In this limit, we expect to recover the dynamics of a single cell, since $V$ is effectively forced to be constant across the domain $[-\ell , \ell]$. Indeed, for small $\ell$, one obtains
\begin{align}
    P_{\ce{K}}(V; \ell)=& -\ell f_{\ce{K}}(V) + \order{\ell^2} \\
    P_{\ce{Na}} (V; -\ell ) =&  \ell h_\infty(V) f_{\ce{Na}}(V) + \order{\ell^2}
\end{align}
Equating $P_{\ce{K}}(V; \ell) = P_{\ce{Na}} (V; -\ell )$ yields  
\begin{align}
    h_\text{eq}(V) = h_\infty(V) 
\end{align}
which is the fixed point equation for a single cell described by Eqs.~(\ref{eq:Mfn1}-\ref{eq:Mfn2}). 
    
Second, we consider $\ell \to \infty$. In this limit,  $P_{\ce{K}}(V;\ell)$ and $P_{\ce{Na}}(V;-\ell)$ approach the separatices of the Hamiltonian flow, as shown in Fig.~\ref{fig:topo}a-b. These separatrices are given explicitly by: 
    \begin{align}
    P_{\ce{K}}(V) =& \sqrt{ -2 \int_{V_{\ce{K}}}^{V} f_{\ce{K}} (V') \dd V'  } \\ 
    P_{\ce{Na}}(V) =& \sqrt{ 2 \int_{V}^{V_{\ce{Na}}} h_\infty( V') f_{\ce{Na}} (V') \dd V' }
\end{align}
Since $f_{\ce{K}}$ and $f_{\ce{Na}} $ only have one zero crossing each, the separatrices are monotonic and they will only intersect once. Hence, in the limit $\ell \to \infty$, the stationary solution $V_0(x)$ is unique. Until this point, we have only required that $h_\infty(V)$ be non-negative on the interval
$[V_{\ce{K}}, V_{\ce{Na}}]$. For $h_\infty(V) = r \, \Theta(V_*-V)$,  the solution $V_0(x)$ is the curve that matches between the left- and right-hand side while reaching $\partial_x V=0$ at the voltage $V_*$, as shown in Fig.~\ref{fig:topo}c. We will let $x_*$ denote the solution to the equation $V_0(x) = V_*$. We note that the intersection construction in Fig.~\ref{fig:topo}a-b is equivalent to the hill picture provided in Fig.~\ref{fig:isochrones}a in which $V$ is the position of the ball and $P$ is its momentum.

\begin{figure}[t!]
    \centering
    \includegraphics[width=0.8\columnwidth]{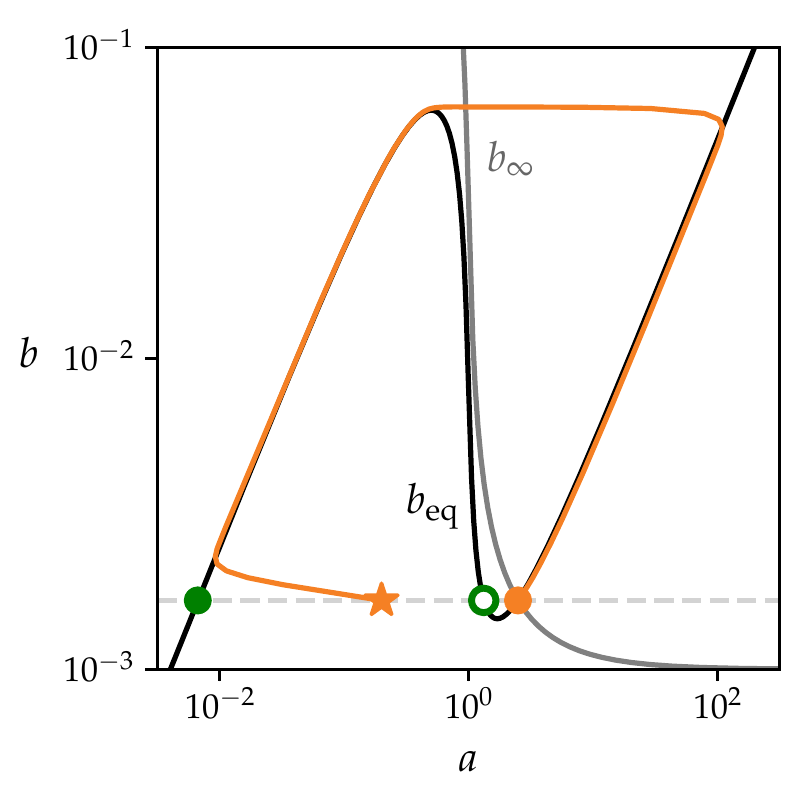}
    \caption{{\bf A phase portrait of an excitable chemical reaction.}~A phase portrait for Eqs.~(\ref{eq:yred}-\ref{eq:zred}) with $m_1=2$, $m_2=10^{-3}$, $\tau=10^{4}$.  The black curve is $b_\text{eq}(a)$ [Eq.~(\ref{eq:beq})]  and the grey curve is $b_\infty(a)$ [Eq.~(\ref{eq:zinf})]. The solid and filled green circles represent critical points of the fast dynamics (c.f.~Fig.~\ref{fig:theorySketch}). The orange curve is an example trajectory starting at the star and ending at the orange solid circle and exhibiting a spike. The same parameters are used to integrate Eqs.~(\ref{eq:Mybulk}-\ref{eq:Mzbulk}) in Fig.~\ref{fig:overview}c. Notably, the interface exhibits oscillation, rather than excitability, due to the presence of diffusion. }
    \label{fig:chemphase}
\end{figure}

Having found the stationary solution $V_0(x)$,  we examine the fast dynamics with $h$ frozen to $h_\infty(V_0(x))$. 
The fast dynamics are governed by:
\begin{align}
    \dot V = \nabla^2 V + F(V,x)  =- \fdv{\Phi}{V} \label{eq:tauinf}
\end{align}
where $F(V,x) = f(V, h_\infty(V_0(x)), x)$. As in Methods~\S\ref{sec:boundary}, 
we seek to find the critical points of $\Phi$, i.e. solve Eq.~(\ref{eq:tauinf}) with $\dot V =0$. This amounts to solving the system:
\begin{align}
    \partial_x V =& P \label{eq:hamil3} \\
    \partial_x P =& - F(V,x) \label{eq:hamil4}
\end{align}
with boundary conditions $P(\pm \ell)=0$. 

In the limit $\ell \to \infty$ we construct the solutions to Eqs.~(\ref{eq:hamil3}-\ref{eq:hamil4}) as follows: 
\begin{enumerate}
\item First define
\begin{align}
P_{\ce{Na}}(V,V_R) = \sqrt{ 2 r \int_{V}^{V_R} f_{\ce{Na}}(V') \dd V' }
\end{align}
\item Second, define $V_1(V_R)$ to be the point of intersection between the left separatrix and the right solution curves, i.e.  $P_{\ce K}(V_1) = P_{\ce{Na}}(V_1, V_R)$, as shown in Fig.~\ref{fig:topo}d. 
\item Thirdly, compute 
\begin{align}
    X (V_R) = \int_{V_1(V_R)}^{V_R} \frac{1}{P_{\ce{Na}}(V, V_R)} \dd V \label{eq:areas}
\end{align}
which is the distance in real space between the locations where the voltage crosses $V_1$ and $V_R$.
\item Finally, as illustrated in Fig.~\ref{fig:topo}e, critical points [i.e. solutions of Eq.~(\ref{eq:hamil3}-\ref{eq:hamil4})] correspond to solutions of the equation $X(V_R) = x_*$. 
\end{enumerate}

As illustrated in Fig.~\ref{fig:topo}f, the construction of $X(V)$ can be visualized in real space: for a range of trial solutions with different slopes at the interface, the points at which each curve first achieves $\partial_x V = 0$ forms the graph $(X(V),V)$. Critical points of $\Phi$ are the curves for which $X(V) = x_*$.
In the S.I., we analyze a special case in which $f_{\ce{K}}$ and $f_{\ce{Na}}$ are piecewise linear, allowing $X(V)$ and $V_0(x)$ to be computed analytically in terms of trigonometric functions.

We note that the precise shape of $X(V)$ depends on the functions $f_{\ce{K}}$ and $f_{\ce{Na}}$ and the parameter $r$. Nevertheless, the hypotheses $f_{\ce{K}} (V) <0$ and $f_{\ce{Na}}(V) >0$ for  $V \in (V_{\ce{K}}, V_{\ce{Na}} )$ imply that $X(V_{\ce{K}}) =0$ and $X(V_{\ce{Na}}) = \infty $. If we specialize to the situation in which $X(V)$ is ``N" shaped, we can   
use the same stability arguments as in Methods~\S\ref{sec:boundary} to obtain the spiking phase diagrams shown in Fig.~\ref{fig:phase}e,h,k: If $X(V_*)$ has degeneracy 1, then the system cannot spike (see Fig.~\ref{fig:topo}g). 
If $X(V_*)$ has degeneracy 3, then the system can spike. If $V_*$ lies on an increasing branch of $X$, then $V_0(x)$ is stable and the system is excitable (see Fig.~\ref{fig:topo}h). If $V_*$ lies on the decreasing branch of $X$, then $V_0(x)$ is unstable and the system exhibits a oscillations (see Fig.~\ref{fig:topo}i). 

\begin{figure*}[t!]
    \centering
    \includegraphics[width=\textwidth]{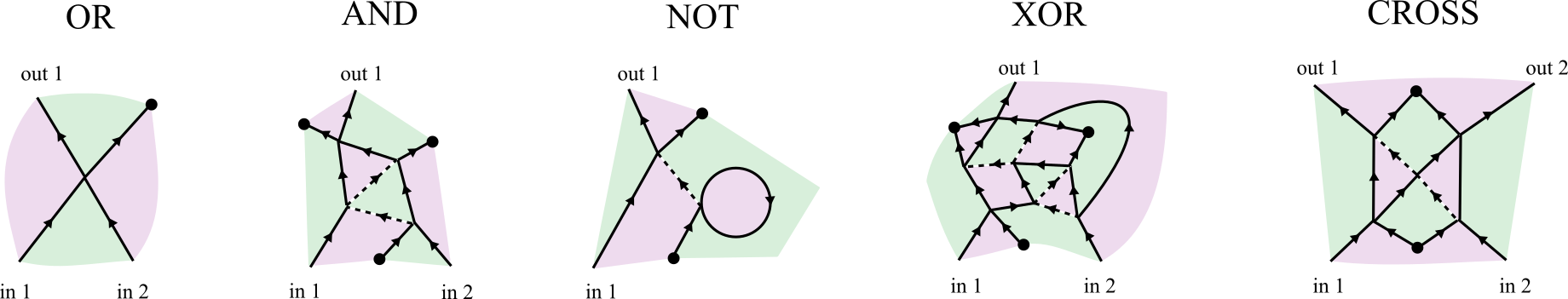}
    \caption{{\bf Examples of OR, AND, NOT, XOR, and CROSS logic gates.} The black lines represent interfaces between distinct materials (purple and green). Solid lines fit an integer number of wavelengths, while dashed lines fit a half integer number of wavelengths. The black circles represent termination points. }
    \label{fig:gates}
\end{figure*}

\subsubsection{Limits: $r \to 0$, $r \to \infty$}

If the potassium channels are much stronger than the sodium channels, then one  expects that the interface will behave as a Dirichlet boundary with $V(0) = V_{\ce{K}}$. 
This can be shown mathematically by taking the limit $r \to 0$. 
From the definition of $V_1(V_R)$, namely $P_{\ce{K}}(V_1) = P_{\ce{Na}}(V_1, V_R) $, we obtain the relationship
\begin{align}
    \int_{V_{\ce{K}}}^{V_1} f_{\ce {K}}(V) \dd V + r \int_{V_1}^{V_R} f_{\ce{Na}}(V) \dd V = 0 \label{eq:areas}
\end{align}
From Eq.~(\ref{eq:areas}) one sees that $V_1 \to V_K$ as $r \to 0$. In this limit, the expression for $X(V_R)$ [Eq.~(\ref{eq:areas})] becomes:
\begin{align}
    X_\text{low}(V_R) = \frac1{\sqrt {2r}} \int_{V_{\ce K}}^{V_R} \frac{1}{\sqrt{U_{\ce{Na}}(V)-U_{\ce{Na}}(V_R)}} \dd V \label{eq:MXlow}
\end{align}
where $U_{\ce{Na}}$ is an antiderivative of $f_{\ce{Na}}$. 
Notice that Eq.~(\ref{eq:MXlow}) has the same functional form as Eq.~(\ref{eq:xdef}). Crucially, $r$ only appears as a multiplicative prefactor of $X(V_R)$, implying that the phase boundaries become independent of $r$ as $r \to 0$.  Furthermore, the form of $X(V)$ and hence the phase diagram at low $r$ only depends on the functional form $f_{\ce {Na}}$ and on the zero crossing $V_{\ce{K}}$, but not on the detailed functional form of $f_{\ce{K}}$. 

Similarly, in the limit that $r \to \infty$, we may write $V_1=V_R-\delta(r)$ where $\delta(r) \ll 1$. From Eq.~(\ref{eq:areas}), one concludes
\begin{align}
    \delta = \frac{U_{\ce K} (V_R) - U_{\ce K } (V_{\ce K})}{ r f_{\ce{Na}}(V_R) } + \order{1/r^2}
\end{align}
where $U_{\ce K}$ is an antiderivative of $f_{\ce {K}}$. 
Then the expression for $X(V_R)$ [Eq.~(\ref{eq:areas})] becomes 
\begin{align}
    X_\text{high} (V_R) &= \sqrt{\frac{2\delta}{f_{\ce{Na}} (V_R) }}  \\
    &= \frac{\sqrt 2}{  r}  \frac{ \sqrt{ U_{\ce K} (V_{\ce K}) - U_{\ce K} (V_R)  } }{f_{\ce{Na}} (V_R)} \label{eq:MXhigh}
\end{align}
Once again, $r$ appears a multiplicative prefactor, implying that the phase boundaries become vertical in the $V_*$-$r$ plane. Notice that Eq.~(\ref{eq:MXhigh}) depends on the functional form of both $f_{\ce{Na}}$ and $f_{\ce{K}}$, implying that $X_\text{high}(V_R)$ and $X_\text{low}(V_R)$ can be tailored independently of each other.

\subsubsection{Shape of unstable modes}
 
Here we discuss why the spatial profile of the edge spike is often localized near the interface or boundary (see Fig.~\ref{fig:overview}a-b for example).  
As illustrated in Fig.~\ref{fig:topo}f, the critical points of $\Phi$ (i.e.\ the curves intersecting green circles) are monotonically increasing. 
However, when $\tau$ is finite, the voltage profile $V(x,t)$ approaches the critical points, but does not completely reach them because the voltage takes a non-negligible amount of time to diffuse out to the boundaries (especially for large systems).  
Instead, the spatial extent of the spike is better approximated by the shape of the unstable mode associated with the unstable critical point. 
Given a critical point $p$, with spatial profile $V_p(x)$, the linearized dynamics obey: 
\begin{align}
    \partial_t \delta V = - \HH \delta V \label{eq:schrodinger} 
\end{align}
where $\delta V (x) = V(x) - V_p(x)$ and 
\begin{align}
    \HH(x) = -\nabla^2 + U(x) 
\end{align}
with $U(x) =- \eval{\pdv{F}{V}}_{V_p (x)}$. Notice that Eq.~(\ref{eq:schrodinger}) is a 1D Schr\"odinger equation with potential $U(x)$. As an illustration, Fig.~\ref{fig:linearization}a show the voltage profile for a biolectric interface for which $\Phi$ has three critical points, $V_0(x)$, $V_1(x)$, and $V_2(x)$. Figure~\ref{fig:linearization}b-d shows the effective potential for each of these three critical points. (In this example,  $f_{\ce{K}}$ and $f_{\ce{Na}}$ are chosen to be piecewise linear, and hence each $U_p(x)$ is a square well). 
Negative eigenvalues of $\HH$ correspond to unstable modes. Since we require $\eval{\partial_x V}_{\pm \infty } = 0$ and $f_{\ce{K}}(V)$ and $f_{\ce{Na}}(V)$ are both decreasing at their zero crossings, it follows that $U(x\to \pm \infty)$ approaches non-negative constants  $C_1=- h_\infty(V_{\ce{Na}}) \partial_V f_V $ and $C_2=-\partial_V f_K$.
Therefore negative energy states, and hence the unstable modes, correspond to bound states of $U(x)$. 
Naturally, these bound states are confined to the well formed by $U(x)$, which coincides with the region in which $V_p(x)$ interpolates between its two asymptotic values.  Examples of the linear spectrum for each critical point are shown Fig.~\ref{fig:linearization}b-d. In accordance with the stability of each critical point, only panel c has a negative energy state, which is highlighted in red.

\subsubsection{Example ion channels}
Here we provide the expression for the ion channels used in  Fig.~\ref{fig:phase}.
For panels (c-e), we use:
\begin{align}
f_{\ce{K}} (V) =& \frac{(V_{\ce{K}}-V) e^{-2.5 (V-V_{\ce{K} } ) }}{N_1}  \\
f_{\ce{Na}} (V) =& \frac{(V_{\ce{Na}}-V) e^{-2.5 (V-V_{\ce{Na}} )}}{N_2 }
\end{align}
For panels (f-h), we use: 
\begin{align}
f_{\ce{K}} (V) =& \frac{- e^{-(V-V_{\ce{K}})} +e^{ - 2 (V-V_{\ce{K}}) }}{N_3} \\
f_{\ce{Na}} (V) =& \frac{e^{(V-V_{\ce{Na}})} -e^{  2 (V-V_{\ce{Na}}) } }{N_4}
\end{align}
For panels (i-k), we use:
\begin{align}
f_{\ce{K}} (V) =& \frac{0.3 (1-e^{ (V-V_{\ce{K}})})}{N_5} \\
f_{\ce{Na}} (V) =& \frac{e^{2.5 V} (V_{\ce{Na}}-V)}{N_6}
\end{align}
with $V_{\ce K} =-1$ and $V_{\ce {Na}} = 1$. The normalization constants $N_1, \dots, N_6$ are chosen such that the local minimum or maximum of each ion channel is normalized to $-1$ and $1$, respectively.  

\begin{figure}[t!]
    \centering
    \includegraphics[width=\columnwidth]{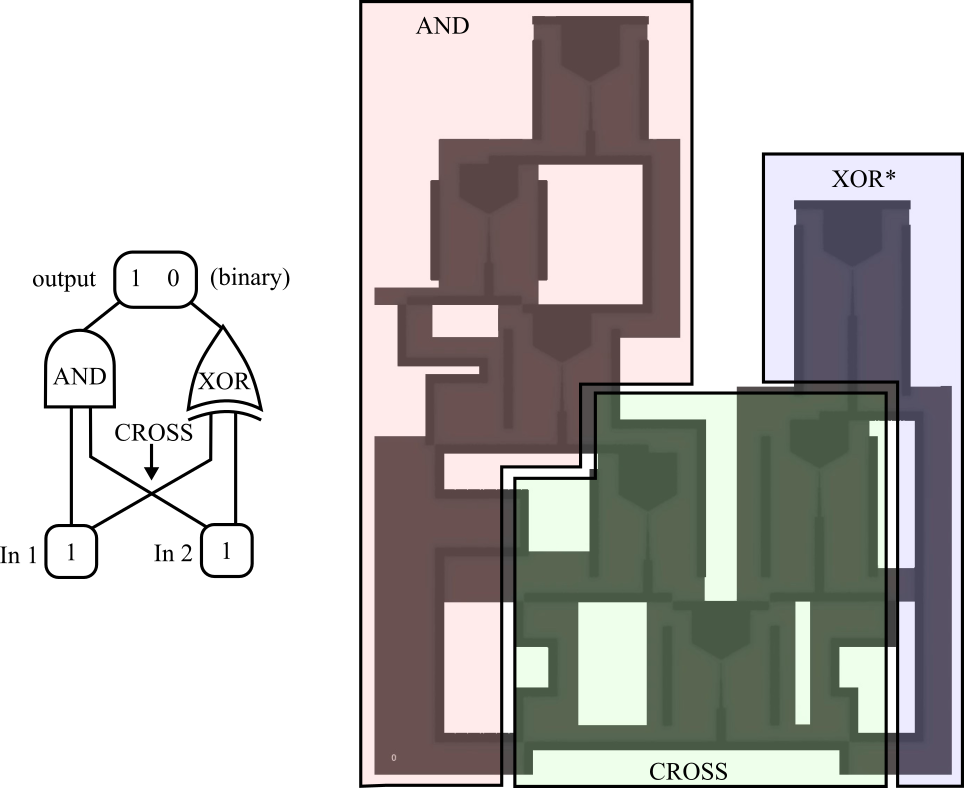}
    \caption{{\bf Decomposition of binary half adder.}~A binary half adder can be decomposed into and AND gate and an XOR gates. Since the XOR gate generates an output of the system, it can be simplified to the XOR$^*$ gate.  
    }
    \label{fig:decomp}
\end{figure}

\subsection{Oscillating chemical reactions} 
\label{sec:chem}
The chemical dynamics we study involve two chemical reservoirs undergoing distinct chemical reactions. 
In the rightmost chamber, we consider a prototypical chemical reaction network known as the Oregonator~\cite{Field1972Oscillations}. The Oregonator can be summarized by 5 elementary reactions
\begin{align} 
\ce{R +A ->[$k_1$] & C + P} \label{eq:step1} \tag{R1}  \\
\ce{C + A ->[$k_2$] & 2P } \label{eq:step2} \tag{R2} \\
\ce{R + C ->[$k_3$] & 2C + 2B} \tag{R3} \\
\ce{2 C ->[$k_4$] & R + P} \label{eq:step4} \tag{R4} \\
\ce{B ->[$k_5$] & $m_1$ A} \label{eq:organic} \tag{R5} 
\end{align} 
with rate constants $k_i$. Here $\ce{R}$ is a reactant, $\ce{P}$ is the product, $\ce{A}$, $\ce{B}$, and $\ce{C}$ are catalysts, and $m_1$ is a stoichiometric coefficient. The Oregonator was originally proposed as a minimal model for the Belousov-Zhabotinsky reaction~\cite{Tyson1976Belousov}. In this context, the variables can be roughly interpreted as: \ce{R = BrO_3^-}, \ce{P = HOBr},  \ce{A= Br-}, \ce{B=Ce^{4+}}, and \ce{C= HBrO_2}, and suitable rate constants can be determined from experiments. See Ref.~\cite{Tyson1976Belousov} for a detailed introduction.

In the right chamber, we assume that the reactant \ce{R} is abundant, so its concentration can be treated as constant. Moreover, the product \ce{P} is assumed to exit the reaction and not affect the subsequent dynamics. Hence, we need only consider the kinetic equations for the intermediates $\ce{A}$, $\ce{B}$, and $\ce{C}$:
\begin{align}
    \dot A =& - k_1 R\, A - k_2 C\, A + m_1 k_5 B \\
    \dot B =& 2 k_3 R\, C - k_5 B \\
    \dot C =& k_1 R\, A - k_2 C\, A + k_3 RC - k_4 C^2 
\end{align}
Here, $A$ denotes the concentration of component $\ce{A}$, etc. 
 We nondimensionalize the kinetic equations by introducing a time scale $t_0=\frac{k_4}{k_3 k_2 R}$ and defining
\begin{align}
    a=& \frac{k_2}{R k_3} A &
    b =& \frac{k_4 k_5}{2 (R k_3)^2 } B  \\
        c =& \frac{k_4}{R k_3} C & 
    \tilde t =& t/t_0 
\end{align}
resulting in
\begin{align}
    \dot a =& 2 m_1 b- a (c+ m_2)  \label{eq:y} \\
    \dot b =& \frac{ c - b }{\tau } \label{eq:z} \\
    \epsilon \dot c =& a(m_2-c) + c(1 -  c)  \label{eq:x} 
\end{align}
with the parameters
\begin{align} 
    \varepsilon =& \frac{ k_2}{k_4}  &
    \tau =& \frac{k_2 k_3 R }{k_4 k_5}  &
    m_2 = &  \frac{k_1 k_4}{k_2 k_3} 
\end{align}
We are interested in the regime $\varepsilon \ll 1$ and $\tau \gg 1$. 
With $\varepsilon \ll 1$, we may integrate out Eq.~(\ref{eq:x}) and define 
\begin{align}
  c(a)= b_\infty(a) \equiv \frac12 \qty[ (1-a) + \sqrt{ (1-a)^2 + 4 m_2 a } ] \label{eq:zinf}
\end{align}
Thus the dynamics become 
\begin{align}
    \dot a =& 2 m_1 b- a (b_\infty(a) + m_2)  \label{eq:yred} \\
    \dot b =& \frac{ b_\infty(a) - b }{\tau } \label{eq:zred} 
\end{align}
The $b$-nullcline is given by $b = b_\infty$  and the $a$-nullcline $\dot a =0 $ is given by
\begin{align} 
b = b_\text{eq} (a) \equiv \frac{ a \qty( b_\infty(a) + m_2 )}{2 m_1} \label{eq:beq}
\end{align}

Depending on the values of $m_1$ and $m_2$, Eqs.~(\ref{eq:yred}-\ref{eq:zred}) can exhibit no-spiking, excitable, and oscillating phases. Our interest is in the regime in which $m_1$ is sufficiently large to create local excitability, as shown in Fig.~\ref{fig:chemphase}.

For a spatially extended system, Eqs.~(\ref{eq:x}-\ref{eq:z}) become
\begin{align}
    \dot a =& \nabla^2 a+ 2 m_1 b- a (c+ m_2)  \label{eq:rdy} \\
    \dot b =& \beta \nabla^2 b + \frac{ c - b }{\tau } \label{eq:rdz} \\
        \varepsilon \dot c =& \alpha \nabla^2 c +  a(m_2-c) + c(1 -  c)  \label{eq:rdx}
\end{align}
In Eqs.~(\ref{eq:rdy}-\ref{eq:rdz}), we nondimentionalize length using $x \to x/\sqrt{D}$  where $D = D_A t_0$ and $D_A$ is the diffusion constant for species \ce{A}.  
In Eq.~(\ref{eq:rdx}) and Eq.~(\ref{eq:rdz}), $\alpha = \varepsilon \frac{D_C}{D_A}$ and $\beta = \frac{D_B}{D_A}$, where $D_C$ and $D_B$ are the diffusion constants for \ce{C} and \ce{B}, respectively. Since $\varepsilon \ll 1$, we have $\alpha \ll 1$ and therefore $c$ can be locally integrated out using Eq.~(\ref{eq:zinf}). 

In the leftmost chamber, we assume that the catalyst $\ce{A}$ is rapidly converted into a product and exits the reaction:
\begin{align}
    \ce{A ->  P} \label{eq:left1} \tag{L1}
\end{align}
When the species $\ce{A}$ is allowed to diffuse across the interface, the governing equations take the form:
\begin{align} 
\dot a =& \nabla^2 a + \begin{cases} 
 - a  & x \in [- \ell ,0] \label{eq:Mybulk} \\
2 m_1 b- a [b_\infty(a)  +m_2] & x \in [0, \ell] 
\end{cases}\\
\dot b =&  \frac{b_\infty(a) -b}{\tau } \label{eq:Mzbulk}
\end{align} 
For Fig.~\ref{fig:overview}c, we choose $m_1=2$ and $m_2=10^{-3}$. In this case, the $x > 0$ reservoir displays excitability (see the phase portrait in Fig.~\ref{fig:chemphase}), and the $x<0$ reservoir exhibits no spiking. Crucially, however, when the catalyst $\ce{A}$ is allowed to diffuse, the full interfacial system exhibits oscillations.

\subsection{Two-dimensional media: interfacial trigger waves}
\label{sec:trigger}
In two dimensions, an interface between two media forms a 1D line. If the interface is excitable, then the 1D line can host a trigger wave, as illustrated by the bioelectric experiments in Fig.~\ref{fig:overview}a.  
In the notation of Eqs.~(\ref{eq:broad1}-\ref{eq:broad2}),
a trigger wave along the interface is described by a profile $A(x,y,t)=A(x,y-c\,t)$ where $x$ runs transverse to the interface, $y$ runs parallel to the interface, and $c$ is the wave speed. 
In a $\tau \to \infty$ approximation, the sharp front along the interface is described by:
\begin{align}
    D\partial_y^2 A = - c \partial_y A + \fdv{\Phi}{A} \label{eq:along}
\end{align}
which is a higher dimensional version of the profile equation for standard trigger waves~\cite{Talia2022Waves}. 
When the interface is excitable, $\Phi$ has two minima, the stationary solution $A_0(x)$ and an additional minimum $A_1(x)$. In the simplest approximation, $A_0(x)$ and $A_1(x)$ are the boundary conditions of Eq.~(\ref{eq:along}) as $y \to + \infty$ and $-\infty$, respectively. 
The conditions for trigger-wave propagation can then be understood by a classic rolling ball analogy~\cite{Vergassola2018Mitotic}: Eq.~(\ref{eq:along}) describes a ball of mass $D$ moving with damping $c$ between two maxima [$A_0(x)$ and $A_1(x)$] of a potential $-\Phi$. 
The front moves in the direction that expands the low potential (larger $\Phi$) region, and 
the wave speed $c$ corresponds to the (unique) value of dissipation that allows the ball to arrive at rest at the top of the lower peak of $-\Phi$. Notably, an undisturbed system will initially exhibit a uniform profile $A(x,y)=A_0(x)$. Consequently,  if $\Phi[A_1] > \Phi[A_0]$, then a sufficiently intense local perturbation will cause the $A_1$ region to expand, implying that a trigger wave will propagate outward in both direction from the initial perturbation. If $\Phi[A_1] < \Phi[A_0]$, then the initial perturbation will close and the trigger wave will not propagate.

\subsection{Design of the binary half adder}
\label{sec:adder}

Here we comment on the design of the binary half adder in Fig.~\ref{fig:waveguide}c-d. Each input channel encodes a Boolean value: \emph{true} corresponds to the presence of a wave train, and \emph{false} corresponds to the absence of a wave train.  Within this paradigm, a self-contained logic gate receives input wave trains, subjects them to nonlinear interference, and produces output wave trains. Figure~\ref{fig:gates} shows examples of canonical  OR, AND, NOT, and XOR logic gates. In each of these diagrams, a solid line indicates an interface of length $ n \lambda$ and a dashed line indicates an interface length $(n+1/2) \lambda$, where $n$ is an integer, and $\lambda = T c$, where $T$ is the period of the input pulses and $c$ is the wave velocity along the interface. 
Each of these logic gates is constructed using only two materials, which implies that each junction must comprise at least four interfaces. 
The device shown in Fig.~\ref{fig:waveguide}c-d is known as a binary half adder, which takes the sum of two binary numbers. 
As shown in Fig.~\ref{fig:decomp}, the binary half adder can be constructed out of XOR and AND gates.   
In order to embed the logic gates in the 2D plane, an additional trivial gate $\operatorname{CROSS}(x,y) = (x,y)$ must be implemented that allows two pulses to cross each other without performing a computation. 
To prevent undesired back scatter at interference junctions, a needle geometry shown in Fig.~\ref{fig:waveguide}a-b is used. 

In order for the gates in Fig.~\ref{fig:gates} to be composable, the phase of the output wave train must be independent of the choice of logically equivalent inputs. For example $\text{XOR}(T,F)=T$ and $\text{XOR}(F,T)=T$, so both inputs $(T,F)$ and $(F,T)$ must produce an output wave train with the same phase.  However, if the output of the logic gate is the final output for the device, then the phase will not be read so certain logic gates, such as XOR, can be simplified. For example, in the binary half adder  the XOR logic gate is simplified to XOR$^*$, as shown in Fig.~\ref{fig:decomp}.

\onecolumngrid

\clearpage

\renewcommand{\theequation}{S\arabic{equation}}
\setcounter{equation}{0}
\setcounter{figure}{0}
\renewcommand{\thetable}{S\arabic{table}}  
\renewcommand{\thefigure}{S\arabic{figure}}
\renewcommand\figurename{Fig.}
\renewcommand{\thesection}{S\arabic{section}}
\setcounter{subsection}{0}

\section*{Supplementary Information}

\subsection{Supplementary videos}

\noindent {\bf Supplementary Video S1: Experimental observation of spiking at a bioelectric interface.} Two tissues of human embryonic  kidney (HEK293) cells were genetically modified to express either sodium (Na$_\text{V}$1.5) or potassium (K$_\text{ir}$2.1) channels. Neither tissue alone is able to spike. However, upon stimulation by a laser, an action potential is observed to propagate along their interface, as revealed by a voltage sensitive red dye. Adapted from Ref.~\cite{Ori2023Observation}.

\

\noindent {\bf Supplementary Video S2: 2D nonlinear waveguides built from excitable interfaces.} A network of excitable intefaces forms a two-dimensional nonlinear waveguide capable of performing computations. Simulations of the binary half adder in Fig.~\ref{fig:waveguide}c-d are shown. See \S\ref{sec:numer} for details of the numerics.

\clearpage

\subsection{Numerics} 
\label{sec:numer} 
\noindent Here we provide the details for the numerical simulations presented in main text. 

\vspace{5mm} 

\noindent {\bf Figure \ref{fig:overview}b.}~We integrate 
\begin{align}
    \dot p =& \nabla^2 p +\begin{cases}
        - p & x \in [-\ell, 0] \\
        n \, p \, \scr(p) & x \in (0, \ell]
    \end{cases} \label{eq:sim1bp} \\
    \dot n =& \frac{n}{\tau }\qty[ 1-n - \frac{p}{q} ] \label{eq:sim1bn}
\end{align}
with no-flux boundary conditions at $x = \pm \ell$. The parameters used are $\tau=100$, $q=0.5$, $\ell =100$.  
The function $k(p)$ is given by Eq.~(\ref{eq:kp}) with $p_*=0.7$ and $n_*=3$. An initial condition 
\begin{align}
p(x,0) =&  0.5 (\tanh(x/10)+1.1) \\
n(x,0) =& 0.25 ( \tanh(-x/10) +1.1)
\end{align} 
is used and a transient is allowed to pass prior to the time interval shown in Fig.~\ref{fig:overview}b. 
The equations are discretized on a 1D lattice and integrated in Python using \texttt{scipy.integrate.solve\_ivp}.

\vspace{5mm} 

\noindent {\bf Figure \ref{fig:overview}c.}~We integrate Eqs.~(\ref{eq:Mybulk}-\ref{eq:Mzbulk}) with parameters $\ell =100$, $\tau=500$, $m_1=2.0$ and $m_2=0.001$. Our initial conditions are $a(x,0)= b(x,0) =0.1$. An initial transient is allowed to pass prior to the time interval shown in Fig.~\ref{fig:overview}c. The equations are discretized on a 1D lattice and integrated in Python using \texttt{scipy.integrate.solve\_ivp}.

\vspace{5mm}

\noindent {\bf Figure~\ref{fig:coupling}e-g.}~The equations integrated are
\begin{align}
    \dot p_i = &  \epsilon( p_{i-1}+p_{i+1} - 2 p_i  ) + n_i \, p_i \, k(p_i)  \\
    \dot n_i = & \frac{n_i}{\tau } \qty(1 - n_i -\frac{p_i}{q} )
\end{align}
for $i = 1,\dots, N$. Here $p_0=0$ represents the desert and setting $p_{N+1}\equiv p_N$ implements a no-flux boundary condition. 
The parameters used are $\epsilon =4$, $q=0.5$, $\tau =20$. The function $k(p)$ is given by Eq.~(\ref{eq:kp}) with $p_*=0.7$ and $n_*=3$. The initial conditions are given by $p_i=n_i=0.6$. The equations are integrated in Python using \texttt{scipy.integrate.solve\_ivp}.

\vspace{5mm}

\noindent {\bf Figure~\ref{fig:isochrones}g-i.}~We integrate the equations 
\begin{align}
    \dot V = & \nabla^2 V+ \begin{cases}
        f_{{\ce K}} (V)  & x \in [- \ell ,0] \\
        h f_{{Na}} (V) & x \in (0, \ell] 
    \end{cases} \label{eq:numer1} \\
    \dot h =& \frac{r \Theta(V_*-V) - h }{\tau } \label{eq:numer2}
\end{align}
with 
\begin{align}
    f_{\ce{ K }} (V) =& - V \label{eq:piecelin1} \\ 
    f_{\ce{Na}}(V) =& 
\begin{cases}
    0 & V< V_a \\
    \frac{ b (V - V_a)  }{ V_b - V_a} &  V_a \le V < V_b \\
    \frac{ (1-b) (V-V_b) }{ W_{\ce{Na}} - V_b } +b& V_b  \le V < W_{\ce{Na}}  \\
    \frac{V-V_{\ce{Na}}}{ W_{\ce{Na}} - V_{\ce{Na}}} & V \ge  W_{\ce{Na}}
\end{cases} \label{eq:piecelin2}
\end{align}
In all three kymographs, we use $V_a=-0.1$, $V_b = 0.25$, $W_{\ce{Na}} = 0.5$, $b=0.1$, $V_{\ce{Na}}=1$,  $r =0.1$, $\ell =100$,  and $\tau =1000$. For panels g, h, and i, we set $V_*=-0.1$, $V_*=0.2$, $V_*=0.3$, respectively.
For panel g, the initial conditions are given by $V(x,0)= 0.3$, $h(x,0)=0.0$. 
For panel h, we first initialize the voltage profile to 
$V (x) = 0.1 \qty( \tanh(x/10)+1.1)$. We then perform a relaxation according to
\begin{align}
    \dot V =& \nabla^2 V +  \begin{cases}
        f_{{\ce K}} (V)  & x \in [- \ell ,0] \\
         r \Theta(V_*-V) f_{\ce{Na}} (V) & x  \in [0, \ell] 
    \end{cases} \\
    h =& r \Theta(V_*-V) 
\end{align}
to obtain stationary profiles $(V_0(x), h_0(x))$. Finally, we integrate Eqs.~(\ref{eq:numer1}-\ref{eq:numer2}) using the initial conditions  $V(x,0) = V_0(x) + 1.1$ and $h(x,0)= h_0(x)$.
For panel i, the initial conditions are given by:
\begin{align}
    V(x,0)=& 0.1 \qty(\tan(x/10) +1.1 ) \\
    h(x,0) =& 0.5
\end{align}
For all panels, the equations are discretized onto a 1D lattice and integrated in Python using \texttt{scipy.integrate.solve\_ivp}.

\vspace{5mm} 

\noindent {\bf Figure \ref{fig:waveguide}.}~We integrate the equations:
\begin{align}
    \dot V =& \nabla^2 V + s(x,y,t) + \begin{cases}
        f_{{\ce K}} (V)  & (x,y) \in \text{Region 1} \\
        h f_{\ce{Na}} (V) & (x,y) \in \text{Region 2} 
    \end{cases} \label{eq:2DV} \\
    \dot h =& \frac{r \Theta(V_*-V) - h }{\tau } \label{eq:2Dh}
\end{align}
where $f_{\ce{K}}$ and $f_{\ce{Na}}$ are given by Eqs.~(\ref{eq:piecelin1} -\ref{eq:piecelin2}) 
with $V_a=-0.1$, $V_b = 0.25$, $W_{\ce{Na}} = 0.5$, $b=0.1$, $V_{\ce{Na}}=1$, $r=0.1$, $V_*=0.15$, and $\tau =300$. 
The wave trains are generated by repeated Gaussian pulses
\begin{align}
    s(x,y,t) = \sum_{\alpha}  A_\alpha(t) \exp{  \frac{ (x-x_\alpha)^2 - (y- y_\alpha)^2 }{400}}
\end{align}
where $(x_\alpha, y_\alpha)$ are the points at which the pulses are initialized. Here, the amplitude is a periodic square wave:
\begin{align}
    A_\alpha(t) = \sum_{n \in \ZZ} \Theta( 1- t - n T -D_\alpha) - \Theta(t-nT)
\end{align}
where $T = 2810$ is the period and $D_\alpha$ is a phase shift (only used to produce the interfering wave trains in Fig.~\ref{fig:waveguide}b). The Laplacian is discretized onto a triangular mesh. Initial conditions are obtained by initializing $V(x,y,0)= 0.16$ then performing a relaxation 
\begin{align}
    \dot V =& \nabla^2 V +  \begin{cases}
        f_{{\ce K}} (V)  & (x,y) \in \text{Region 1} \\
         r \Theta(V_*-V) f_{\ce{Na}} (V) & (x,y) \in \text{Region 2} 
    \end{cases} \\
    h =& r \Theta(V_*-V) 
\end{align}
over a time interval $t \in [0,400]$. The full dynamics, Eqs.~(\ref{eq:2DV}-\ref{eq:2Dh}), were then integrated over a time interval $t \in [0,40000]$. Integration is performed in Python using the \texttt{scipy.integrate.solve\_ivp} function. 

\clearpage 

\subsection{Explicitly solvable bioelectric interface model}

\label{sec:exact}

\begin{figure}
    \centering
    \includegraphics[width=0.3\textwidth]{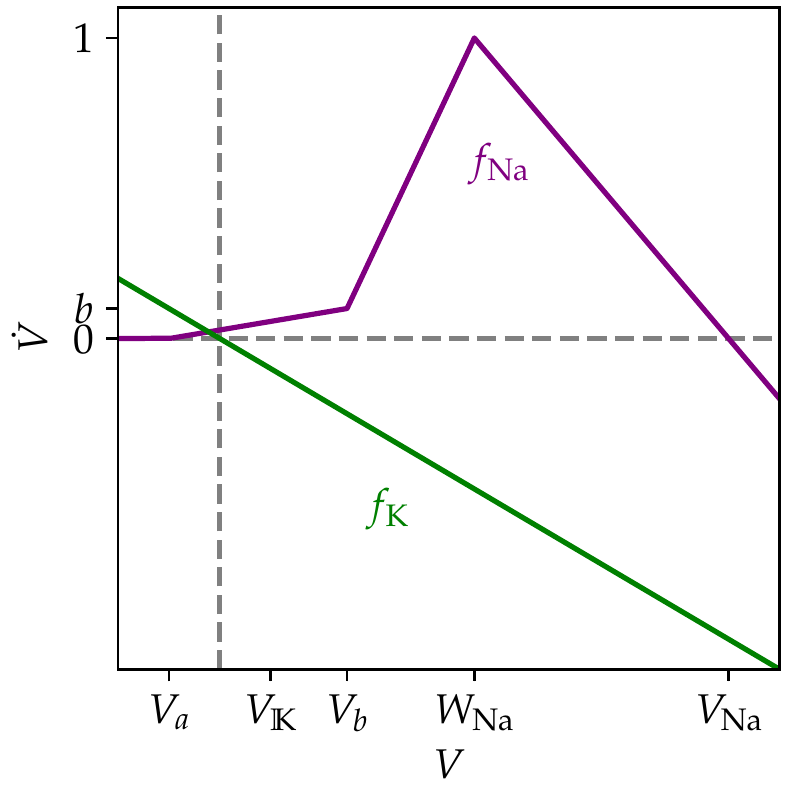}
    \caption{{\bf Piecewise linear ion channels.} }
    \label{fig:piecewise}
\end{figure}

We now present an example of Eqs.~(\ref{eq:big1}-\ref{eq:big2}) for which the stationary solutions and critical points can be computed explicitly. We note that explicit solutions are readily obtained whenever one can solve 
\begin{align}
    \nabla^2 V =& -f_{\ce K}(V) \label{eq:kk} \\ 
    \nabla^2 V =& -r f_{\ce{Na}} (V) \label{eq:nn} 
\end{align}
analytically for arbitrary initial conditions.  
One example is to take $f_{\ce K}$ and $f_{\ce Na}$ to be piecewise linear in $V$, as shown in Fig.~\ref{fig:piecewise}. 
Namely, we take:
\begin{align}
    f_{\ce K}(V) =& - m V \label{eq:kform}  \\
    f_{\ce{Na}}(V) =& \frac{b(V-V_a)}{V_b-V_a} \Theta_{[V_a,V_b)} (V) + \qty( \frac{1-b}{W_{\ce{Na}}-V_b} (V-V_b)  +b) \Theta_{[V_b,W_{\ce {Na}})} (V) + \frac{V-V_{\ce{Na}}}{W_{\ce{Na}}-V_{\ce{Na}}} \Theta_{[W_{\ce{Na}}, V_{\ce{Na}}]}(V)
\end{align}
where for a given interval $I$, $\Theta_I(V) = 1$ if $V \in I$ and $\Theta_I(V) = 0 $ if $V \not \in I$.
We will use the shorthand $\Theta_1 = \Theta_{[V_a,V_b)}$, $\Theta_2 = \Theta_{[V_b,W_N)}$, and $\Theta_3 =\Theta_{[W_{\ce {Na}},V_{\ce {Na}})} $. We require that $V_a<0< V_b < W_{\ce{Na}} < V_{\ce{Na}}$.

Given the form of Eq.~(\ref{eq:kform}), the solution to Eq.~(\ref{eq:kk}) is given by:
\begin{align}
    V(x) = V(0) e^{\sqrt{m}x}
\end{align}
where we have enforced that $\eval{\partial_x V}_{-\infty} =0$. Hence, the left separatrix is given by  $P_{\ce{K}}(V) = \sqrt{m} V$. 
The solution to Eq.~(\ref{eq:nn}) is slightly more subtle. Since $f_{\ce{Na}}$ is piecewise linear in $V$, it will be the more convenient to solve for the inverse $x(V)$. 
Given the initial conditions $V(0) = V_I$ and $\eval{\partial_x V}_{0} =P_I$, one obtains
\begin{align}
    x(V;V_I,P_I) = \int_{V_I}^{V} \frac{1}{P(V', V_I, P_I)} \dd V' 
\end{align}
where 
\begin{align}
    P(V; V_I, P_I) = \sqrt{2r \qty[F_{\ce{Na}} (V_I) - F_{\ce{Na}} (V) ] + P_I^2 }
\end{align}
and $F_{\ce{Na}}$ is an antiderivative of $f_{\ce{Na}}$. Since $f_{\ce{Na}}$ is piecewise linear, the antiderivative is straightforward to compute: 
\begin{align}
    F_{\ce {Na}} (V) = \frac12 \sum_i  \qty[A_i \qty(V-B_i)^2 + C_i ] \Theta_{i} (V) 
\end{align}
where 
\begin{align}
    A_1 =& \frac{b}{V_b- V_a} & B_1 =& V_a & C_1=&0 \\
    A_2 =& \frac{1-b}{W_{\ce{Na}}-V_b} & B_2 =& \frac{b W_{\ce{Na}} - V_b}{ b-1} & C_2 =& \frac{b(V_a - b V_a -V_b + b W_{\ce{Na}})}{b-1} \\
    A_3 =& \frac{1}{W_{\ce{Na}} - V_{\ce{Na}}} & B_3 =& V_{\ce{Na}} & C_3 =& V_{\ce{Na}}- V_b  + b(W_{\ce{Na}} -V_a) 
\end{align}
Therefore, we have:
\begin{align}
    P(V; V_I, P_I ) =  \sum_i \sqrt{ D_i - r A_i (V - B_2 )^2 } \, \Theta_i(V) 
\end{align}
where 
\begin{align}
    D_i = r \qty[A_1 (V_I - b_1)^2 +C_1 - C_i ] + P_I \label{eq:di}
\end{align}
In Eq.~(\ref{eq:di}) and onwards, we assume for simplicity that $V_I \in [V_a, V_b]$. Next, we recall the following antiderivative
\begin{align}
    \int \frac1{\sqrt{D - rA (V -B)^2}} \dd V = \frac{1}{\sqrt{rA}} \atan\qty[ \sqrt{\frac{ r A }{D - r A (V-B)^2 } } \qty(V-B)  ] 
\end{align}
Thus we obtain the solution
\begin{align}
    x(V; V_I,P_I ) = \sum_i X_i (V; V_I, P_I) \Theta_i (V)
\end{align}
where 
\begin{align}
    X_1 (V; V_I, P_I) =& \frac{1}{\sqrt{r A_1}} \qty( \atan \qty[\sqrt{\frac{ r A_1 }{D_1 - r A_1 (V-B_1)^2 } } \qty(V-B_1) ] - \atan \qty[\sqrt{\frac{ r A_1 }{D_1 - r A_1 (V_I-B_1)^2 } } \qty(V_I-B_1) ] ) \\
    X_2 (V; V_I, P_I) =& \frac{1}{\sqrt{r A_2}} \qty( \atan \qty[\sqrt{\frac{ r A_2 }{D_2 - r A_2 (V-B_2)^2 } } \qty(V-B_2) ] - \atan \qty[\sqrt{\frac{ r A_2 }{D_2 - r A_2 (V_b-B_2)^2 } } \qty(V_b-B_2) ] ) \nonumber  \\
    &+ X_1(V_b; V_I, P_I ) \\
    X_3 (V; V_I, P_I) =& \frac{1}{\sqrt{r A_3}} \qty( \atan \qty[\sqrt{\frac{ r A_3 }{D_3 - r A_3 (V-B_3)^2 } } \qty(V-B_3) ] - \atan \qty[\sqrt{\frac{ r A_3 }{D_3 - r A_3 (W_{\ce{Na}}-B_3)^2 } } \qty(W_N-B_3) ] ) \nonumber \\
    &+ X_2(W_{\ce{Na}}; V_I, P_I ) 
\end{align}

Since we now know the solution $x(V;V_I, P_I)$ for arbitrary initial conditions, we can explicitly compute the curves highlighted in the main text. For example, suppose we wish to compute $X(V)$ from Eq.~(\ref{eq:areas}). 
First we note that $V_1 (V) $ is given by solving $P(V;V_1, \sqrt{M} V_1) =0$, yielding
\begin{align}
    V_1 (V) = \sum_i  \frac{ r A_1 B_1 + \sqrt{r^2 A_1^2 B_1^2 - r(m+r A_1) ( A_1 B_1^2 - A_i (V-B_i)^2 +C_1 -C_i ) }}{r A_1 + m } \Theta_i(V)  \label{eq:v1explicit} 
\end{align}
In Eq.~(\ref{eq:v1explicit}), we have assumed that $r< \frac{m V_b^2}{V_{\ce {Na}} + b W_{\ce {Na}} - V_b(1+b)} $, so that  $V_1 \in [0, V_b ]$. Then we obtain $X(V)$ by substituting
\begin{align}
    X(V) = x\qty[V; V_1(V), \sqrt m V_1(V)] 
\end{align}
More explicitly:
\begin{align}
    X(V \in [0, V_b]) =& \frac{1}{\sqrt{r A_1}} \qty( \frac\pi2  - \atan \qty[\sqrt{\frac{ r A_1 }{D_1 - r A_1 (V_1-B_1)^2 } } \qty(V_1 -B_1) ] ) \\
    X(V \in [V_b,W_{\ce{Na}}]) =& \frac{1}{\sqrt{r A_2}} \qty( \frac \pi 2 - \atan \qty[\sqrt{\frac{ r A_2 }{D_2 - r A_2 (V_b-B_2)^2 } } \qty(V_b-B_2) ] ) + X_1(V_b; V_1, \sqrt m V_1 ) \\
    X(V \in [W_{\ce{Na}},V_{\ce{Na}}]) =& \frac{1}{\sqrt{r A_3}} \qty( \frac \pi 2  - \atan \qty[\sqrt{\frac{ r A_3 }{D_3 - r A_3 (W_{\ce{Na}}-B_3)^2 } } \qty(W_{\ce{Na}}-B_3) ] ) + X_2(W_{\ce{Na}}; V_1  , \sqrt m V_1 ) 
\end{align}
where the $V$ dependence enters through $D_i[V_1(V)]$ and $V_1(V)$.

\clearpage 

\subsection{Preliminaries on the Conley index} 
\label{sec:conley} 

\subsubsection{Homotopy Conley index} 
Here we provide supplemental background on the Conley index~\cite{conley1978isolated,Mischaikow2002Conley} and provide derivations of the associated results used in Methods~\S\ref{sec:boundary}, \ref{sec:FN}. 
Let $X$ be a metric space. A flow $\phi$ on $X$ is a continuous map $\phi: X \times \RR \to X$ obeying the following conditions $\phi(x,0)=x$ and  $\phi(x,a+b) = \phi(\phi(x, a), b) $ for all $x \in X$ and $a, b \in \RR$. 
An invariant set $S \subset X$ is one which obeys $\phi(S, \RR) = S$. In other words, it is any union of solution curves of the flow. Given a set $N \subset X$, we define $\inv(N) = \{ x \in N : \phi(x, \RR) \subset N \} $. A compact set $N$ is known as an isolating set if $\inv(N) \cap \partial N = \emptyset $. We say that an isolating set $N$ is an isolating neighborhood of the invariant set $S$ if $S = \inv(N) $. An invariant set $S$ is an isolated invariant set if it permits an isolating neighborhood. A pair of compact sets $(N,L)$ with $L \subset N$ is known as an index pair if they satisfy the three following conditions:
\begin{enumerate}
    \item $\cl(N\setminus L)$ is an isolating set
    \item $L$ is forward invariant in $N$. That is $N \cap \phi(L, [0, \infty)) \subset L $. 
    \item $L$ is an exit set for $N$. That is for all $x \in N$, if $\phi(x, [0, \infty)) \not \subset N$, then there exists  $ t >0 $ such that $\phi(x, [0, t)) \subset N$ and $ \phi(x, t) \in L$. 
\end{enumerate}
We say that $(N, L)$ is an index pair for an isolated invariant set $S$ if $S = \inv(N\setminus L)$. One can show that for every isolating set $N$, there exists an index pair for $\inv(N)$.
Given an index pair, we define the homotopy Conley index as follows 
\begin{align}
    h(N,L) = \overline{(N/L ,[L])}
\end{align}
Here, the notation $(N/L, [L])$ denotes the pointed topological space formed by the quotient of $N$ and $L$, i.e.\ the space formed by identifying $L$ to a single point. 
Given a pointed topological space $(Y,y)$, the notation $\overline{(Y,y)}$ denotes the equivalence class of all topological spaces that are homotopically equivalent to $Y$ with fixed point $y$. If $(N,L)$ and $(N', L')$ are both index pairs for the isolated invariant set $S$, then one can show that $h(N,L) = h(N', L')$. Hence, given an isolated invariant set $S$, it is unambiguous to write $h(S)$ and refer to the Conley index as an intrinsic property of the isolated invariant set. 

The Conley index obeys two crucial properties typical of topological indices. First, it obeys an addition formula: If $S_1$ and $S_2$ are disjoint isolated invariant sets, then
\begin{align}
    h(S_1 \sqcup S_2) = h(S_1) \vee h(S_2)
\end{align}
where $\sqcup$ is the disjoint union and $\vee$ is the wedge sum, which glues two pointed spaces together by identifying their distinguished points. 
Second, $h(S)$ has a continuity property. Suppose $\phi_\lambda$ has a continuous parameter $\lambda \in \Lambda$. If $N_\lambda$ is continuous in $\lambda$ and $N_\lambda$ is an isolating neighborhood for all $\lambda \in \Lambda$, then $h(\inv(N_\lambda)) = h(\inv(N_{\lambda'} ))$ for all $\lambda, \lambda' \in \Lambda$. Hence, we can deform flows and isolating neighborhoods, and so long as the isolated invariant sets do not intersect the boundaries of the proposed isolating neighborhoods, the index is invariant throughout the deformation. 

In practice, the following facts are frequently used. For a hyperbolic fixed point $p$, $h(p) = \overline{\Sigma^d}$, where $\Sigma^d$ is the pointed $d$-sphere and $d$ is the unstable dimension of the fixed point. 
Also, if an isolating neighborhood $N$ does not contain an isolated invariant set, then $\inv(N) =\emptyset$, and $h(\emptyset)= \overline{0}$, where $\overline{0}$ denotes the space with a single point. This yields a seemingly simple but useful existence theorem: if $h(\inv(N)) \neq \overline{0}$, then $N$ must contain an isolated invariant set. 
This is useful because computing $h(\inv(N))$ only requires information about $\phi$ on the boundary of $N$. Hence, one can choose $N$ wisely to simplify computations of $h(\inv(N))$. 
Finally, we note that for a hyperbolic fixed point $p$, the unstable dimension $d$ is also known as the Morse index. For this reason, the Conley index may be thought of as generalization of the Morse index that is defined for any isolated invariant set, not just fixed points.

\subsubsection{Homological Conley index}
Homology theory is a set of algebraic tools for studying homotopy type classes.  
Given topological spaces $X \subset Y$, their relative homology refers to a sequence
\begin{align}
    H(Y,X)= \qty( H_0(Y,X), \,  H_1(Y,X), \, \dots )
\end{align}
where each $H_n$ is an abelian group. For a pointed space $(Y,y)$, the reduced homology of $Y$ is defined as $ H(Y) \equiv H(Y,y)$. There are multiple varieties of homology (i.e.\ different ways of assigning the sequence of abelian groups to topological spaces) that encode different types of topological information. Alexander-Spanier homology is a standard choice for Conley index theory. (We will simply state the facts that we need without offering a full definition.)   
All types of homology have the following two properties (among others). First, given two homotopically equivalent pointed topological space $Y$ and $Y'$, $H(Y) \cong H(Y')$, where $\cong$ denotes isomorphism. 
Hence, we can unambiguously talk about $H$ as being a map from homotopy equivalence classes to graded abelian groups. 
Next, there is a useful addition property:
\begin{align}
    H(Y \vee Y') \cong H(Y) \oplus H(Y')
\end{align}
Finally, if $X \subset  Y \subset Z $, then 
\begin{align}
\dots  \to    H_n(Y,X) \to H_n(Z,X) \to H_n(Z,Y) \to H_{n-1} (Y,X) \to \cdots \label{eq:exact}
\end{align}
is an exact sequence. By \emph{exact sequence}, we mean that there exists homomorphisms between subsequent groups in the sequence and the image of one homomorphism is equal to the kernel of the next. Exact sequences are a useful construction for the following reason: if Eq.~(\ref{eq:exact}) contains a subsequence of the form $0 \to A \to B \to 0$, then the exactness property implies $A \cong B$. 
For our purposes, the only homology groups that we will explicitly need are those of spheres:
\begin{align}
    H_n(\Sigma^d) = \begin{cases}
    0 & n \neq d \\
    \ZZ & n = d
    \end{cases} \label{eq:spherehomo} 
\end{align}
and for the single point $H(\overline 0 )  =0$.
In the context of Conley index theory, homology theory is used as follows. For any isolated invariant set $S$, one can always find an index pair $(N,L)$ such that $H(N,L) = H(N/L, [L])= H(h(S))$. The quantity $H(h(S))$ is known as the homological Conley index.

\subsubsection{Morse decompositions}
For $x \in X$, define $\alpha(x) = \bigcap_{t}  \cl(\phi(x, (-\infty, -t)))$ and $\omega(x) = \bigcap_t \cl(\phi(x, (t, \infty)))$.
Given an isolated invariant set $S$, a Morse decomposition of $S$ is a finite list $(S_1, S_2, \dots, S_n)$ of isolated invariant sets with $S_i \subset S$ such that for all $x \in S \setminus \sqcup_i S_i$ there exists $j < k$ such that $\omega(x) \subset S_j$ and $\alpha(x) \subset S_k$. For every Morse decomposition, there exists a sequence of isolating sets $(N_0, N_1, \dots, N_n)$ such that $(N_n, N_0)$ is an index pair for $S$ and $(N_i, N_{i-1})$ is an index pair for $S_i$. Such a sequence is known as a Morse filtration. The notion of a Morse filtration is useful in conjunction with Eq.~(\ref{eq:exact}). Given a subset $ (N_{i-1}, N_i, N_{i+1})$, one then obtains the exact sequence:
\begin{align}
\dots \to    H_n( h(S_{i-1}))  \to H_n (N_{i+1},N_{i-1}) \to H_n( h(S_{i})) \to H_{n-1} (h (S_{i-1})) \to \cdots 
\end{align}
It turns out that $(N_{i+1}, N_{i-1})$ is itself an index pair.

\subsubsection{Stability lemma} 
Following~Ref.~\citeSup{conley1983algebraic}, we will use the above structure to prove the following lemma. 
Suppose $N$ is an isolating set with respect to gradient flow $\phi$.  
Suppose $h(\inv(N))=\overline{0}$ and $N$ contains exactly two fixed points $a$ and $b$, and let $d_a$ and $d_b$ denote their unstable dimensions. 
Then $d_a = d_b \pm 1$ and there exists a heteroclinic orbit from $a$ to $b$ if the $-$ is taken, and from $b$ to $a$ if the $+$ is taken. 

To prove this, notice that $h(a) \vee h(b) = \overline{\Sigma^{d_a} \vee \Sigma^{d_b}} \neq \overline{0}$. Hence, there must be additional isolated invariant sets in $N$. Since we are considering gradient flow, the only possible additional isolated invariant set is a heteroclinic orbit flowing from $a$ to $b$ or from $b$ to $a$. Without loss of generality, we assume the flow runs from $a$ to $b$. Then $ (b, a)$ constitutes a Morse decomposition, so there exists a Morse filtration $(N_0, N_1, N)$. This Morse filtration gives rise to the exact sequence 
\begin{align}
    \cdots \to H_n(\Sigma^{d_b}) \to H_n(\overline{0}) \to H_n(\Sigma^{d_a} )  \to H_{n-1} (\Sigma^{d_b}) \to \cdots 
\end{align}
But noting that $H_n(\overline{0}) =0$ and using Eq.~(\ref{eq:spherehomo}), we obtain the short exact sequence:
\begin{align}
    0 \to \ZZ \to H_{d_a-1} (\Sigma^{d_b}) \to 0 \label{eq:short}
\end{align}
The short exact sequence in Eq.~(\ref{eq:short}) implies that $\ZZ \cong H_{d_a-1} (\Sigma^{d_b}) $, which implies through Eq.~(\ref{eq:spherehomo}) that $d_a = d_b+1$, as desired.


\begin{thebibliography}{81}%
\makeatletter
\providecommand \@ifxundefined [1]{%
 \@ifx{#1\undefined}
}%
\providecommand \@ifnum [1]{%
 \ifnum #1\expandafter \@firstoftwo
 \else \expandafter \@secondoftwo
 \fi
}%
\providecommand \@ifx [1]{%
 \ifx #1\expandafter \@firstoftwo
 \else \expandafter \@secondoftwo
 \fi
}%
\providecommand \natexlab [1]{#1}%
\providecommand \enquote  [1]{``#1''}%
\providecommand \bibnamefont  [1]{#1}%
\providecommand \bibfnamefont [1]{#1}%
\providecommand \citenamefont [1]{#1}%
\providecommand \href@noop [0]{\@secondoftwo}%
\providecommand \href [0]{\begingroup \@sanitize@url \@href}%
\providecommand \@href[1]{\@@startlink{#1}\@@href}%
\providecommand \@@href[1]{\endgroup#1\@@endlink}%
\providecommand \@sanitize@url [0]{\catcode `\\12\catcode `\$12\catcode
  `\&12\catcode `\#12\catcode `\^12\catcode `\_12\catcode `\%12\relax}%
\providecommand \@@startlink[1]{}%
\providecommand \@@endlink[0]{}%
\providecommand \url  [0]{\begingroup\@sanitize@url \@url }%
\providecommand \@url [1]{\endgroup\@href {#1}{\urlprefix }}%
\providecommand \urlprefix  [0]{URL }%
\providecommand \Eprint [0]{\href }%
\providecommand \doibase [0]{https://doi.org/}%
\providecommand \selectlanguage [0]{\@gobble}%
\providecommand \bibinfo  [0]{\@secondoftwo}%
\providecommand \bibfield  [0]{\@secondoftwo}%
\providecommand \translation [1]{[#1]}%
\providecommand \BibitemOpen [0]{}%
\providecommand \bibitemStop [0]{}%
\providecommand \bibitemNoStop [0]{.\EOS\space}%
\providecommand \EOS [0]{\spacefactor3000\relax}%
\providecommand \BibitemShut  [1]{\csname bibitem#1\endcsname}%
\let\auto@bib@innerbib\@empty
\bibitem [{\citenamefont
  {Winfree}(1994{\natexlab{a}})}]{Winfree1994Electrical}%
  \BibitemOpen
  \bibfield  {author} {\bibinfo {author} {\bibfnamefont {A.~T.}\ \bibnamefont
  {Winfree}},\ }\bibfield  {title} {\bibinfo {title} {Electrical turbulence in
  three-dimensional heart muscle},\ }\href
  {https://doi.org/10.1126/science.7973648} {\bibfield  {journal} {\bibinfo
  {journal} {Science}\ }\textbf {\bibinfo {volume} {266}},\ \bibinfo {pages}
  {1003} (\bibinfo {year} {1994}{\natexlab{a}})}\BibitemShut {NoStop}%
\bibitem [{\citenamefont {Bers}(2002)}]{Bers2002Cardiac}%
  \BibitemOpen
  \bibfield  {author} {\bibinfo {author} {\bibfnamefont {D.~M.}\ \bibnamefont
  {Bers}},\ }\bibfield  {title} {\bibinfo {title} {Cardiac
  excitation--contraction coupling},\ }\href {https://doi.org/10.1038/415198a}
  {\bibfield  {journal} {\bibinfo  {journal} {Nature}\ }\textbf {\bibinfo
  {volume} {415}},\ \bibinfo {pages} {198} (\bibinfo {year}
  {2002})}\BibitemShut {NoStop}%
\bibitem [{\citenamefont {ten Tusscher}\ \emph {et~al.}(2004)\citenamefont {ten
  Tusscher}, \citenamefont {Noble}, \citenamefont {Noble},\ and\ \citenamefont
  {Panfilov}}]{tusscher2004model}%
  \BibitemOpen
  \bibfield  {author} {\bibinfo {author} {\bibfnamefont {K.~H. W.~J.}\
  \bibnamefont {ten Tusscher}}, \bibinfo {author} {\bibfnamefont
  {D.}~\bibnamefont {Noble}}, \bibinfo {author} {\bibfnamefont {P.~J.}\
  \bibnamefont {Noble}},\ and\ \bibinfo {author} {\bibfnamefont {A.~V.}\
  \bibnamefont {Panfilov}},\ }\bibfield  {title} {\bibinfo {title} {A model for
  human ventricular tissue},\ }\href
  {https://doi.org/10.1152/ajpheart.00794.2003} {\bibfield  {journal} {\bibinfo
   {journal} {American Journal of Physiology-Heart and Circulatory Physiology}\
  }\textbf {\bibinfo {volume} {286}},\ \bibinfo {pages} {H1573} (\bibinfo
  {year} {2004})},\ \bibinfo {note} {pMID: 14656705}\BibitemShut {NoStop}%
\bibitem [{\citenamefont {Cheng}\ \emph {et~al.}(1993)\citenamefont {Cheng},
  \citenamefont {Lederer},\ and\ \citenamefont {Cannell}}]{Cheng1993Calcium}%
  \BibitemOpen
  \bibfield  {author} {\bibinfo {author} {\bibfnamefont {H.}~\bibnamefont
  {Cheng}}, \bibinfo {author} {\bibfnamefont {W.~J.}\ \bibnamefont {Lederer}},\
  and\ \bibinfo {author} {\bibfnamefont {M.~B.}\ \bibnamefont {Cannell}},\
  }\bibfield  {title} {\bibinfo {title} {Calcium sparks: Elementary events
  underlying excitation-contraction coupling in heart muscle},\ }\href
  {https://doi.org/10.1126/science.8235594} {\bibfield  {journal} {\bibinfo
  {journal} {Science}\ }\textbf {\bibinfo {volume} {262}},\ \bibinfo {pages}
  {740} (\bibinfo {year} {1993})}\BibitemShut {NoStop}%
\bibitem [{\citenamefont {Stern}(1992)}]{Stern1992Theory}%
  \BibitemOpen
  \bibfield  {author} {\bibinfo {author} {\bibfnamefont {M.}~\bibnamefont
  {Stern}},\ }\bibfield  {title} {\bibinfo {title} {Theory of
  excitation-contraction coupling in cardiac muscle},\ }\href
  {https://doi.org/https://doi.org/10.1016/S0006-3495(92)81615-6} {\bibfield
  {journal} {\bibinfo  {journal} {Biophysical Journal}\ }\textbf {\bibinfo
  {volume} {63}},\ \bibinfo {pages} {497} (\bibinfo {year} {1992})}\BibitemShut
  {NoStop}%
\bibitem [{\citenamefont {Witkowski}\ \emph {et~al.}(1998)\citenamefont
  {Witkowski}, \citenamefont {Leon}, \citenamefont {Penkoske}, \citenamefont
  {Giles}, \citenamefont {Spano}, \citenamefont {Ditto},\ and\ \citenamefont
  {Winfree}}]{Witkowski1998Spatiotemporal}%
  \BibitemOpen
  \bibfield  {author} {\bibinfo {author} {\bibfnamefont {F.~X.}\ \bibnamefont
  {Witkowski}}, \bibinfo {author} {\bibfnamefont {L.~J.}\ \bibnamefont {Leon}},
  \bibinfo {author} {\bibfnamefont {P.~A.}\ \bibnamefont {Penkoske}}, \bibinfo
  {author} {\bibfnamefont {W.~R.}\ \bibnamefont {Giles}}, \bibinfo {author}
  {\bibfnamefont {M.~L.}\ \bibnamefont {Spano}}, \bibinfo {author}
  {\bibfnamefont {W.~L.}\ \bibnamefont {Ditto}},\ and\ \bibinfo {author}
  {\bibfnamefont {A.~T.}\ \bibnamefont {Winfree}},\ }\bibfield  {title}
  {\bibinfo {title} {Spatiotemporal evolution of ventricular fibrillation},\
  }\href {https://doi.org/10.1038/32170} {\bibfield  {journal} {\bibinfo
  {journal} {Nature}\ }\textbf {\bibinfo {volume} {392}},\ \bibinfo {pages}
  {78} (\bibinfo {year} {1998})}\BibitemShut {NoStop}%
\bibitem [{\citenamefont {Rieke}\ \emph {et~al.}(1997)\citenamefont {Rieke},
  \citenamefont {Warland}, \citenamefont {De~Ruyter~van Steveninck},\ and\
  \citenamefont {Bialek}}]{rieke1997spikes}%
  \BibitemOpen
  \bibfield  {author} {\bibinfo {author} {\bibfnamefont {F.}~\bibnamefont
  {Rieke}}, \bibinfo {author} {\bibfnamefont {D.}~\bibnamefont {Warland}},
  \bibinfo {author} {\bibfnamefont {R.}~\bibnamefont {De~Ruyter~van
  Steveninck}},\ and\ \bibinfo {author} {\bibfnamefont {W.}~\bibnamefont
  {Bialek}},\ }\href {https://books.google.com/books?id=0xmDcOLZGu0C} {\emph
  {\bibinfo {title} {Spikes: Exploring the Neural Code}}},\ Bradford book\
  (\bibinfo  {publisher} {MIT Press},\ \bibinfo {year} {1997})\BibitemShut
  {NoStop}%
\bibitem [{\citenamefont {Drossel}\ and\ \citenamefont
  {Schwabl}(1992)}]{Drossel1992Self}%
  \BibitemOpen
  \bibfield  {author} {\bibinfo {author} {\bibfnamefont {B.}~\bibnamefont
  {Drossel}}\ and\ \bibinfo {author} {\bibfnamefont {F.}~\bibnamefont
  {Schwabl}},\ }\bibfield  {title} {\bibinfo {title} {Self-organized critical
  forest-fire model},\ }\href {https://doi.org/10.1103/PhysRevLett.69.1629}
  {\bibfield  {journal} {\bibinfo  {journal} {Phys. Rev. Lett.}\ }\textbf
  {\bibinfo {volume} {69}},\ \bibinfo {pages} {1629} (\bibinfo {year}
  {1992})}\BibitemShut {NoStop}%
\bibitem [{\citenamefont {Anderson}\ and\ \citenamefont
  {May}(1979)}]{Anderson1979Population}%
  \BibitemOpen
  \bibfield  {author} {\bibinfo {author} {\bibfnamefont {R.~M.}\ \bibnamefont
  {Anderson}}\ and\ \bibinfo {author} {\bibfnamefont {R.~M.}\ \bibnamefont
  {May}},\ }\bibfield  {title} {\bibinfo {title} {Population biology of
  infectious diseases: Part i},\ }\href {https://doi.org/10.1038/280361a0}
  {\bibfield  {journal} {\bibinfo  {journal} {Nature}\ }\textbf {\bibinfo
  {volume} {280}},\ \bibinfo {pages} {361} (\bibinfo {year}
  {1979})}\BibitemShut {NoStop}%
\bibitem [{\citenamefont {Murray}\ \emph {et~al.}(1986)\citenamefont {Murray},
  \citenamefont {Stanley},\ and\ \citenamefont {Brown}}]{Murray1986spatial}%
  \BibitemOpen
  \bibfield  {author} {\bibinfo {author} {\bibfnamefont {J.~D.}\ \bibnamefont
  {Murray}}, \bibinfo {author} {\bibfnamefont {E.~A.}\ \bibnamefont
  {Stanley}},\ and\ \bibinfo {author} {\bibfnamefont {D.~L.}\ \bibnamefont
  {Brown}},\ }\bibfield  {title} {\bibinfo {title} {On the spatial spread of
  rabies among foxes},\ }\href {https://doi.org/10.1098/rspb.1986.0078}
  {\bibfield  {journal} {\bibinfo  {journal} {Proceedings of the Royal Society
  of London. Series B. Biological Sciences}\ }\textbf {\bibinfo {volume}
  {229}},\ \bibinfo {pages} {111} (\bibinfo {year} {1986})}\BibitemShut
  {NoStop}%
\bibitem [{\citenamefont {Anderson}\ \emph {et~al.}(1981)\citenamefont
  {Anderson}, \citenamefont {Jackson}, \citenamefont {May},\ and\ \citenamefont
  {Smith}}]{Anderson1981Population}%
  \BibitemOpen
  \bibfield  {author} {\bibinfo {author} {\bibfnamefont {R.~M.}\ \bibnamefont
  {Anderson}}, \bibinfo {author} {\bibfnamefont {H.~C.}\ \bibnamefont
  {Jackson}}, \bibinfo {author} {\bibfnamefont {R.~M.}\ \bibnamefont {May}},\
  and\ \bibinfo {author} {\bibfnamefont {A.~M.}\ \bibnamefont {Smith}},\
  }\bibfield  {title} {\bibinfo {title} {Population dynamics of fox rabies in
  europe},\ }\href {https://doi.org/10.1038/289765a0} {\bibfield  {journal}
  {\bibinfo  {journal} {Nature}\ }\textbf {\bibinfo {volume} {289}},\ \bibinfo
  {pages} {765} (\bibinfo {year} {1981})}\BibitemShut {NoStop}%
\bibitem [{\citenamefont {Rohani}\ \emph {et~al.}(1999)\citenamefont {Rohani},
  \citenamefont {Earn},\ and\ \citenamefont {Grenfell}}]{Rohani1999Opposite}%
  \BibitemOpen
  \bibfield  {author} {\bibinfo {author} {\bibfnamefont {P.}~\bibnamefont
  {Rohani}}, \bibinfo {author} {\bibfnamefont {D.~J.~D.}\ \bibnamefont
  {Earn}},\ and\ \bibinfo {author} {\bibfnamefont {B.~T.}\ \bibnamefont
  {Grenfell}},\ }\bibfield  {title} {\bibinfo {title} {Opposite patterns of
  synchrony in sympatric disease metapopulations},\ }\href
  {https://doi.org/10.1126/science.286.5441.968} {\bibfield  {journal}
  {\bibinfo  {journal} {Science}\ }\textbf {\bibinfo {volume} {286}},\ \bibinfo
  {pages} {968} (\bibinfo {year} {1999})}\BibitemShut {NoStop}%
\bibitem [{\citenamefont {Kondo}\ and\ \citenamefont
  {Miura}(2010)}]{Kondo2010Reaction}%
  \BibitemOpen
  \bibfield  {author} {\bibinfo {author} {\bibfnamefont {S.}~\bibnamefont
  {Kondo}}\ and\ \bibinfo {author} {\bibfnamefont {T.}~\bibnamefont {Miura}},\
  }\bibfield  {title} {\bibinfo {title} {Reaction-diffusion model as a
  framework for understanding biological pattern formation},\ }\href
  {https://doi.org/10.1126/science.1179047} {\bibfield  {journal} {\bibinfo
  {journal} {Science}\ }\textbf {\bibinfo {volume} {329}},\ \bibinfo {pages}
  {1616} (\bibinfo {year} {2010})}\BibitemShut {NoStop}%
\bibitem [{\citenamefont {Bourret}\ \emph {et~al.}(1969)\citenamefont
  {Bourret}, \citenamefont {Lincoln},\ and\ \citenamefont
  {Carpenter}}]{Bourret1969Fungal}%
  \BibitemOpen
  \bibfield  {author} {\bibinfo {author} {\bibfnamefont {J.~A.}\ \bibnamefont
  {Bourret}}, \bibinfo {author} {\bibfnamefont {R.~G.}\ \bibnamefont
  {Lincoln}},\ and\ \bibinfo {author} {\bibfnamefont {B.~H.}\ \bibnamefont
  {Carpenter}},\ }\bibfield  {title} {\bibinfo {title} {Fungal endogenous
  rhythms expressed by spiral figures},\ }\href
  {https://doi.org/10.1126/science.166.3906.763} {\bibfield  {journal}
  {\bibinfo  {journal} {Science}\ }\textbf {\bibinfo {volume} {166}},\ \bibinfo
  {pages} {763} (\bibinfo {year} {1969})}\BibitemShut {NoStop}%
\bibitem [{\citenamefont {Loose}\ \emph {et~al.}(2008)\citenamefont {Loose},
  \citenamefont {Fischer-Friedrich}, \citenamefont {Ries}, \citenamefont
  {Kruse},\ and\ \citenamefont {Schwille}}]{Loose2008Spatial}%
  \BibitemOpen
  \bibfield  {author} {\bibinfo {author} {\bibfnamefont {M.}~\bibnamefont
  {Loose}}, \bibinfo {author} {\bibfnamefont {E.}~\bibnamefont
  {Fischer-Friedrich}}, \bibinfo {author} {\bibfnamefont {J.}~\bibnamefont
  {Ries}}, \bibinfo {author} {\bibfnamefont {K.}~\bibnamefont {Kruse}},\ and\
  \bibinfo {author} {\bibfnamefont {P.}~\bibnamefont {Schwille}},\ }\bibfield
  {title} {\bibinfo {title} {Spatial regulators for bacterial cell division
  self-organize into surface waves in vitro},\ }\href
  {https://doi.org/10.1126/science.1154413} {\bibfield  {journal} {\bibinfo
  {journal} {Science}\ }\textbf {\bibinfo {volume} {320}},\ \bibinfo {pages}
  {789} (\bibinfo {year} {2008})}\BibitemShut {NoStop}%
\bibitem [{\citenamefont {Tompkins}\ \emph {et~al.}(2014)\citenamefont
  {Tompkins}, \citenamefont {Li}, \citenamefont {Girabawe}, \citenamefont
  {Heymann}, \citenamefont {Ermentrout}, \citenamefont {Epstein},\ and\
  \citenamefont {Fraden}}]{Tompkins2014Testing}%
  \BibitemOpen
  \bibfield  {author} {\bibinfo {author} {\bibfnamefont {N.}~\bibnamefont
  {Tompkins}}, \bibinfo {author} {\bibfnamefont {N.}~\bibnamefont {Li}},
  \bibinfo {author} {\bibfnamefont {C.}~\bibnamefont {Girabawe}}, \bibinfo
  {author} {\bibfnamefont {M.}~\bibnamefont {Heymann}}, \bibinfo {author}
  {\bibfnamefont {G.~B.}\ \bibnamefont {Ermentrout}}, \bibinfo {author}
  {\bibfnamefont {I.~R.}\ \bibnamefont {Epstein}},\ and\ \bibinfo {author}
  {\bibfnamefont {S.}~\bibnamefont {Fraden}},\ }\bibfield  {title} {\bibinfo
  {title} {Testing turing's theory of morphogenesis in chemical cells},\ }\href
  {https://doi.org/10.1073/pnas.1322005111} {\bibfield  {journal} {\bibinfo
  {journal} {Proceedings of the National Academy of Sciences}\ }\textbf
  {\bibinfo {volume} {111}},\ \bibinfo {pages} {4397} (\bibinfo {year}
  {2014})}\BibitemShut {NoStop}%
\bibitem [{\citenamefont {Rotermund}\ \emph {et~al.}(1990)\citenamefont
  {Rotermund}, \citenamefont {Engel}, \citenamefont {Kordesch},\ and\
  \citenamefont {Ertl}}]{Rotermund1990imaging}%
  \BibitemOpen
  \bibfield  {author} {\bibinfo {author} {\bibfnamefont {H.~H.}\ \bibnamefont
  {Rotermund}}, \bibinfo {author} {\bibfnamefont {W.}~\bibnamefont {Engel}},
  \bibinfo {author} {\bibfnamefont {M.}~\bibnamefont {Kordesch}},\ and\
  \bibinfo {author} {\bibfnamefont {G.}~\bibnamefont {Ertl}},\ }\bibfield
  {title} {\bibinfo {title} {Imaging of spatio-temporal pattern evolution
  during carbon monoxide oxidation on platinum},\ }\href
  {https://doi.org/10.1038/343355a0} {\bibfield  {journal} {\bibinfo  {journal}
  {Nature}\ }\textbf {\bibinfo {volume} {343}},\ \bibinfo {pages} {355}
  (\bibinfo {year} {1990})}\BibitemShut {NoStop}%
\bibitem [{\citenamefont {Steinbock}\ \emph
  {et~al.}(1995{\natexlab{a}})\citenamefont {Steinbock}, \citenamefont
  {Kettunen},\ and\ \citenamefont {Showalter}}]{Anisotropy1995Science}%
  \BibitemOpen
  \bibfield  {author} {\bibinfo {author} {\bibfnamefont {O.}~\bibnamefont
  {Steinbock}}, \bibinfo {author} {\bibfnamefont {P.}~\bibnamefont
  {Kettunen}},\ and\ \bibinfo {author} {\bibfnamefont {K.}~\bibnamefont
  {Showalter}},\ }\bibfield  {title} {\bibinfo {title} {Anisotropy and spiral
  organizing centers in patterned excitable media},\ }\href
  {https://doi.org/10.1126/science.269.5232.1857} {\bibfield  {journal}
  {\bibinfo  {journal} {Science}\ }\textbf {\bibinfo {volume} {269}},\ \bibinfo
  {pages} {1857} (\bibinfo {year} {1995}{\natexlab{a}})}\BibitemShut {NoStop}%
\bibitem [{\citenamefont {Vinson}\ \emph {et~al.}(1997)\citenamefont {Vinson},
  \citenamefont {Mironov}, \citenamefont {Mulvey},\ and\ \citenamefont
  {Pertsov}}]{Vinson1997Control}%
  \BibitemOpen
  \bibfield  {author} {\bibinfo {author} {\bibfnamefont {M.}~\bibnamefont
  {Vinson}}, \bibinfo {author} {\bibfnamefont {S.}~\bibnamefont {Mironov}},
  \bibinfo {author} {\bibfnamefont {S.}~\bibnamefont {Mulvey}},\ and\ \bibinfo
  {author} {\bibfnamefont {A.}~\bibnamefont {Pertsov}},\ }\bibfield  {title}
  {\bibinfo {title} {Control of spatial orientation and lifetime of scroll
  rings in excitable media},\ }\href {https://doi.org/10.1038/386477a0}
  {\bibfield  {journal} {\bibinfo  {journal} {Nature}\ }\textbf {\bibinfo
  {volume} {386}},\ \bibinfo {pages} {477} (\bibinfo {year}
  {1997})}\BibitemShut {NoStop}%
\bibitem [{\citenamefont {Fuseya}\ \emph {et~al.}(2021)\citenamefont {Fuseya},
  \citenamefont {Katsuno}, \citenamefont {Behnia},\ and\ \citenamefont
  {Kapitulnik}}]{Fuseya2021Nanoscale}%
  \BibitemOpen
  \bibfield  {author} {\bibinfo {author} {\bibfnamefont {Y.}~\bibnamefont
  {Fuseya}}, \bibinfo {author} {\bibfnamefont {H.}~\bibnamefont {Katsuno}},
  \bibinfo {author} {\bibfnamefont {K.}~\bibnamefont {Behnia}},\ and\ \bibinfo
  {author} {\bibfnamefont {A.}~\bibnamefont {Kapitulnik}},\ }\bibfield  {title}
  {\bibinfo {title} {Nanoscale turing patterns in a bismuth monolayer},\ }\href
  {https://doi.org/10.1038/s41567-021-01288-y} {\bibfield  {journal} {\bibinfo
  {journal} {Nature Physics}\ }\textbf {\bibinfo {volume} {17}},\ \bibinfo
  {pages} {1031} (\bibinfo {year} {2021})}\BibitemShut {NoStop}%
\bibitem [{\citenamefont {Tan}\ \emph {et~al.}(2020)\citenamefont {Tan},
  \citenamefont {Liu}, \citenamefont {Miller}, \citenamefont {Tekant},
  \citenamefont {Dunkel},\ and\ \citenamefont {Fakhri}}]{Tan2020Topological}%
  \BibitemOpen
  \bibfield  {author} {\bibinfo {author} {\bibfnamefont {T.~H.}\ \bibnamefont
  {Tan}}, \bibinfo {author} {\bibfnamefont {J.}~\bibnamefont {Liu}}, \bibinfo
  {author} {\bibfnamefont {P.~W.}\ \bibnamefont {Miller}}, \bibinfo {author}
  {\bibfnamefont {M.}~\bibnamefont {Tekant}}, \bibinfo {author} {\bibfnamefont
  {J.}~\bibnamefont {Dunkel}},\ and\ \bibinfo {author} {\bibfnamefont
  {N.}~\bibnamefont {Fakhri}},\ }\bibfield  {title} {\bibinfo {title}
  {Topological turbulence in the membrane of a living cell},\ }\href
  {https://doi.org/10.1038/s41567-020-0841-9} {\bibfield  {journal} {\bibinfo
  {journal} {Nature Physics}\ }\textbf {\bibinfo {volume} {16}},\ \bibinfo
  {pages} {657} (\bibinfo {year} {2020})}\BibitemShut {NoStop}%
\bibitem [{\citenamefont
  {Winfree}(1994{\natexlab{b}})}]{Winfree1994Persistent}%
  \BibitemOpen
  \bibfield  {author} {\bibinfo {author} {\bibfnamefont {A.~T.}\ \bibnamefont
  {Winfree}},\ }\bibfield  {title} {\bibinfo {title} {Persistent tangled vortex
  rings in generic excitable media},\ }\href {https://doi.org/10.1038/371233a0}
  {\bibfield  {journal} {\bibinfo  {journal} {Nature}\ }\textbf {\bibinfo
  {volume} {371}},\ \bibinfo {pages} {233} (\bibinfo {year}
  {1994}{\natexlab{b}})}\BibitemShut {NoStop}%
\bibitem [{\citenamefont {Turing}(1952)}]{Turing1952chemical}%
  \BibitemOpen
  \bibfield  {author} {\bibinfo {author} {\bibfnamefont {A.~M.}\ \bibnamefont
  {Turing}},\ }\bibfield  {title} {\bibinfo {title} {The chemical basis of
  morphogenesis},\ }\href {https://doi.org/10.1098/rstb.1952.0012} {\bibfield
  {journal} {\bibinfo  {journal} {Philosophical Transactions of the Royal
  Society of London. Series B, Biological Sciences}\ }\textbf {\bibinfo
  {volume} {237}},\ \bibinfo {pages} {37} (\bibinfo {year} {1952})}\BibitemShut
  {NoStop}%
\bibitem [{\citenamefont {Nakamasu}\ \emph {et~al.}(2009)\citenamefont
  {Nakamasu}, \citenamefont {Takahashi}, \citenamefont {Kanbe},\ and\
  \citenamefont {Kondo}}]{Nakamasu2009Interactions}%
  \BibitemOpen
  \bibfield  {author} {\bibinfo {author} {\bibfnamefont {A.}~\bibnamefont
  {Nakamasu}}, \bibinfo {author} {\bibfnamefont {G.}~\bibnamefont {Takahashi}},
  \bibinfo {author} {\bibfnamefont {A.}~\bibnamefont {Kanbe}},\ and\ \bibinfo
  {author} {\bibfnamefont {S.}~\bibnamefont {Kondo}},\ }\bibfield  {title}
  {\bibinfo {title} {Interactions between zebrafish pigment cells responsible
  for the generation of turing patterns},\ }\href
  {https://doi.org/10.1073/pnas.0808622106} {\bibfield  {journal} {\bibinfo
  {journal} {Proceedings of the National Academy of Sciences}\ }\textbf
  {\bibinfo {volume} {106}},\ \bibinfo {pages} {8429} (\bibinfo {year}
  {2009})}\BibitemShut {NoStop}%
\bibitem [{\citenamefont {Kondo}\ and\ \citenamefont
  {Asai}(1995)}]{kondo1995reaction}%
  \BibitemOpen
  \bibfield  {author} {\bibinfo {author} {\bibfnamefont {S.}~\bibnamefont
  {Kondo}}\ and\ \bibinfo {author} {\bibfnamefont {R.}~\bibnamefont {Asai}},\
  }\bibfield  {title} {\bibinfo {title} {A reaction--diffusion wave on the skin
  of the marine angelfish pomacanthus},\ }\href
  {https://doi.org/10.1038/376765a0} {\bibfield  {journal} {\bibinfo  {journal}
  {Nature}\ }\textbf {\bibinfo {volume} {376}},\ \bibinfo {pages} {765}
  (\bibinfo {year} {1995})}\BibitemShut {NoStop}%
\bibitem [{\citenamefont {Newman}\ and\ \citenamefont
  {Frisch}(1979)}]{Newman1979Dynamics}%
  \BibitemOpen
  \bibfield  {author} {\bibinfo {author} {\bibfnamefont {S.~A.}\ \bibnamefont
  {Newman}}\ and\ \bibinfo {author} {\bibfnamefont {H.~L.}\ \bibnamefont
  {Frisch}},\ }\bibfield  {title} {\bibinfo {title} {Dynamics of skeletal
  pattern formation in developing chick limb},\ }\href
  {https://doi.org/10.1126/science.462174} {\bibfield  {journal} {\bibinfo
  {journal} {Science}\ }\textbf {\bibinfo {volume} {205}},\ \bibinfo {pages}
  {662} (\bibinfo {year} {1979})}\BibitemShut {NoStop}%
\bibitem [{\citenamefont {Mitchell}\ \emph {et~al.}(2022)\citenamefont
  {Mitchell}, \citenamefont {Cislo}, \citenamefont {Shankar}, \citenamefont
  {Lin}, \citenamefont {Shraiman},\ and\ \citenamefont
  {Streichan}}]{Mitchell2022Visceral}%
  \BibitemOpen
  \bibfield  {author} {\bibinfo {author} {\bibfnamefont {N.~P.}\ \bibnamefont
  {Mitchell}}, \bibinfo {author} {\bibfnamefont {D.~J.}\ \bibnamefont {Cislo}},
  \bibinfo {author} {\bibfnamefont {S.}~\bibnamefont {Shankar}}, \bibinfo
  {author} {\bibfnamefont {Y.}~\bibnamefont {Lin}}, \bibinfo {author}
  {\bibfnamefont {B.~I.}\ \bibnamefont {Shraiman}},\ and\ \bibinfo {author}
  {\bibfnamefont {S.~J.}\ \bibnamefont {Streichan}},\ }\bibfield  {title}
  {\bibinfo {title} {Visceral organ morphogenesis via calcium-patterned muscle
  constrictions},\ }\href {https://doi.org/10.7554/eLife.77355} {\bibfield
  {journal} {\bibinfo  {journal} {eLife}\ }\textbf {\bibinfo {volume} {11}},\
  \bibinfo {pages} {e77355} (\bibinfo {year} {2022})}\BibitemShut {NoStop}%
\bibitem [{\citenamefont {Wigbers}\ \emph {et~al.}(2021)\citenamefont
  {Wigbers}, \citenamefont {Tan}, \citenamefont {Brauns}, \citenamefont {Liu},
  \citenamefont {Swartz}, \citenamefont {Frey},\ and\ \citenamefont
  {Fakhri}}]{Wigbers2021hierarchy}%
  \BibitemOpen
  \bibfield  {author} {\bibinfo {author} {\bibfnamefont {M.~C.}\ \bibnamefont
  {Wigbers}}, \bibinfo {author} {\bibfnamefont {T.~H.}\ \bibnamefont {Tan}},
  \bibinfo {author} {\bibfnamefont {F.}~\bibnamefont {Brauns}}, \bibinfo
  {author} {\bibfnamefont {J.}~\bibnamefont {Liu}}, \bibinfo {author}
  {\bibfnamefont {S.~Z.}\ \bibnamefont {Swartz}}, \bibinfo {author}
  {\bibfnamefont {E.}~\bibnamefont {Frey}},\ and\ \bibinfo {author}
  {\bibfnamefont {N.}~\bibnamefont {Fakhri}},\ }\bibfield  {title} {\bibinfo
  {title} {A hierarchy of protein patterns robustly decodes cell shape
  information},\ }\href {https://doi.org/10.1038/s41567-021-01164-9} {\bibfield
   {journal} {\bibinfo  {journal} {Nature Physics}\ }\textbf {\bibinfo {volume}
  {17}},\ \bibinfo {pages} {578} (\bibinfo {year} {2021})}\BibitemShut
  {NoStop}%
\bibitem [{\citenamefont {Di~Talia}\ and\ \citenamefont
  {Vergassola}(2022)}]{Talia2022Waves}%
  \BibitemOpen
  \bibfield  {author} {\bibinfo {author} {\bibfnamefont {S.}~\bibnamefont
  {Di~Talia}}\ and\ \bibinfo {author} {\bibfnamefont {M.}~\bibnamefont
  {Vergassola}},\ }\bibfield  {title} {\bibinfo {title} {Waves in embryonic
  development},\ }\href {https://doi.org/10.1146/annurev-biophys-111521-102500}
  {\bibfield  {journal} {\bibinfo  {journal} {Annual Review of Biophysics}\
  }\textbf {\bibinfo {volume} {51}},\ \bibinfo {pages} {327} (\bibinfo {year}
  {2022})},\ \bibinfo {note} {pMID: 35119944}\BibitemShut {NoStop}%
\bibitem [{\citenamefont {Vergassola}\ \emph {et~al.}(2018)\citenamefont
  {Vergassola}, \citenamefont {Deneke},\ and\ \citenamefont
  {Talia}}]{Vergassola2018Mitotic}%
  \BibitemOpen
  \bibfield  {author} {\bibinfo {author} {\bibfnamefont {M.}~\bibnamefont
  {Vergassola}}, \bibinfo {author} {\bibfnamefont {V.~E.}\ \bibnamefont
  {Deneke}},\ and\ \bibinfo {author} {\bibfnamefont {S.~D.}\ \bibnamefont
  {Talia}},\ }\bibfield  {title} {\bibinfo {title} {Mitotic waves in the early
  embryogenesis of \emph{Drosophila}: Bistability traded for speed},\ }\href
  {https://doi.org/10.1073/pnas.1714873115} {\bibfield  {journal} {\bibinfo
  {journal} {Proceedings of the National Academy of Sciences}\ }\textbf
  {\bibinfo {volume} {115}},\ \bibinfo {pages} {E2165} (\bibinfo {year}
  {2018})}\BibitemShut {NoStop}%
\bibitem [{\citenamefont {Lechleiter}\ \emph {et~al.}(1991)\citenamefont
  {Lechleiter}, \citenamefont {Girard}, \citenamefont {Peralta},\ and\
  \citenamefont {Clapham}}]{Lechleiter1991Spiral}%
  \BibitemOpen
  \bibfield  {author} {\bibinfo {author} {\bibfnamefont {J.}~\bibnamefont
  {Lechleiter}}, \bibinfo {author} {\bibfnamefont {S.}~\bibnamefont {Girard}},
  \bibinfo {author} {\bibfnamefont {E.}~\bibnamefont {Peralta}},\ and\ \bibinfo
  {author} {\bibfnamefont {D.}~\bibnamefont {Clapham}},\ }\bibfield  {title}
  {\bibinfo {title} {Spiral calcium wave propagation and annihilation in
  \emph{Xenopus laevis} oocytes},\ }\href
  {https://doi.org/10.1126/science.2011747} {\bibfield  {journal} {\bibinfo
  {journal} {Science}\ }\textbf {\bibinfo {volume} {252}},\ \bibinfo {pages}
  {123} (\bibinfo {year} {1991})}\BibitemShut {NoStop}%
\bibitem [{\citenamefont {Michaux}\ \emph {et~al.}(2018)\citenamefont
  {Michaux}, \citenamefont {Robin}, \citenamefont {McFadden},\ and\
  \citenamefont {Munro}}]{McFadden2018excitable}%
  \BibitemOpen
  \bibfield  {author} {\bibinfo {author} {\bibfnamefont {J.~B.}\ \bibnamefont
  {Michaux}}, \bibinfo {author} {\bibfnamefont {F.~B.}\ \bibnamefont {Robin}},
  \bibinfo {author} {\bibfnamefont {W.~M.}\ \bibnamefont {McFadden}},\ and\
  \bibinfo {author} {\bibfnamefont {E.~M.}\ \bibnamefont {Munro}},\ }\bibfield
  {title} {\bibinfo {title} {{Excitable RhoA dynamics drive pulsed contractions
  in the early C. elegans embryo}},\ }\href
  {https://doi.org/10.1083/jcb.201806161} {\bibfield  {journal} {\bibinfo
  {journal} {Journal of Cell Biology}\ }\textbf {\bibinfo {volume} {217}},\
  \bibinfo {pages} {4230} (\bibinfo {year} {2018})}\BibitemShut {NoStop}%
\bibitem [{\citenamefont {Chang}\ and\ \citenamefont
  {Ferrell~Jr}(2013)}]{Change2013mitotic}%
  \BibitemOpen
  \bibfield  {author} {\bibinfo {author} {\bibfnamefont {J.~B.}\ \bibnamefont
  {Chang}}\ and\ \bibinfo {author} {\bibfnamefont {J.~E.}\ \bibnamefont
  {Ferrell~Jr}},\ }\bibfield  {title} {\bibinfo {title} {Mitotic trigger waves
  and the spatial coordination of the xenopus cell cycle},\ }\href
  {https://doi.org/10.1038/nature12321} {\bibfield  {journal} {\bibinfo
  {journal} {Nature}\ }\textbf {\bibinfo {volume} {500}},\ \bibinfo {pages}
  {603} (\bibinfo {year} {2013})}\BibitemShut {NoStop}%
\bibitem [{\citenamefont {Davidenko}\ \emph {et~al.}(1992)\citenamefont
  {Davidenko}, \citenamefont {Pertsov}, \citenamefont {Salomonsz},
  \citenamefont {Baxter},\ and\ \citenamefont
  {Jalife}}]{Davidenko1992Stationary}%
  \BibitemOpen
  \bibfield  {author} {\bibinfo {author} {\bibfnamefont {J.~M.}\ \bibnamefont
  {Davidenko}}, \bibinfo {author} {\bibfnamefont {A.~V.}\ \bibnamefont
  {Pertsov}}, \bibinfo {author} {\bibfnamefont {R.}~\bibnamefont {Salomonsz}},
  \bibinfo {author} {\bibfnamefont {W.}~\bibnamefont {Baxter}},\ and\ \bibinfo
  {author} {\bibfnamefont {J.}~\bibnamefont {Jalife}},\ }\bibfield  {title}
  {\bibinfo {title} {Stationary and drifting spiral waves of excitation in
  isolated cardiac muscle},\ }\href {https://doi.org/10.1038/355349a0}
  {\bibfield  {journal} {\bibinfo  {journal} {Nature}\ }\textbf {\bibinfo
  {volume} {355}},\ \bibinfo {pages} {349} (\bibinfo {year}
  {1992})}\BibitemShut {NoStop}%
\bibitem [{\citenamefont {Gray}\ \emph {et~al.}(1998)\citenamefont {Gray},
  \citenamefont {Pertsov},\ and\ \citenamefont {Jalife}}]{Gray1998Spatial}%
  \BibitemOpen
  \bibfield  {author} {\bibinfo {author} {\bibfnamefont {R.~A.}\ \bibnamefont
  {Gray}}, \bibinfo {author} {\bibfnamefont {A.~M.}\ \bibnamefont {Pertsov}},\
  and\ \bibinfo {author} {\bibfnamefont {J.}~\bibnamefont {Jalife}},\
  }\bibfield  {title} {\bibinfo {title} {Spatial and temporal organization
  during cardiac fibrillation},\ }\href {https://doi.org/10.1038/32164}
  {\bibfield  {journal} {\bibinfo  {journal} {Nature}\ }\textbf {\bibinfo
  {volume} {392}},\ \bibinfo {pages} {75} (\bibinfo {year} {1998})}\BibitemShut
  {NoStop}%
\bibitem [{\citenamefont {Halatek}\ and\ \citenamefont
  {Frey}(2018)}]{Halatek2018Rethinking}%
  \BibitemOpen
  \bibfield  {author} {\bibinfo {author} {\bibfnamefont {J.}~\bibnamefont
  {Halatek}}\ and\ \bibinfo {author} {\bibfnamefont {E.}~\bibnamefont {Frey}},\
  }\bibfield  {title} {\bibinfo {title} {Rethinking pattern formation in
  reaction--diffusion systems},\ }\href
  {https://doi.org/10.1038/s41567-017-0040-5} {\bibfield  {journal} {\bibinfo
  {journal} {Nature Physics}\ }\textbf {\bibinfo {volume} {14}},\ \bibinfo
  {pages} {507} (\bibinfo {year} {2018})}\BibitemShut {NoStop}%
\bibitem [{\citenamefont {Cross}\ and\ \citenamefont
  {Hohenberg}(1993)}]{Cross1993Pattern}%
  \BibitemOpen
  \bibfield  {author} {\bibinfo {author} {\bibfnamefont {M.~C.}\ \bibnamefont
  {Cross}}\ and\ \bibinfo {author} {\bibfnamefont {P.~C.}\ \bibnamefont
  {Hohenberg}},\ }\bibfield  {title} {\bibinfo {title} {Pattern formation
  outside of equilibrium},\ }\href {https://doi.org/10.1103/RevModPhys.65.851}
  {\bibfield  {journal} {\bibinfo  {journal} {Rev. Mod. Phys.}\ }\textbf
  {\bibinfo {volume} {65}},\ \bibinfo {pages} {851} (\bibinfo {year}
  {1993})}\BibitemShut {NoStop}%
\bibitem [{\citenamefont {Kim}\ \emph {et~al.}(2001)\citenamefont {Kim},
  \citenamefont {Bertram}, \citenamefont {Pollmann}, \citenamefont {von
  Oertzen}, \citenamefont {Mikhailov}, \citenamefont {Rotermund},\ and\
  \citenamefont {Ertl}}]{Kim2001Controlling}%
  \BibitemOpen
  \bibfield  {author} {\bibinfo {author} {\bibfnamefont {M.}~\bibnamefont
  {Kim}}, \bibinfo {author} {\bibfnamefont {M.}~\bibnamefont {Bertram}},
  \bibinfo {author} {\bibfnamefont {M.}~\bibnamefont {Pollmann}}, \bibinfo
  {author} {\bibfnamefont {A.}~\bibnamefont {von Oertzen}}, \bibinfo {author}
  {\bibfnamefont {A.~S.}\ \bibnamefont {Mikhailov}}, \bibinfo {author}
  {\bibfnamefont {H.~H.}\ \bibnamefont {Rotermund}},\ and\ \bibinfo {author}
  {\bibfnamefont {G.}~\bibnamefont {Ertl}},\ }\bibfield  {title} {\bibinfo
  {title} {Controlling chemical turbulence by global delayed feedback: Pattern
  formation in catalytic co oxidation on pt(110)},\ }\href
  {https://doi.org/10.1126/science.1059478} {\bibfield  {journal} {\bibinfo
  {journal} {Science}\ }\textbf {\bibinfo {volume} {292}},\ \bibinfo {pages}
  {1357} (\bibinfo {year} {2001})}\BibitemShut {NoStop}%
\bibitem [{\citenamefont {Brauns}\ \emph {et~al.}(2020)\citenamefont {Brauns},
  \citenamefont {Halatek},\ and\ \citenamefont {Frey}}]{Brauns2020Phase}%
  \BibitemOpen
  \bibfield  {author} {\bibinfo {author} {\bibfnamefont {F.}~\bibnamefont
  {Brauns}}, \bibinfo {author} {\bibfnamefont {J.}~\bibnamefont {Halatek}},\
  and\ \bibinfo {author} {\bibfnamefont {E.}~\bibnamefont {Frey}},\ }\bibfield
  {title} {\bibinfo {title} {Phase-space geometry of mass-conserving
  reaction-diffusion dynamics},\ }\href
  {https://doi.org/10.1103/PhysRevX.10.041036} {\bibfield  {journal} {\bibinfo
  {journal} {Phys. Rev. X}\ }\textbf {\bibinfo {volume} {10}},\ \bibinfo
  {pages} {041036} (\bibinfo {year} {2020})}\BibitemShut {NoStop}%
\bibitem [{\citenamefont {Alonso}\ \emph {et~al.}(2003)\citenamefont {Alonso},
  \citenamefont {Sagu{\'e}s},\ and\ \citenamefont
  {Mikhailov}}]{Alonso2003Taming}%
  \BibitemOpen
  \bibfield  {author} {\bibinfo {author} {\bibfnamefont {S.}~\bibnamefont
  {Alonso}}, \bibinfo {author} {\bibfnamefont {F.}~\bibnamefont {Sagu{\'e}s}},\
  and\ \bibinfo {author} {\bibfnamefont {A.~S.}\ \bibnamefont {Mikhailov}},\
  }\bibfield  {title} {\bibinfo {title} {Taming winfree turbulence of scroll
  waves in excitable media},\ }\href {https://doi.org/10.1126/science.1080207}
  {\bibfield  {journal} {\bibinfo  {journal} {Science}\ }\textbf {\bibinfo
  {volume} {299}},\ \bibinfo {pages} {1722} (\bibinfo {year}
  {2003})}\BibitemShut {NoStop}%
\bibitem [{\citenamefont {McNamara}\ \emph {et~al.}(2020)\citenamefont
  {McNamara}, \citenamefont {Salegame}, \citenamefont {Tanoury}, \citenamefont
  {Xu}, \citenamefont {Begum}, \citenamefont {Ortiz}, \citenamefont
  {Pourquie},\ and\ \citenamefont {Cohen}}]{McNamara2020bioelectrical}%
  \BibitemOpen
  \bibfield  {author} {\bibinfo {author} {\bibfnamefont {H.~M.}\ \bibnamefont
  {McNamara}}, \bibinfo {author} {\bibfnamefont {R.}~\bibnamefont {Salegame}},
  \bibinfo {author} {\bibfnamefont {Z.~A.}\ \bibnamefont {Tanoury}}, \bibinfo
  {author} {\bibfnamefont {H.}~\bibnamefont {Xu}}, \bibinfo {author}
  {\bibfnamefont {S.}~\bibnamefont {Begum}}, \bibinfo {author} {\bibfnamefont
  {G.}~\bibnamefont {Ortiz}}, \bibinfo {author} {\bibfnamefont
  {O.}~\bibnamefont {Pourquie}},\ and\ \bibinfo {author} {\bibfnamefont
  {A.~E.}\ \bibnamefont {Cohen}},\ }\bibfield  {title} {\bibinfo {title}
  {Bioelectrical domain walls in homogeneous tissues},\ }\href
  {https://doi.org/10.1038/s41567-019-0765-4} {\bibfield  {journal} {\bibinfo
  {journal} {Nature Physics}\ }\textbf {\bibinfo {volume} {16}},\ \bibinfo
  {pages} {357} (\bibinfo {year} {2020})}\BibitemShut {NoStop}%
\bibitem [{\citenamefont {Eckstein}\ \emph {et~al.}(2020)\citenamefont
  {Eckstein}, \citenamefont {Vidal-Henriquez},\ and\ \citenamefont
  {Gholami}}]{Eckstein2020Experimental}%
  \BibitemOpen
  \bibfield  {author} {\bibinfo {author} {\bibfnamefont {T.}~\bibnamefont
  {Eckstein}}, \bibinfo {author} {\bibfnamefont {E.}~\bibnamefont
  {Vidal-Henriquez}},\ and\ \bibinfo {author} {\bibfnamefont {A.}~\bibnamefont
  {Gholami}},\ }\bibfield  {title} {\bibinfo {title} {Experimental observation
  of boundary-driven oscillations in a reaction--diffusion--advection system},\
  }\href {https://doi.org/10.1039/C9SM02291K} {\bibfield  {journal} {\bibinfo
  {journal} {Soft Matter}\ }\textbf {\bibinfo {volume} {16}},\ \bibinfo {pages}
  {4243} (\bibinfo {year} {2020})}\BibitemShut {NoStop}%
\bibitem [{\citenamefont {Vidal-Henriquez}\ \emph {et~al.}(2017)\citenamefont
  {Vidal-Henriquez}, \citenamefont {Zykov}, \citenamefont {Bodenschatz},\ and\
  \citenamefont {Gholami}}]{VidalHenriquez2017Convective}%
  \BibitemOpen
  \bibfield  {author} {\bibinfo {author} {\bibfnamefont {E.}~\bibnamefont
  {Vidal-Henriquez}}, \bibinfo {author} {\bibfnamefont {V.}~\bibnamefont
  {Zykov}}, \bibinfo {author} {\bibfnamefont {E.}~\bibnamefont {Bodenschatz}},\
  and\ \bibinfo {author} {\bibfnamefont {A.}~\bibnamefont {Gholami}},\
  }\bibfield  {title} {\bibinfo {title} {Convective instability and boundary
  driven oscillations in a reaction-diffusion-advection model},\ }\href
  {https://doi.org/10.1063/1.4986153} {\bibfield  {journal} {\bibinfo
  {journal} {Chaos: An Interdisciplinary Journal of Nonlinear Science}\
  }\textbf {\bibinfo {volume} {27}},\ \bibinfo {pages} {103110} (\bibinfo
  {year} {2017})}\BibitemShut {NoStop}%
\bibitem [{\citenamefont {Ni}\ and\ \citenamefont
  {Wei}(1995)}]{Ni1995location}%
  \BibitemOpen
  \bibfield  {author} {\bibinfo {author} {\bibfnamefont {W.-M.}\ \bibnamefont
  {Ni}}\ and\ \bibinfo {author} {\bibfnamefont {J.}~\bibnamefont {Wei}},\
  }\bibfield  {title} {\bibinfo {title} {On the location and profile of
  spike-layer solutions to singularly perturbed semilinear dirichlet
  problems},\ }\href {https://doi.org/https://doi.org/10.1002/cpa.3160480704}
  {\bibfield  {journal} {\bibinfo  {journal} {Communications on Pure and
  Applied Mathematics}\ }\textbf {\bibinfo {volume} {48}},\ \bibinfo {pages}
  {731} (\bibinfo {year} {1995})}\BibitemShut {NoStop}%
\bibitem [{\citenamefont {Bub}\ \emph {et~al.}(2002{\natexlab{a}})\citenamefont
  {Bub}, \citenamefont {Shrier},\ and\ \citenamefont {Glass}}]{Bub2002Spiral}%
  \BibitemOpen
  \bibfield  {author} {\bibinfo {author} {\bibfnamefont {G.}~\bibnamefont
  {Bub}}, \bibinfo {author} {\bibfnamefont {A.}~\bibnamefont {Shrier}},\ and\
  \bibinfo {author} {\bibfnamefont {L.}~\bibnamefont {Glass}},\ }\bibfield
  {title} {\bibinfo {title} {Spiral wave generation in heterogeneous excitable
  media},\ }\href {https://doi.org/10.1103/PhysRevLett.88.058101} {\bibfield
  {journal} {\bibinfo  {journal} {Phys. Rev. Lett.}\ }\textbf {\bibinfo
  {volume} {88}},\ \bibinfo {pages} {058101} (\bibinfo {year}
  {2002}{\natexlab{a}})}\BibitemShut {NoStop}%
\bibitem [{\citenamefont {Mainen}\ and\ \citenamefont
  {Sejnowski}(1996)}]{Mainen1996influence}%
  \BibitemOpen
  \bibfield  {author} {\bibinfo {author} {\bibfnamefont {Z.~F.}\ \bibnamefont
  {Mainen}}\ and\ \bibinfo {author} {\bibfnamefont {T.~J.}\ \bibnamefont
  {Sejnowski}},\ }\bibfield  {title} {\bibinfo {title} {Influence of dendritic
  structure on firing pattern in model neocortical neurons},\ }\href
  {https://doi.org/10.1038/382363a0} {\bibfield  {journal} {\bibinfo  {journal}
  {Nature}\ }\textbf {\bibinfo {volume} {382}},\ \bibinfo {pages} {363}
  (\bibinfo {year} {1996})}\BibitemShut {NoStop}%
\bibitem [{\citenamefont {Wigbers}\ \emph {et~al.}(2020)\citenamefont
  {Wigbers}, \citenamefont {Brauns}, \citenamefont {Hermann},\ and\
  \citenamefont {Frey}}]{Wigbers2020pattern}%
  \BibitemOpen
  \bibfield  {author} {\bibinfo {author} {\bibfnamefont {M.~C.}\ \bibnamefont
  {Wigbers}}, \bibinfo {author} {\bibfnamefont {F.}~\bibnamefont {Brauns}},
  \bibinfo {author} {\bibfnamefont {T.}~\bibnamefont {Hermann}},\ and\ \bibinfo
  {author} {\bibfnamefont {E.}~\bibnamefont {Frey}},\ }\bibfield  {title}
  {\bibinfo {title} {Pattern localization to a domain edge},\ }\href
  {https://doi.org/10.1103/PhysRevE.101.022414} {\bibfield  {journal} {\bibinfo
   {journal} {Phys. Rev. E}\ }\textbf {\bibinfo {volume} {101}},\ \bibinfo
  {pages} {022414} (\bibinfo {year} {2020})}\BibitemShut {NoStop}%
\bibitem [{\citenamefont {Brauns}\ \emph {et~al.}(2021)\citenamefont {Brauns},
  \citenamefont {Halatek},\ and\ \citenamefont {Frey}}]{Brauns2021Diffusive}%
  \BibitemOpen
  \bibfield  {author} {\bibinfo {author} {\bibfnamefont {F.}~\bibnamefont
  {Brauns}}, \bibinfo {author} {\bibfnamefont {J.}~\bibnamefont {Halatek}},\
  and\ \bibinfo {author} {\bibfnamefont {E.}~\bibnamefont {Frey}},\ }\bibfield
  {title} {\bibinfo {title} {Diffusive coupling of two well-mixed compartments
  elucidates elementary principles of protein-based pattern formation},\ }\href
  {https://doi.org/10.1103/PhysRevResearch.3.013258} {\bibfield  {journal}
  {\bibinfo  {journal} {Phys. Rev. Res.}\ }\textbf {\bibinfo {volume} {3}},\
  \bibinfo {pages} {013258} (\bibinfo {year} {2021})}\BibitemShut {NoStop}%
\bibitem [{\citenamefont {Bub}\ \emph {et~al.}(2002{\natexlab{b}})\citenamefont
  {Bub}, \citenamefont {Shrier},\ and\ \citenamefont {Glass}}]{Bub200Spiral}%
  \BibitemOpen
  \bibfield  {author} {\bibinfo {author} {\bibfnamefont {G.}~\bibnamefont
  {Bub}}, \bibinfo {author} {\bibfnamefont {A.}~\bibnamefont {Shrier}},\ and\
  \bibinfo {author} {\bibfnamefont {L.}~\bibnamefont {Glass}},\ }\bibfield
  {title} {\bibinfo {title} {Spiral wave generation in heterogeneous excitable
  media},\ }\href {https://doi.org/10.1103/PhysRevLett.88.058101} {\bibfield
  {journal} {\bibinfo  {journal} {Phys. Rev. Lett.}\ }\textbf {\bibinfo
  {volume} {88}},\ \bibinfo {pages} {058101} (\bibinfo {year}
  {2002}{\natexlab{b}})}\BibitemShut {NoStop}%
\bibitem [{\citenamefont {Agladze}\ \emph {et~al.}(1994)\citenamefont
  {Agladze}, \citenamefont {Keener}, \citenamefont {M{\"u}ller},\ and\
  \citenamefont {Panfilov}}]{Agladze1994Rotating}%
  \BibitemOpen
  \bibfield  {author} {\bibinfo {author} {\bibfnamefont {K.}~\bibnamefont
  {Agladze}}, \bibinfo {author} {\bibfnamefont {J.~P.}\ \bibnamefont {Keener}},
  \bibinfo {author} {\bibfnamefont {S.~C.}\ \bibnamefont {M{\"u}ller}},\ and\
  \bibinfo {author} {\bibfnamefont {A.}~\bibnamefont {Panfilov}},\ }\bibfield
  {title} {\bibinfo {title} {Rotating spiral waves created by geometry},\
  }\href {https://doi.org/10.1126/science.264.5166.1746} {\bibfield  {journal}
  {\bibinfo  {journal} {Science}\ }\textbf {\bibinfo {volume} {264}},\ \bibinfo
  {pages} {1746} (\bibinfo {year} {1994})}\BibitemShut {NoStop}%
\bibitem [{\citenamefont {Staddon}\ \emph {et~al.}(2022)\citenamefont
  {Staddon}, \citenamefont {Munro},\ and\ \citenamefont
  {Banerjee}}]{Staddon2022Pulsatile}%
  \BibitemOpen
  \bibfield  {author} {\bibinfo {author} {\bibfnamefont {M.~F.}\ \bibnamefont
  {Staddon}}, \bibinfo {author} {\bibfnamefont {E.~M.}\ \bibnamefont {Munro}},\
  and\ \bibinfo {author} {\bibfnamefont {S.}~\bibnamefont {Banerjee}},\
  }\bibfield  {title} {\bibinfo {title} {Pulsatile contractions and pattern
  formation in excitable actomyosin cortex},\ }\href
  {https://doi.org/10.1371/journal.pcbi.1009981} {\bibfield  {journal}
  {\bibinfo  {journal} {PLOS Computational Biology}\ }\textbf {\bibinfo
  {volume} {18}},\ \bibinfo {pages} {1} (\bibinfo {year} {2022})}\BibitemShut
  {NoStop}%
\bibitem [{\citenamefont {Murugan}\ and\ \citenamefont
  {Vaikuntanathan}(2017)}]{Murugan2017Topologically}%
  \BibitemOpen
  \bibfield  {author} {\bibinfo {author} {\bibfnamefont {A.}~\bibnamefont
  {Murugan}}\ and\ \bibinfo {author} {\bibfnamefont {S.}~\bibnamefont
  {Vaikuntanathan}},\ }\bibfield  {title} {\bibinfo {title} {Topologically
  protected modes in non-equilibrium stochastic systems},\ }\href
  {https://doi.org/10.1038/ncomms13881} {\bibfield  {journal} {\bibinfo
  {journal} {Nature Communications}\ }\textbf {\bibinfo {volume} {8}},\
  \bibinfo {pages} {13881} (\bibinfo {year} {2017})}\BibitemShut {NoStop}%
\bibitem [{\citenamefont {Kane}\ and\ \citenamefont
  {Lubensky}(2014)}]{Kane2014Topologically}%
  \BibitemOpen
  \bibfield  {author} {\bibinfo {author} {\bibfnamefont {C.~L.}\ \bibnamefont
  {Kane}}\ and\ \bibinfo {author} {\bibfnamefont {T.~C.}\ \bibnamefont
  {Lubensky}},\ }\bibfield  {title} {\bibinfo {title} {Topological boundary
  modes in isostatic lattices},\ }\href {https://doi.org/10.1038/nphys2835}
  {\bibfield  {journal} {\bibinfo  {journal} {Nature Physics}\ }\textbf
  {\bibinfo {volume} {10}},\ \bibinfo {pages} {39} (\bibinfo {year}
  {2014})}\BibitemShut {NoStop}%
\bibitem [{\citenamefont {Hasan}\ and\ \citenamefont
  {Kane}(2010)}]{Hasan2010Colloquium}%
  \BibitemOpen
  \bibfield  {author} {\bibinfo {author} {\bibfnamefont {M.~Z.}\ \bibnamefont
  {Hasan}}\ and\ \bibinfo {author} {\bibfnamefont {C.~L.}\ \bibnamefont
  {Kane}},\ }\bibfield  {title} {\bibinfo {title} {Colloquium: Topological
  insulators},\ }\href {https://doi.org/10.1103/RevModPhys.82.3045} {\bibfield
  {journal} {\bibinfo  {journal} {Rev. Mod. Phys.}\ }\textbf {\bibinfo {volume}
  {82}},\ \bibinfo {pages} {3045} (\bibinfo {year} {2010})}\BibitemShut
  {NoStop}%
\bibitem [{\citenamefont {Shankar}\ \emph {et~al.}(2022)\citenamefont
  {Shankar}, \citenamefont {Souslov}, \citenamefont {Bowick}, \citenamefont
  {Marchetti},\ and\ \citenamefont {Vitelli}}]{Shankar2022Topological}%
  \BibitemOpen
  \bibfield  {author} {\bibinfo {author} {\bibfnamefont {S.}~\bibnamefont
  {Shankar}}, \bibinfo {author} {\bibfnamefont {A.}~\bibnamefont {Souslov}},
  \bibinfo {author} {\bibfnamefont {M.~J.}\ \bibnamefont {Bowick}}, \bibinfo
  {author} {\bibfnamefont {M.~C.}\ \bibnamefont {Marchetti}},\ and\ \bibinfo
  {author} {\bibfnamefont {V.}~\bibnamefont {Vitelli}},\ }\bibfield  {title}
  {\bibinfo {title} {Topological active matter},\ }\href
  {https://doi.org/10.1038/s42254-022-00445-3} {\bibfield  {journal} {\bibinfo
  {journal} {Nature Reviews Physics}\ }\textbf {\bibinfo {volume} {4}},\
  \bibinfo {pages} {380} (\bibinfo {year} {2022})}\BibitemShut {NoStop}%
\bibitem [{\citenamefont {ge~Chen}\ \emph {et~al.}(2014)\citenamefont
  {ge~Chen}, \citenamefont {Upadhyaya},\ and\ \citenamefont
  {Vitelli}}]{chen2014Nonlinear}%
  \BibitemOpen
  \bibfield  {author} {\bibinfo {author} {\bibfnamefont {B.~G.}\ \bibnamefont
  {ge~Chen}}, \bibinfo {author} {\bibfnamefont {N.}~\bibnamefont {Upadhyaya}},\
  and\ \bibinfo {author} {\bibfnamefont {V.}~\bibnamefont {Vitelli}},\
  }\bibfield  {title} {\bibinfo {title} {Nonlinear conduction via solitons in a
  topological mechanical insulator},\ }\href
  {https://doi.org/10.1073/pnas.1405969111} {\bibfield  {journal} {\bibinfo
  {journal} {Proceedings of the National Academy of Sciences}\ }\textbf
  {\bibinfo {volume} {111}},\ \bibinfo {pages} {13004} (\bibinfo {year}
  {2014})}\BibitemShut {NoStop}%
\bibitem [{\citenamefont {Mao}\ and\ \citenamefont
  {Lubensky}(2018)}]{Mao2018Maxwell}%
  \BibitemOpen
  \bibfield  {author} {\bibinfo {author} {\bibfnamefont {X.}~\bibnamefont
  {Mao}}\ and\ \bibinfo {author} {\bibfnamefont {T.~C.}\ \bibnamefont
  {Lubensky}},\ }\bibfield  {title} {\bibinfo {title} {Maxwell lattices and
  topological mechanics},\ }\href
  {https://doi.org/10.1146/annurev-conmatphys-033117-054235} {\bibfield
  {journal} {\bibinfo  {journal} {Annual Review of Condensed Matter Physics}\
  }\textbf {\bibinfo {volume} {9}},\ \bibinfo {pages} {413} (\bibinfo {year}
  {2018})}\BibitemShut {NoStop}%
\bibitem [{\citenamefont {Huber}(2016)}]{Huber2016Topological}%
  \BibitemOpen
  \bibfield  {author} {\bibinfo {author} {\bibfnamefont {S.~D.}\ \bibnamefont
  {Huber}},\ }\bibfield  {title} {\bibinfo {title} {Topological mechanics},\
  }\href {https://doi.org/10.1038/nphys3801} {\bibfield  {journal} {\bibinfo
  {journal} {Nature Physics}\ }\textbf {\bibinfo {volume} {12}},\ \bibinfo
  {pages} {621} (\bibinfo {year} {2016})}\BibitemShut {NoStop}%
\bibitem [{\citenamefont {Ori}\ \emph {et~al.}(2023)\citenamefont {Ori},
  \citenamefont {Duque}, \citenamefont {Frank~Hayward}, \citenamefont
  {Scheibner}, \citenamefont {Tian}, \citenamefont {Ortiz}, \citenamefont
  {Vitelli},\ and\ \citenamefont {Cohen}}]{Ori2023Observation}%
  \BibitemOpen
  \bibfield  {author} {\bibinfo {author} {\bibfnamefont {H.}~\bibnamefont
  {Ori}}, \bibinfo {author} {\bibfnamefont {M.}~\bibnamefont {Duque}}, \bibinfo
  {author} {\bibfnamefont {R.}~\bibnamefont {Frank~Hayward}}, \bibinfo {author}
  {\bibfnamefont {C.}~\bibnamefont {Scheibner}}, \bibinfo {author}
  {\bibfnamefont {H.}~\bibnamefont {Tian}}, \bibinfo {author} {\bibfnamefont
  {G.}~\bibnamefont {Ortiz}}, \bibinfo {author} {\bibfnamefont
  {V.}~\bibnamefont {Vitelli}},\ and\ \bibinfo {author} {\bibfnamefont {A.~E.}\
  \bibnamefont {Cohen}},\ }\bibfield  {title} {\bibinfo {title} {Observation of
  topological action potentials in engineered tissues},\ }\href
  {https://doi.org/10.1038/s41567-022-01853-z} {\bibfield  {journal} {\bibinfo
  {journal} {Nature Physics}\ }\textbf {\bibinfo {volume} {19}},\ \bibinfo
  {pages} {290} (\bibinfo {year} {2023})}\BibitemShut {NoStop}%
\bibitem [{Note1()}]{Note1}%
  \BibitemOpen
  \bibinfo {note} {Notice that Eqs.~(\ref {eq:lumber1}-\ref {eq:lumber2}) do
  not contain advective transport, which has also been shown to give rise
  oscillations near Dirichlet boundaries, for example in models of and
  experiments on \protect \emph {Dictyostelium discoideum}~\cite
  {Eckstein2020Experimental,VidalHenriquez2017Convective}.}\BibitemShut {Stop}%
\bibitem [{Note2()}]{Note2}%
  \BibitemOpen
  \bibinfo {note} {For simplicity, in this example we are using the same
  hopping rate $\epsilon $ within the forest as between the forest and desert.
  This distinction becomes irrelevant in the continuum limit (large $\epsilon $
  and large $N$), because this subextensive heterogeneity is absorbed into the
  Dirichlet boundary condition at an edge or into the continuity requirements
  across an interface.}\BibitemShut {Stop}%
\bibitem [{\citenamefont {FitzHugh}(1961)}]{fitzhugh1961impluses}%
  \BibitemOpen
  \bibfield  {author} {\bibinfo {author} {\bibfnamefont {R.}~\bibnamefont
  {FitzHugh}},\ }\bibfield  {title} {\bibinfo {title} {Impulses and
  physiological states in theoretical models of nerve membrane},\ }\href
  {https://doi.org/https://doi.org/10.1016/S0006-3495(61)86902-6} {\bibfield
  {journal} {\bibinfo  {journal} {Biophysical Journal}\ }\textbf {\bibinfo
  {volume} {1}},\ \bibinfo {pages} {445} (\bibinfo {year} {1961})}\BibitemShut
  {NoStop}%
\bibitem [{\citenamefont {Xu}\ \emph {et~al.}(2020)\citenamefont {Xu},
  \citenamefont {Lu}, \citenamefont {Gamal El-Din}, \citenamefont {Pei},
  \citenamefont {Johnson}, \citenamefont {Uyeda}, \citenamefont {Bick},
  \citenamefont {Xu}, \citenamefont {Jiang}, \citenamefont {Bai}, \citenamefont
  {Reggiano}, \citenamefont {Hsia}, \citenamefont {Brunette}, \citenamefont
  {Dou}, \citenamefont {Ma}, \citenamefont {Lynch}, \citenamefont {Boyken},
  \citenamefont {Huang}, \citenamefont {Stewart}, \citenamefont {DiMaio},
  \citenamefont {Kollman}, \citenamefont {Luisi}, \citenamefont {Matsuura},
  \citenamefont {Catterall},\ and\ \citenamefont
  {Baker}}]{Xu2020Computational}%
  \BibitemOpen
  \bibfield  {author} {\bibinfo {author} {\bibfnamefont {C.}~\bibnamefont
  {Xu}}, \bibinfo {author} {\bibfnamefont {P.}~\bibnamefont {Lu}}, \bibinfo
  {author} {\bibfnamefont {T.~M.}\ \bibnamefont {Gamal El-Din}}, \bibinfo
  {author} {\bibfnamefont {X.~Y.}\ \bibnamefont {Pei}}, \bibinfo {author}
  {\bibfnamefont {M.~C.}\ \bibnamefont {Johnson}}, \bibinfo {author}
  {\bibfnamefont {A.}~\bibnamefont {Uyeda}}, \bibinfo {author} {\bibfnamefont
  {M.~J.}\ \bibnamefont {Bick}}, \bibinfo {author} {\bibfnamefont
  {Q.}~\bibnamefont {Xu}}, \bibinfo {author} {\bibfnamefont {D.}~\bibnamefont
  {Jiang}}, \bibinfo {author} {\bibfnamefont {H.}~\bibnamefont {Bai}}, \bibinfo
  {author} {\bibfnamefont {G.}~\bibnamefont {Reggiano}}, \bibinfo {author}
  {\bibfnamefont {Y.}~\bibnamefont {Hsia}}, \bibinfo {author} {\bibfnamefont
  {T.~J.}\ \bibnamefont {Brunette}}, \bibinfo {author} {\bibfnamefont
  {J.}~\bibnamefont {Dou}}, \bibinfo {author} {\bibfnamefont {D.}~\bibnamefont
  {Ma}}, \bibinfo {author} {\bibfnamefont {E.~M.}\ \bibnamefont {Lynch}},
  \bibinfo {author} {\bibfnamefont {S.~E.}\ \bibnamefont {Boyken}}, \bibinfo
  {author} {\bibfnamefont {P.-S.}\ \bibnamefont {Huang}}, \bibinfo {author}
  {\bibfnamefont {L.}~\bibnamefont {Stewart}}, \bibinfo {author} {\bibfnamefont
  {F.}~\bibnamefont {DiMaio}}, \bibinfo {author} {\bibfnamefont {J.~M.}\
  \bibnamefont {Kollman}}, \bibinfo {author} {\bibfnamefont {B.~F.}\
  \bibnamefont {Luisi}}, \bibinfo {author} {\bibfnamefont {T.}~\bibnamefont
  {Matsuura}}, \bibinfo {author} {\bibfnamefont {W.~A.}\ \bibnamefont
  {Catterall}},\ and\ \bibinfo {author} {\bibfnamefont {D.}~\bibnamefont
  {Baker}},\ }\bibfield  {title} {\bibinfo {title} {Computational design of
  transmembrane pores},\ }\href {https://doi.org/10.1038/s41586-020-2646-5}
  {\bibfield  {journal} {\bibinfo  {journal} {Nature}\ }\textbf {\bibinfo
  {volume} {585}},\ \bibinfo {pages} {129} (\bibinfo {year}
  {2020})}\BibitemShut {NoStop}%
\bibitem [{\citenamefont {Payandeh}\ \emph {et~al.}(2011)\citenamefont
  {Payandeh}, \citenamefont {Scheuer}, \citenamefont {Zheng},\ and\
  \citenamefont {Catterall}}]{Payandeh2011crystal}%
  \BibitemOpen
  \bibfield  {author} {\bibinfo {author} {\bibfnamefont {J.}~\bibnamefont
  {Payandeh}}, \bibinfo {author} {\bibfnamefont {T.}~\bibnamefont {Scheuer}},
  \bibinfo {author} {\bibfnamefont {N.}~\bibnamefont {Zheng}},\ and\ \bibinfo
  {author} {\bibfnamefont {W.~A.}\ \bibnamefont {Catterall}},\ }\bibfield
  {title} {\bibinfo {title} {The crystal structure of a voltage-gated sodium
  channel},\ }\href {https://doi.org/10.1038/nature10238} {\bibfield  {journal}
  {\bibinfo  {journal} {Nature}\ }\textbf {\bibinfo {volume} {475}},\ \bibinfo
  {pages} {353} (\bibinfo {year} {2011})}\BibitemShut {NoStop}%
\bibitem [{\citenamefont {Conley}\ and\ \citenamefont
  {Smoller}(1983)}]{conley1983algebraic}%
  \BibitemOpen
  \bibfield  {author} {\bibinfo {author} {\bibfnamefont {C.~C.}\ \bibnamefont
  {Conley}}\ and\ \bibinfo {author} {\bibfnamefont {J.~A.}\ \bibnamefont
  {Smoller}},\ }\bibinfo {title} {Algebraic and topological invariants for
  reaction-diffusion equations},\ in\ \href
  {https://doi.org/10.1007/978-94-009-7189-9_1} {\emph {\bibinfo {booktitle}
  {Systems of Nonlinear Partial Differential Equations}}},\ \bibinfo {editor}
  {edited by\ \bibinfo {editor} {\bibfnamefont {J.~M.}\ \bibnamefont {Ball}}}\
  (\bibinfo  {publisher} {Springer Netherlands},\ \bibinfo {address}
  {Dordrecht},\ \bibinfo {year} {1983})\ pp.\ \bibinfo {pages}
  {3--24}\BibitemShut {NoStop}%
\bibitem [{\citenamefont {Mischaikow}\ and\ \citenamefont
  {Mrozek}(2002)}]{Mischaikow2002Conley}%
  \BibitemOpen
  \bibfield  {author} {\bibinfo {author} {\bibfnamefont {K.}~\bibnamefont
  {Mischaikow}}\ and\ \bibinfo {author} {\bibfnamefont {M.}~\bibnamefont
  {Mrozek}},\ }\bibfield  {title} {\bibinfo {title} {Conley index},\ }in\ \href
  {https://www.elsevier.com/books/handbook-of-dynamical-systems/fiedler/978-0-444-50168-4}
  {\emph {\bibinfo {booktitle} {Handbook of Dynamical Systems}}},\
  Vol.~\bibinfo {volume} {2},\ \bibinfo {editor} {edited by\ \bibinfo {editor}
  {\bibfnamefont {B.}~\bibnamefont {Fiedler}}}\ (\bibinfo  {publisher}
  {Elsevier Science},\ \bibinfo {year} {2002})\ Chap.~\bibinfo {chapter} {9},
  pp.\ \bibinfo {pages} {393--460}\BibitemShut {NoStop}%
\bibitem [{\citenamefont {Tyson}(1976)}]{Tyson1976Belousov}%
  \BibitemOpen
  \bibfield  {author} {\bibinfo {author} {\bibfnamefont {J.~J.}\ \bibnamefont
  {Tyson}},\ }\href@noop {} {\emph {\bibinfo {title} {The Belousov-Zhabotinskii
  reaction}}},\ Lecture notes in biomathematics\ (\bibinfo  {publisher}
  {Springer-Verlag},\ \bibinfo {address} {Berlin; New York},\ \bibinfo {year}
  {1976})\BibitemShut {NoStop}%
\bibitem [{\citenamefont {Budroni}\ \emph {et~al.}(2016)\citenamefont
  {Budroni}, \citenamefont {Lemaigre}, \citenamefont {Escala}, \citenamefont
  {Mu{\~n}uzuri},\ and\ \citenamefont {De~Wit}}]{Budroni2016spatially}%
  \BibitemOpen
  \bibfield  {author} {\bibinfo {author} {\bibfnamefont {M.~A.}\ \bibnamefont
  {Budroni}}, \bibinfo {author} {\bibfnamefont {L.}~\bibnamefont {Lemaigre}},
  \bibinfo {author} {\bibfnamefont {D.~M.}\ \bibnamefont {Escala}}, \bibinfo
  {author} {\bibfnamefont {A.~P.}\ \bibnamefont {Mu{\~n}uzuri}},\ and\ \bibinfo
  {author} {\bibfnamefont {A.}~\bibnamefont {De~Wit}},\ }\bibfield  {title}
  {\bibinfo {title} {Spatially localized chemical patterns around an a + b
  →oscillator front},\ }\href {https://doi.org/10.1021/acs.jpca.5b10802}
  {\bibfield  {journal} {\bibinfo  {journal} {The Journal of Physical Chemistry
  A}\ }\textbf {\bibinfo {volume} {120}},\ \bibinfo {pages} {851} (\bibinfo
  {year} {2016})}\BibitemShut {NoStop}%
\bibitem [{\citenamefont {D{\'u}zs}\ \emph {et~al.}(2019)\citenamefont
  {D{\'u}zs}, \citenamefont {De~Kepper},\ and\ \citenamefont
  {Szalai}}]{Duzs2019Turing}%
  \BibitemOpen
  \bibfield  {author} {\bibinfo {author} {\bibfnamefont {B.}~\bibnamefont
  {D{\'u}zs}}, \bibinfo {author} {\bibfnamefont {P.}~\bibnamefont
  {De~Kepper}},\ and\ \bibinfo {author} {\bibfnamefont {I.}~\bibnamefont
  {Szalai}},\ }\bibfield  {title} {\bibinfo {title} {Turing patterns and waves
  in closed two-layer gel reactors},\ }\href
  {https://doi.org/10.1021/acsomega.8b02997} {\bibfield  {journal} {\bibinfo
  {journal} {ACS Omega}\ }\textbf {\bibinfo {volume} {4}},\ \bibinfo {pages}
  {3213} (\bibinfo {year} {2019})}\BibitemShut {NoStop}%
\bibitem [{\citenamefont {Semenov}\ \emph {et~al.}(2016)\citenamefont
  {Semenov}, \citenamefont {Kraft}, \citenamefont {Ainla}, \citenamefont
  {Zhao}, \citenamefont {Baghbanzadeh}, \citenamefont {Campbell}, \citenamefont
  {Kang}, \citenamefont {Fox},\ and\ \citenamefont
  {Whitesides}}]{semenov2016autocatalytic}%
  \BibitemOpen
  \bibfield  {author} {\bibinfo {author} {\bibfnamefont {S.~N.}\ \bibnamefont
  {Semenov}}, \bibinfo {author} {\bibfnamefont {L.~J.}\ \bibnamefont {Kraft}},
  \bibinfo {author} {\bibfnamefont {A.}~\bibnamefont {Ainla}}, \bibinfo
  {author} {\bibfnamefont {M.}~\bibnamefont {Zhao}}, \bibinfo {author}
  {\bibfnamefont {M.}~\bibnamefont {Baghbanzadeh}}, \bibinfo {author}
  {\bibfnamefont {V.~E.}\ \bibnamefont {Campbell}}, \bibinfo {author}
  {\bibfnamefont {K.}~\bibnamefont {Kang}}, \bibinfo {author} {\bibfnamefont
  {J.~M.}\ \bibnamefont {Fox}},\ and\ \bibinfo {author} {\bibfnamefont {G.~M.}\
  \bibnamefont {Whitesides}},\ }\bibfield  {title} {\bibinfo {title}
  {Autocatalytic, bistable, oscillatory networks of biologically relevant
  organic reactions},\ }\href {https://doi.org/10.1038/nature19776} {\bibfield
  {journal} {\bibinfo  {journal} {Nature}\ }\textbf {\bibinfo {volume} {537}},\
  \bibinfo {pages} {656} (\bibinfo {year} {2016})}\BibitemShut {NoStop}%
\bibitem [{\citenamefont {Yoshida}(2010)}]{Yoshida2010Self}%
  \BibitemOpen
  \bibfield  {author} {\bibinfo {author} {\bibfnamefont {R.}~\bibnamefont
  {Yoshida}},\ }\bibfield  {title} {\bibinfo {title} {Self-oscillating gels
  driven by the belousov--zhabotinsky reaction as novel smart materials},\
  }\href {https://doi.org/https://doi.org/10.1002/adma.200904075} {\bibfield
  {journal} {\bibinfo  {journal} {Advanced Materials}\ }\textbf {\bibinfo
  {volume} {22}},\ \bibinfo {pages} {3463} (\bibinfo {year}
  {2010})}\BibitemShut {NoStop}%
\bibitem [{\citenamefont {Rabai}\ \emph {et~al.}(1989)\citenamefont {Rabai},
  \citenamefont {Kustin},\ and\ \citenamefont
  {Epstein}}]{Rabai1989systematically}%
  \BibitemOpen
  \bibfield  {author} {\bibinfo {author} {\bibfnamefont {G.}~\bibnamefont
  {Rabai}}, \bibinfo {author} {\bibfnamefont {K.}~\bibnamefont {Kustin}},\ and\
  \bibinfo {author} {\bibfnamefont {I.~R.}\ \bibnamefont {Epstein}},\
  }\bibfield  {title} {\bibinfo {title} {A systematically designed ph
  oscillator: the hydrogen peroxide-sulfite-ferrocyanide reaction in a
  continuous-flow stirred tank reactor},\ }\href
  {https://doi.org/10.1021/ja00193a018} {\bibfield  {journal} {\bibinfo
  {journal} {Journal of the American Chemical Society}\ }\textbf {\bibinfo
  {volume} {111}},\ \bibinfo {pages} {3870} (\bibinfo {year}
  {1989})}\BibitemShut {NoStop}%
\bibitem [{\citenamefont {Testa}\ \emph {et~al.}(2021)\citenamefont {Testa},
  \citenamefont {Dindo}, \citenamefont {Rebane}, \citenamefont {Nasouri},
  \citenamefont {Style}, \citenamefont {Golestanian}, \citenamefont
  {Dufresne},\ and\ \citenamefont {Laurino}}]{Testa2021Sustained}%
  \BibitemOpen
  \bibfield  {author} {\bibinfo {author} {\bibfnamefont {A.}~\bibnamefont
  {Testa}}, \bibinfo {author} {\bibfnamefont {M.}~\bibnamefont {Dindo}},
  \bibinfo {author} {\bibfnamefont {A.~A.}\ \bibnamefont {Rebane}}, \bibinfo
  {author} {\bibfnamefont {B.}~\bibnamefont {Nasouri}}, \bibinfo {author}
  {\bibfnamefont {R.~W.}\ \bibnamefont {Style}}, \bibinfo {author}
  {\bibfnamefont {R.}~\bibnamefont {Golestanian}}, \bibinfo {author}
  {\bibfnamefont {E.~R.}\ \bibnamefont {Dufresne}},\ and\ \bibinfo {author}
  {\bibfnamefont {P.}~\bibnamefont {Laurino}},\ }\bibfield  {title} {\bibinfo
  {title} {Sustained enzymatic activity and flow in crowded protein droplets},\
  }\href {https://doi.org/10.1038/s41467-021-26532-0} {\bibfield  {journal}
  {\bibinfo  {journal} {Nature Communications}\ }\textbf {\bibinfo {volume}
  {12}},\ \bibinfo {pages} {6293} (\bibinfo {year} {2021})}\BibitemShut
  {NoStop}%
\bibitem [{\citenamefont {Adamatzky}\ \emph {et~al.}(2005)\citenamefont
  {Adamatzky}, \citenamefont {De~Lacy~Costello},\ and\ \citenamefont
  {Asai}}]{adamatzky2005reaction}%
  \BibitemOpen
  \bibfield  {author} {\bibinfo {author} {\bibfnamefont {A.}~\bibnamefont
  {Adamatzky}}, \bibinfo {author} {\bibfnamefont {B.}~\bibnamefont
  {De~Lacy~Costello}},\ and\ \bibinfo {author} {\bibfnamefont {T.}~\bibnamefont
  {Asai}},\ }\href {https://books.google.com/books?id=9naAL-AkHXcC} {\emph
  {\bibinfo {title} {Reaction-Diffusion Computers}}}\ (\bibinfo  {publisher}
  {Elsevier Science},\ \bibinfo {year} {2005})\BibitemShut {NoStop}%
\bibitem [{\citenamefont {Holley}\ \emph {et~al.}(2011)\citenamefont {Holley},
  \citenamefont {Jahan}, \citenamefont {De~Lacy~Costello}, \citenamefont
  {Bull},\ and\ \citenamefont {Adamatzky}}]{Holley2011Logical}%
  \BibitemOpen
  \bibfield  {author} {\bibinfo {author} {\bibfnamefont {J.}~\bibnamefont
  {Holley}}, \bibinfo {author} {\bibfnamefont {I.}~\bibnamefont {Jahan}},
  \bibinfo {author} {\bibfnamefont {B.}~\bibnamefont {De~Lacy~Costello}},
  \bibinfo {author} {\bibfnamefont {L.}~\bibnamefont {Bull}},\ and\ \bibinfo
  {author} {\bibfnamefont {A.}~\bibnamefont {Adamatzky}},\ }\bibfield  {title}
  {\bibinfo {title} {Logical and arithmetic circuits in belousov-zhabotinsky
  encapsulated disks},\ }\href {https://doi.org/10.1103/PhysRevE.84.056110}
  {\bibfield  {journal} {\bibinfo  {journal} {Phys. Rev. E}\ }\textbf {\bibinfo
  {volume} {84}},\ \bibinfo {pages} {056110} (\bibinfo {year}
  {2011})}\BibitemShut {NoStop}%
\bibitem [{\citenamefont {T{\'o}th}\ and\ \citenamefont
  {Showalter}(1995)}]{Toth1995Logic}%
  \BibitemOpen
  \bibfield  {author} {\bibinfo {author} {\bibfnamefont {{\'A}.}~\bibnamefont
  {T{\'o}th}}\ and\ \bibinfo {author} {\bibfnamefont {K.}~\bibnamefont
  {Showalter}},\ }\bibfield  {title} {\bibinfo {title} {Logic gates in
  excitable media},\ }\href {https://doi.org/10.1063/1.469732} {\bibfield
  {journal} {\bibinfo  {journal} {The Journal of Chemical Physics}\ }\textbf
  {\bibinfo {volume} {103}},\ \bibinfo {pages} {2058} (\bibinfo {year}
  {1995})}\BibitemShut {NoStop}%
\bibitem [{\citenamefont {Steinbock}\ \emph
  {et~al.}(1995{\natexlab{b}})\citenamefont {Steinbock}, \citenamefont
  {T{\'o}th},\ and\ \citenamefont {Showalter}}]{Steinbock1995Navigating}%
  \BibitemOpen
  \bibfield  {author} {\bibinfo {author} {\bibfnamefont {O.}~\bibnamefont
  {Steinbock}}, \bibinfo {author} {\bibfnamefont {{\'A}.}~\bibnamefont
  {T{\'o}th}},\ and\ \bibinfo {author} {\bibfnamefont {K.}~\bibnamefont
  {Showalter}},\ }\bibfield  {title} {\bibinfo {title} {Navigating complex
  labyrinths: Optimal paths from chemical waves},\ }\href
  {https://doi.org/10.1126/science.267.5199.868} {\bibfield  {journal}
  {\bibinfo  {journal} {Science}\ }\textbf {\bibinfo {volume} {267}},\ \bibinfo
  {pages} {868} (\bibinfo {year} {1995}{\natexlab{b}})}\BibitemShut {NoStop}%
\bibitem [{\citenamefont {McNamara}\ \emph {et~al.}(2018)\citenamefont
  {McNamara}, \citenamefont {Dodson}, \citenamefont {Huang}, \citenamefont
  {Miller}, \citenamefont {Sandstede},\ and\ \citenamefont
  {Cohen}}]{McNamara2018Geometry}%
  \BibitemOpen
  \bibfield  {author} {\bibinfo {author} {\bibfnamefont {H.~M.}\ \bibnamefont
  {McNamara}}, \bibinfo {author} {\bibfnamefont {S.}~\bibnamefont {Dodson}},
  \bibinfo {author} {\bibfnamefont {Y.-L.}\ \bibnamefont {Huang}}, \bibinfo
  {author} {\bibfnamefont {E.~W.}\ \bibnamefont {Miller}}, \bibinfo {author}
  {\bibfnamefont {B.}~\bibnamefont {Sandstede}},\ and\ \bibinfo {author}
  {\bibfnamefont {A.~E.}\ \bibnamefont {Cohen}},\ }\bibfield  {title} {\bibinfo
  {title} {Geometry-dependent arrhythmias in electrically excitable tissues},\
  }\href {https://doi.org/https://doi.org/10.1016/j.cels.2018.08.013}
  {\bibfield  {journal} {\bibinfo  {journal} {Cell Systems}\ }\textbf {\bibinfo
  {volume} {7}},\ \bibinfo {pages} {359} (\bibinfo {year} {2018})}\BibitemShut
  {NoStop}%
\bibitem [{\citenamefont {Murray}(2013)}]{murray2013mathematical}%
  \BibitemOpen
  \bibfield  {author} {\bibinfo {author} {\bibfnamefont {J.}~\bibnamefont
  {Murray}},\ }\href {https://books.google.com/books?id=uuBHvQEACAAJ} {\emph
  {\bibinfo {title} {Mathematical Biology: I. An Introduction}}},\
  Interdisciplinary Applied Mathematics\ (\bibinfo  {publisher} {Springer New
  York},\ \bibinfo {year} {2013})\BibitemShut {NoStop}%
\bibitem [{\citenamefont {Field}\ \emph {et~al.}(1972)\citenamefont {Field},
  \citenamefont {Koros},\ and\ \citenamefont {Noyes}}]{Field1972Oscillations}%
  \BibitemOpen
  \bibfield  {author} {\bibinfo {author} {\bibfnamefont {R.~J.}\ \bibnamefont
  {Field}}, \bibinfo {author} {\bibfnamefont {E.}~\bibnamefont {Koros}},\ and\
  \bibinfo {author} {\bibfnamefont {R.~M.}\ \bibnamefont {Noyes}},\ }\bibfield
  {title} {\bibinfo {title} {Oscillations in chemical systems. {II}. {T}horough
  analysis of temporal oscillation in the bromate-cerium-malonic acid system},\
  }\href {https://doi.org/10.1021/ja00780a001} {\bibfield  {journal} {\bibinfo
  {journal} {Journal of the American Chemical Society}\ }\textbf {\bibinfo
  {volume} {94}},\ \bibinfo {pages} {8649} (\bibinfo {year}
  {1972})}\BibitemShut {NoStop}%
\bibitem [{\citenamefont {Conley}\ and\ \citenamefont
  {Sciences}(1978)}]{conley1978isolated}%
  \BibitemOpen
  \bibfield  {author} {\bibinfo {author} {\bibfnamefont {C.}~\bibnamefont
  {Conley}}\ and\ \bibinfo {author} {\bibfnamefont {C.}~\bibnamefont
  {Sciences}},\ }\href {https://books.google.com/books?id=OsmcAwAAQBAJ} {\emph
  {\bibinfo {title} {Isolated Invariant Sets and the Morse Index}}},\ \bibinfo
  {series} {Conference Board of the Mathematical Sciences Series No. 38}\ No.\
  \bibinfo {number} {no. 38}\ (\bibinfo  {publisher} {American Mathematical
  Society},\ \bibinfo {year} {1978})\BibitemShut {NoStop}%
\end{thebibliography}
\end{document}